\documentclass[acmsmall]{acmart}

\AtBeginDocument{%
  \providecommand\BibTeX{{%
    \normalfont B\kern-0.5em{\scshape i\kern-0.25em b}\kern-0.8em\TeX}}}

\setcopyright{acmcopyright}
\copyrightyear{2020}
\acmYear{2020}
\acmDOI{10.1145/1122445.1122456}

\usepackage{rotating}
\usepackage{color, colortbl}
\definecolor{Gray}{gray}{0.9}
\usepackage{multirow}
\usepackage{amssymb}
\usepackage{amsmath,bm}
\usepackage{xcolor}
\usepackage{color, colortbl}
\usepackage{graphicx}
\usepackage{subcaption}
\definecolor{light-gray}{gray}{0.85}
\definecolor{medium-gray}{gray}{0.8}
\definecolor{dark-gray}{gray}{0.75}

\definecolor{gray1}{gray}{0.85}
\definecolor{gray2}{gray}{0.8}
\definecolor{gray3}{gray}{0.75}
\definecolor{gray4}{gray}{0.7}
\definecolor{gray5}{gray}{0.65}
\definecolor{gray6}{gray}{0.6}
\definecolor{gray7}{gray}{0.55}
\usepackage{rotating}

\usepackage{array}
\newcolumntype{?}{!{\vrule width 1pt}}

\newlength{\Oldarrayrulewidth}
\newcommand{\Cline}[2]{
  \noalign{\global\setlength{\Oldarrayrulewidth}{\arrayrulewidth}}
  \noalign{\global\setlength{\arrayrulewidth}{#1}}\cline{#2}
  \noalign{\global\setlength{\arrayrulewidth}{\Oldarrayrulewidth}}
}

\begin{document}

\title{Transparency's Influence on Human-Collective Interactions}

\author{Karina A. Roundtree}
\email{roundtrk@oregonstate.edu}
\affiliation{%
  \institution{Oregon State University}
  \streetaddress{2000 SW Monroe Ave}
  \city{Corvallis}
  \state{Oregon}
  \postcode{97331}
}

\author{Jason R. Cody}
\affiliation{%
  \institution{United States Military Academy}
  \city{West Point}
  \state{New York}}
  
\author{Jennifer Leaf}
\affiliation{%
  \institution{Oregon State University}
  \city{Corvallis}
  \state{Oregon}}

\author{H. Onan Demirel}
\affiliation{%
  \institution{Oregon State University}
  \city{Corvallis}
  \state{Oregon}}

\author{Julie A. Adams}
\affiliation{%
  \institution{Oregon State University}
  \city{Corvallis}
  \state{Oregon}}

\renewcommand{\shortauthors}{Roundtree, Cody, Leaf, Demirel, and Adams}

\begin{abstract}
Collective robotic systems are biologically inspired and advantageous due to their apparent global intelligence and emergent behaviors. Many applications can benefit from the incorporation of collectives, including environmental monitoring, disaster response missions, and infrastructure support. Transparency research has primarily focused on how the design of the models, visualizations, and control mechanisms influence human-collective interactions. Traditionally most evaluations have focused only on one particular system design element, evaluating its respective transparency. This manuscript analyzed two models and visualizations to understand how the system design elements impacted human-collective interactions, to quantify which model and visualization combination provided the best transparency, and provide design guidance, based on remote supervision of collectives. The consensus decision-making and baseline models, as well as an individual agent and abstract visualizations, were analyzed for sequential best-of-n decision-making tasks involving four collectives, composed of 200 entities each. Both models and visualizations provided transparency and influenced human-collective interactions differently. No single combination provided the best transparency.
\end{abstract}




\maketitle

\section{Introduction}

Few evaluations have investigated how transparency, the principle of providing easily exchangeable information to enhance comprehension \cite{Roundtree2019}, influences interactions and behaviors between human operators and spatial swarms (6 evaluations) or colonies (1 evaluation). This manuscript's first objective was to expand the existing transparency literature by assessing how different models (i.e., algorithms) and visualizations influence human-collective interactions and behavior. Collective robotic systems, which are composed of many simple individual entities, exhibit biological behaviors found in spatial swarms \cite{Brambilla2013}, colonies \cite{Gordon1999}, or a combination of both \cite{Seeley2010}. Understanding the influence of system design elements, such as the models \citep{Cody2020}, visualizations \citep{Roundtree2020visual}, and operator control mechanisms, on human-collective team interactions is necessary to ensure desired outcomes (e.g., high performance). Integrating transparency into the system design elements, can mitigate poor operator behaviors, help attain meaningful and insightful information exchanges between the operator and collective, and improve the human-collective's overall effectiveness.

Many of the transparency evaluations that have assessed human-collective interactions and behavior have only focused on the influence of one system design element (e.g., control mechanism). Implementing the best of these identified system design elements together in one collective system design may not always yield optimal results. The human-collective system may become less transparent due to unanticipated and undesired operator behavior that results from the combined system design elements. This manuscript's second objective was to determine whether using the best model and visualization, identified from two previous evaluations \citep{Cody2020,Roundtree2020visual}, provided the best transparency. Understanding how design elements interact with one another to influence human-collective behavior is necessary in order to determine how to quantify transparency. 

The task evaluated in this manuscript was a sequential best-of-\textit{n} decision making problem, similar to that of a bee colony searching for a new hive. A subset of the colony leave the hive and fly to a nearby tree branch, where they wait while scout bees search for a new hive location \cite{Seeley2010}. The bees exhibit spatial swarm behaviors during the initial flight, similar to those found in schools of fish \cite{Couzin2005} and flocks of birds \cite{BallerinietalNAS2008}. Scout bees find possible hive locations and evaluate each option with respect to ideal hive criteria. The scouts return to the waiting colony in order to begin a selection process (i.e., colony behavior) entailing debate and building consensus on the best hive location (i.e., best-of-\textit{n} \cite{Valentini2017}). After completing a consensus decision-making process, the bees travel to the new hive, transitioning from colony based behaviors back to spatial swarm behaviors.

Adding an operator, who may possess information that a collective does not, can positively influence the time to make decisions and ensure the collective selects a higher valued option. The operator's ability to influence the collective's behavior positively will rely on interacting with a transparent system that enables the operator to perceive accurately the collective's state, comprehend what the collective is doing currently, and plans to do in the future. Transparency provided to a human supervisor \citep{Scholtz2003}, was analyzed using two models \citep{Cody2018}, one designed for sequential best-of-\textit{n} decision making task and another that served as a baseline behavior model, as well as two visualizations, a traditional collective representation, Individual Agents \citep{Roundtree20191}, and an abstract Collective \citep{Cody2018} representation. The single human operator-collective system incorporated four hub-based collectives, each tasked with making a series of sequential best-of-\textit{n} decisions \citep{Valentini2017}. Focusing on the model and visualization are necessary, when the means of communicating and interacting with remote collectives will only occur via an interface. Understanding how the design of the model and visualization impact the operator's ability to positively influence the collective's decision-making process will aid this manuscript's third objective, which was to provide additional design guidelines to achieve transparency in human-collective systems.

This manuscript expands on the results of two previous analyses. The first evaluation investigated the performance of two best-of-\textit{n} models, including a new model that compensated for environment bias \citep{Cody2020}, and a baseline model, with and without the influence of a single operator. The second evaluation investigated how different visualizations of the collectives impacted operators using the best-of-\textit{n} decision making model with environmental bias and a baseline model from the first evaluation \citep{Roundtree2020visual}. The assessment of transparency considered the impact on individuals with different capabilities, operator comprehension, usability, and human-collective performance.

Section 2 provides definitions and background information related to collective systems and transparency. 
Section 3 explains the sequential best-of-\textit{n} decision making task, models, and visualizations. The experimental design and procedure are outlined in Section 4. Sections 5-8 present the hypotheses, metrics, results, and discussion for each respective research question. An overall discussion and conclusion are provided in Sections 9 and 10.

\section{Related Work}

Behaviors of spatial swarms and colonies, which constitute collective behaviors, are provided in order to develop an understanding of what collective characteristics may be important to a human teammate and collective system designers. Understanding how collective entities communicate and interact with one another to influence individual entity and global collective state changes is necessary to ground collective system design. A review of transparency research focused on designing and evaluating collective system design elements and understanding their influence on human-collective interactions is also presented. Factors that affect transparency, or are influenced by transparency, such as explainability, usability, and performance, can be measured to assess the models and visualizations. Understanding the transparency factors and how they influence the human-collective system is necessary in order to inform design decisions.  

\subsection{Collective Behavior}

Collectives exhibit biological behaviors found in spatial swarms \cite{Brambilla2013}, colonies \cite{Gordon1999}, or a combination of both \cite{Seeley2010}. Spatial swarm robot systems are inspired by self-organized social animals (e.g., schools of fish) \cite{Brambilla2013}, and exhibit intelligent, emergent behaviors as a unit, by responding to locally available information \cite{Sahin2005}. Spatial swarms rely on distributed, localized, and often implicit communication \citep{StrandburgetalCurrBio2013, Haque2017}. Basic rules of repulsion, attraction, and orientation enable individual spatial swarm entities to position themselves relative to neighboring entities \cite{Couzin2002,Aoki1982}. Robotic colony entities exhibit unique roles, such as foraging, which adapt over time to maintain consistent states in changing conditions \cite{Wilson1984}. Colonies share information in a centralized location, such as honeybees inside a nest \citep{Seeley2010}. Positive feedback loops support gaining a consensus to change the colony's behavior \cite{Sumpter2006} and negative feedback mitigates saturation issues, such as food source exhaustion \cite{Bonabeau1999}. More detailed information about collective behavior is provided in Cody \textit{et al.} \citep{Cody2020} and Roundtree \textit{et al.} \citep{Roundtree2020visual}.

\subsection{Collective System Transparency and Influence on Human-Collective Interactions}

Many of the existing transparency evaluations investigated how control mechanisms influenced human-spatial swarm interactions and behavior (e.g., \citep{Kolling2012,Jung2013}). Two mechanisms were used to control a spatial swarm foraging in simple and complex environments \citep{Kolling2012,Kolling2013}. Selection, influenced a selected subgroup, and beacon, exerted influence on entities within a set range. The highest performance occurred when fully autonomous spatial swarms (i.e., no operator influence) foraged in simple environments, while selection was optimal in complex environments \citep{Kolling2012}. Selection generally outperformed beacon; however, as the spatial swarm size increased, beacon became more advantageous by requiring less operator influence \citep{Kolling2013}. Improvements must be considered in order to reduce the learning curve of using beacon and improve it's effectiveness (i.e., learning where to strategically place beacons). Leader, predator, and mediator control mechanisms were assessed, with regard to spatial swarm manageability and performance \citep{Jung2013}. Leaders attracted entities towards them, predators repelled entities away, and mediators allowed the operator to mold and adapt the spatial swarm. Operators experienced different workload levels and implemented different control strategies depending on the control mechanism. Workload increased when using leaders, decreased with predators, and remained relatively stable with mediators. Operators using leaders gathered all of the spatial swarm entities together and guided them in a particular direction. Spatial subswarms emerged and were pushed in different directions when the operators used predators. Mediators were strategically placed in the environment, which resulted in lower workload, suggesting that this control mechanism may be easier to use. The quantity and quality of operator influence was investigated to identify when the influence begins to have a detrimental effect on human-spatial swarm performance \citep{Walker2013}. Operators moved a spatial swarm around in a variety of environments, at two levels of autonomy using an autonomous dispersion algorithm (high autonomy) and user-defined goto points (low autonomy). Operator influence was required in complex environments containing numerous obstacles and small passageways; however, too much control never allowed the automation to operate, resulting in a performance decline. Two operator interaction strategies emerged: (1) allow the autonomous algorithm to control spatial swarm movement or (2) manually break the spatial swarm into subgroups and guide them to explore different areas of the map.

Only one colony based evaluation assessed operator influence level and information reliability during a best-of-n decision making task \citep{Ashcraft2019}. Operators placed beacons in the environment to attract support in particular locations. The direction of the entities was communicated to operators using a radial display surrounding the hub. Low operator influence resulted in high performance when reliable information was provided, while high influence was best when inaccurate or incomplete site information was provided. Further analysis is required to determine if less operator influence can be achieved when there is imperfect communication. Additional human-colony based system evaluations are also needed to establish a broader understanding of the influence of control mechanisms on human-colony interactions.

Two evaluations assessed the influence of visualizations on human-spatial swarm interactions and behavior. Four methods of displaying current spatial swarm information were assessed based on the operator's ability to predict the spatial swarm's future state \citep{Walker2016}. The full information display showed the position and heading of each individual entity, the centroid/ellipse showed a bounding ellipse at the center of the spatial swarm, the minimum volume enclosing ellipse showed leaders at the edge of the spatial swarm, and random condition clustering showed leaders evenly spaced throughout the spatial swarm. The full information and centroid/ellipse displays enabled the most accurate predictions when estimating spatial qualities, with a preference for the bounding ellipse in low bandwidth situations. The leader-based strategies may be more advantageous for other tasks that have a particular goal, such as the best-of-\textit{n} decision making task. A metacognition model that enabled individual entities to monitor changes in the spatial swarm's state and a visualization that communicated spatial swarm status during a convoy mission were assessed when information was provided in different modalities (spatial, audio, and tactile cues) in order to increase situational awareness of surroundings and improve visual attention \citep{Haas2009}. The primary task required monitoring the spatial swarm and responding to display signals, while performing a secondary robotic planning task. The visualization enabled 99.9\% accuracy of signal detection and recognition. 

Transparency embedded in a traditional visualization, which showed all of the individual entities composing a collective, and an abstract visualization, was evaluated for a single human operator-collective team performing a sequential best-of-\textit{n} decision making task \citep{Roundtree2020visual}. Transparency was assessed by understanding how the visualization impacted operators with different individual capabilities, their comprehension of the information, the usability of the interface, and the human-collective team's performance. The abstract visualization provided better transparency compared to the traditional visualization, because it enabled operators with individual differences and capabilities to perform relatively the same and promoted higher human-collective performance. Additional comprehension and interaction metrics were needed to better assess how the visualization's transparency influenced operator comprehension and system usability. The same abstract visualization was used in an evaluation assessing how different models (2 sequential best-of-\textit{n} decision making models and 1 baseline model) influenced performance with and without a human operator \citep{Cody2020}. The sequential best-of-n decision making model, that compensated for environmental bias, without an operator performed slower, but made 57\% more accurate hard decisions compared the sequential best-of-n model that only assessed a target's value. The addition of an operator using the environmental bias compensated model required less operator influence and achieved 25\% higher accuracy for the hard decisions. Further transparency analysis is needed to determine how models and visualizations influence human-collective interactions and which combination of system design elements promote better transparency with respect to transparency factors. 

\subsection{Transparency Factors}

\begin{figure}[!b]
\centering
	\includegraphics[width=\textwidth]{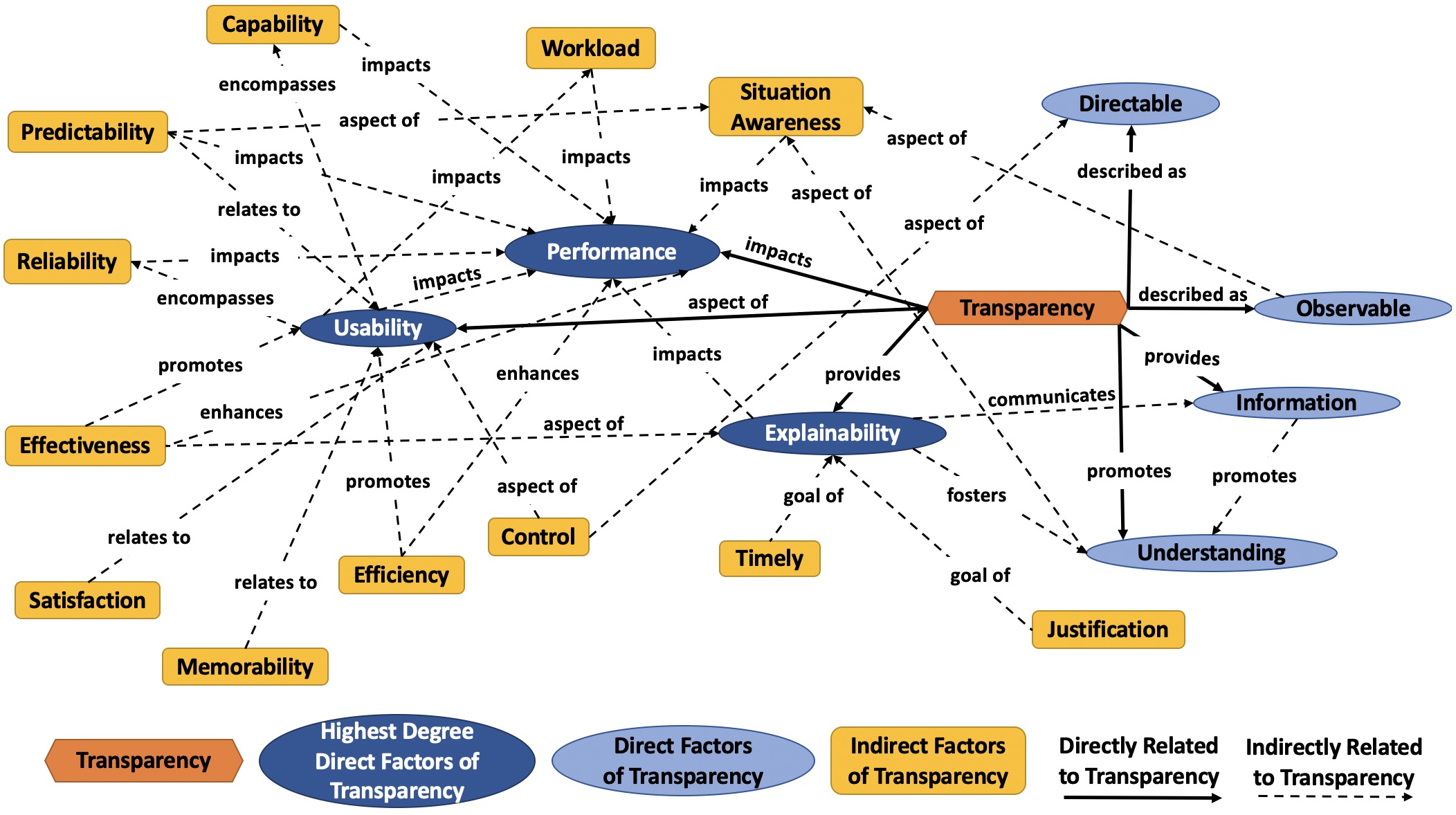}
	\caption{Concept map showing a subset of direct and indirect transparency factors \citep{Roundtree2019} used to assess the influence of transparency embedded in the models and visualizations on human-collective interactions. The indirect control and reliability factors were added from the human-collective interactions literature, and the memorability factor was added in order to assess a human-collective interaction metric.}
    \label{fig: Concept Map}
\end{figure}

Transparency is the principle of providing information that is easy to use in an exchange between human operators' and collectives to promote comprehension of intent, roles (e.g., decision-maker versus information gatherer), interactions, performance, future plans, and reasoning processes \citep{Roundtree2019}. The term principle describes the results of a process of identifying what factors affect and are influenced by transparency, why those factors are important, how the factors may influence one another, and how to design a system to achieve transparency. A subset of human-collective transparency factors \citep{Roundtree2019}, in Figure \ref{fig: Concept Map}, are used to assess the influence of transparency for the different models and visualizations on human-collective interactions in this manuscript. Seven factors impact transparency directly, shown as the blue ovals. The three highest total degree (number of in degree + number of out degree) direct factors are explainability, usability, and performance (dark blue). Information and understanding (light blue) are not considered high degree factors, because \textit{explainability} uses information to communicate and promote understanding. Observable, the ability to be perceived, and directable, the operator's ability to guide problem solving \cite{Chenetal2014}, were not considered high degree factors due to the low number of in and out connections.  Explainability and \textit{usability}, a multifaceted quality that influences the operator's perception of a system, are used for the implementation of transparency in the presented models and visualizations. \textit{Performance} can be used to assess the influence of the models and visualizations by determining how well the human-collective team was able to produce an output when executing a task \citep{Alarcon2017}.

Many factors that impact transparency are embedded in the different models or visualizations indirectly, the yellow rectangles. The timing and quantity of information visualized on an interface, such as collective status or feedback, requires considering the human operator's capability limitations \citep{Entin1996}, system limitations, as well as task and environmental impacts, in order to be explainable \citep{Atoyan2006}. Human-collective efficiency and effectiveness can be improved by enabling both the operator and collective to control particular aspects of the decision-making process, via the model or control mechanisms (evaluated in the existing human-collective interaction transparency literature). Visualizing information, such as predicted collective states, in a clear and cohesive manner that helps alleviate the time and effort an operator must exert when integrating information in order to draw conclusions \citep{StJohn2005} and justify particular actions \citep{Fox2017} is another strategy to promote efficiency and effectiveness. Poor judgements may occur if the operator lacks an understanding of the model's reliability (evaluated in the existing human-collective interaction transparency literature), due to inadequate training prior to interacting with the system, or the model is not memorable. The visualization usability may also contribute to negative behavior by hindering the operator's accurate perception and comprehension of what the collective is currently doing and predictions of future collective behavior. Poor judgements and human-collective interactions may cause operator dissatisfaction. Models (i.e., algorithms) that are not designed to take advantage of the operator's and collective's strengths to achieve a task, visualizations that do not provide needed information, and control mechanisms that do not promote effective influence over collective behavior will hinder human-collective performance, lower operator situational awareness, impact workload negatively, and potentially compromise the safety of the human-collective team. Understanding the relationships between the direct and indirect factors, and their relationship to transparency, is needed in order to assess metrics that can quantify how the transparency embedded in different models and visualizations influence human-collective interactions.

\section{Human-Collective Task}

A single operator supervised and assisted four collectives of 200 simulated Unmanned Aerial Vehicles each performing a sequential best-of-\textit{n} decision-making task. The human-collective team was to choose the best option from a finite set of \textit{n} options \cite{Cody2020} and performed two sequential decisions per collective (moved the collective to a new hub site). The decision-making task required identifying and selecting the highest valued target within a 500 \textit{m} range of the current hub, the collective hub moved to the selected target, and initiated the second target selection decision. 

The four collective hubs were visible at the start of each trial. Targets became visible as each was discovered by a collective's entities. The target's value was assessed by the collectives' entities, who returned to their respective hub to report the target location and value. The collectives were only allowed to discover and occupy targets within their search range, but some targets were within proximity of multiple hubs. A collective's designated search area changed after establishing at a new hub site. The operator was to prevent merging of multiple collectives by not permitting their respective hubs to move to the same target. When a collective moved to a target, the hub moved to the target location, and the target was no longer visible to the operator or available to other collectives. The collective that moved its hub to a target's location first, when two collectives were investigating the same target, established its hub, while the second collective returned to its previous hub location. Both collectives made a decision when a potential merge occurred, even though only one collective moved its hub to the selected target's location.

Two models were used. A sequential best-of-\textit{n} decision-making model ($M_{2}$) adapted an existing model ($M_{1}$), which based decisions on the target's quality (i.e., value) \cite{Reina2015}. Information exchanges between a collective's entities were restricted to occur inside the hub. Episodic queuing cleared messages when the collective entities transitioned to different states, which resulted in more successful and faster decision completion. Interaction delay and interaction frequency were added as bias reduction methods in order to consider a target's distance from the collective hub and increase interactions among the collectives' entities. Interaction delay improved the success of choosing the ground truth best targets, and interaction frequency improved decision time. The baseline model ($M_{3}$) allowed the collective entities to search and investigate potential targets, but the operator was required to influence the consensus-building element and select the final target. 

The interface control mechanisms allowed the operator to alter the collectives' internal states, including their levels of autonomy, throughout the sequential best-of-\textit{n} selection process. The collective's entities were in one of four states. \textit{Uncommitted} entities explored the environment searching for targets, and were recruited by other entities while inside of the collective's hub. Collective entities that \textit{favored} a target reassessed the target's value periodically, and attempted to recruit other entities within the collective's hub to investigate the specified target. Collective entities were \textit{committed} to a particular target once a quorum of support was detected, or after interacting with another committed entity. \textit{Executing} entities moved from the collective's current hub location to the selected target's location. A collective operated at a high level of autonomy by executing actions associated with potential targets independently. The operator was able to influence the collective's actions in order to aid better decision-making, effectively lowering the level of autonomy. Communication from the operator with the collective's entities occurred inside the hub in order to simulate limited real-world communication capabilities. The commands, for influencing the collective, were communicated to the specified hub. 

Two visualizations were used. The Individual Agents (IA) interface displayed the location of all the individual collective entities \citep{Roundtree20191} on a central map, along with the respective hubs, discovered targets, and other associated information. A static central map was used to provide ecological validity and emulate a task searching an urban environment for potential locations of interest. Understanding the environment's topography is necessary for identifying what type of vehicles (e.g., air versus ground vehicles) will be most effective at completing a task, depending on the environmental conditions. The discovered targets were initially white and transitioned to green when at least two individual collective entities evaluated the target. The highest valued targets were a bright opaque green, while lower valued targets had a more translucent green color. Targets within the collective's 500 \textit{m} search range had different colored outlines, depending on the collective's state: explored, but not currently favored; explored and favored; and abandoned targets.

The individual collective entities began each trial by exploring the environment in an uncommitted state, which transitioned to favoring as targets were assessed and supported. The collective committed to a target when 30\% of the collective (60 individual entities) favored it. The collective moved to the selected target's location once 50\% of the collective (100 entities) favored the target. The individual collective entities' state information, uncommitted, favoring, committed, and executing, was conveyed via individual collective entity color coding. The number of individual collective entities in a particular state, or supporting a target was provided via the collective hub and target information pop-up displays that appeared relative to the respective collective's hub or target. The operator was able to move the information displays by dragging the pop-up display.

The operator had the ability to influence an individual collectives' current state via a collective request. The \textit{investigate} command increased a collective's support for an operator specified target by transitioning uncommitted entities (5\% of the population) to the favoring state. Additional support for the same target was achieved by reissuing the investigate command repeatedly. The \textit{abandon} command reduced a collective's support for a specific target by transitioning favoring individual entities to the uncommitted state and only needed to be issued once for the collective to ignore a specified target. A collective's entities stopped exploring alternative targets and moved to the operator selected target when the \textit{decide} command was issued, which was valid when at least 30\% of the collective supported the specified target. An operator was unable to further influence a collective once the decide command was issued.

The collective assignments section logged the operator's issued commands with respect to particular collectives and indicated if the command was active or completed. Once a collective reached a decision, all prior commands were removed from the collective assignments log. The operator was only able to cancel an abandon command. Illegal messages were displayed in the system messages area and occurred when an operator requested an invalid command, which arouse when the operator attempted to issue an investigate command for targets outside of the collective's search region; abandon newly discovered targets that did not have an assigned value (white targets); and issue decide commands when less than 30\% of the collective supported a target. 

The Collective interface provided an abstract visualization that does not represent individual collectives' entities \citep{Cody2018}, and has the same three primary areas as the IA interface. The operator commands were and functioned the same as those in the IA interface. The collectives were represented as rectangles with four quadrants representing the collectives' states (uncommitted (U), favoring (F), committed (C), and executing (X)) and used a brighter white quadrant for a larger number of individual collective entities in a particular state. Targets contained two sections, where the top green section represented the target's value (brighter and more opaque the green, the higher the value) and a bottom blue section indicated the number of individual entities favoring a particular target (brighter and more opaque the blue, the higher the number of collective entities). The Collective interface operated similarly to the IA interface; however, the collective's outline moved from the hub to the target's location to indicate the hub's transition to the selected target. More detailed information about the human-collective task, as well as the IA and Collective interfaces are provided in Cody \textit{et al.} \citep{Cody2020} and Roundtree \textit{et al.} \citep{Roundtree2020visual}. 

\section{Experimental Design}

The primary research question for the within-model and between-visualization analyses was related to the manuscript's second objective: \textit{which model and visualization combination achieved better transparency?} Four secondary questions were developed to investigate how the model and visualization impacted the highest degree direct transparency factors in Figure \ref{fig: Concept Map}. The first research question ($R_{1}$) focused on understanding \textit{how the model and visualization influenced the operator}. Individual differences, such as spatial capability, will impact an operator's ability to interact with the interface effectively and cause different responses (i.e., loss of situational awareness or more workload). A model and visualization that can aid operators with different capabilities are desired. The explainability factor was encompassed in $R_{2}$, which explored whether \textit{the model and visualization promoted operator comprehension}. Perception and comprehension of the visualized information are necessary to inform future actions. Understanding \textit{which model and visualization promoted better usability}, $R_{3}$, will aid designers in promoting effective transparency in human-collective systems. The final research question, $R_{4}$, assessed \textit{which model and visualization promoted better human-collective performance}. A system that performs a task quickly and accurately is ideal. 

The independent variables included the within model variable, $M_{2}$ and $M_{3}$, the between visualization variable, IA versus Collective, and the trial difficulty (overall, easy, and hard). Trials that had a larger number of high valued targets in closer proximity to a collective's hub were deemed \textit{easy}, while \textit{hard} trials placed high valued targets further away from the hub. The dependent variable details are embedded into the sections associated with each research question. 

\subsection{Experimental Procedure}

The experimental procedure required participants to complete a demographic questionnaire and a Mental Rotations test \cite{Vandenberg1978}. The IA participants completed an additional Working Memory Capacity assessment. Upon completion of the demographic data collection, participants received training and practiced using their respective interface. Two practice sessions occurred prior to each trial in order to ensure familiarity with the underlying sequential best-of-\textit{n} ($M_{2}$) and baseline $M_{3}$ models. The $M_{2}$ model trial was always completed first in the IA evaluation, in order to alleviate any learning effects from using the $M_{3}$ model. The Collective evaluation randomized the order of the $M_{1}$ and $M_{2}$ models, which were always presented before the $M_{3}$ model. The participants were instructed that the objective was to aid each collective in selecting and moving to the highest valued target two sequential times. A trial began once the practice session was completed. Each trial was divided into two components (one easy and one hard) of approximately ten minutes each. The simulation environment was reset between the components with 16 new (not initially visible) targets. The easy and hard trial orderings were randomly assigned, and counterbalanced across the participants. The situational awareness (SA) probe questions \cite{Cody2018}, intended to serve as a secondary task, were asked beginning at 50 seconds into the trial and were repeated at one-minute increments. Six SA probes were asked during each trial component, or twelve per trial. The trial ended after eight decisions, two per collective, or once six decisions were made, if the trial length exceeded the ten-minute limit. Decision times were not limited. A post-trial questionnaire was completed after each trial and the post-experiment questionnaire was completed before the evaluation termination. 

\subsection{Participants}

The demographic questionnaire collected information regarding age, gender, education level, and \textit{weekly hours on a desktop or laptop} (0, less than 3, 3-8, and more than 8). The \textit{Mental Rotation Assessment} \cite{Vandenberg1978} required participants to judge three-dimensional object orientation to assess spatial reasoning within a scoring range of 0 (low) to 24 (high). The \textit{Working Memory Capacity} assessment, only completed by IA participants, evaluated higher-order cognitive task performance \citep{Engle2002}. 

Fourteen females and nineteen males (33 total) completed the IA evaluation at Oregon State University. Five participants were excluded due to inconsistent methodology (1) and software failure (4). The mean weekly hours spent on a desktop or laptop was 3.79, with a standard deviation (SD) = 0.5, median = 4, minimum (min) = 2, and maximum (max) = 4. The Mental Rotation Assessment \cite{Vandenberg1978} mean was 12.36 (SD = 5.85, median = 12, min = 3, and max = 24) \cite{Roundtree20191}. The Working Memory Capacity mean was 86.14 (SD = 9.73, median = 89.5, min = 59, and max = 98) \cite{Roundtree20191}. 

Twenty-eight participants, 15 females and 13 males, from Vanderbilt University, completed the Collective evaluation. The weekly hours spent on a desktop or laptop was slightly higher than the IA participants (mean = 3.86, SD = 0.45, median = 4, min = 2, and max = 4) and the Mental Rotations Assessment was slightly lower (mean = 10.93, SD = 5.58, median = 10, min = 1, and max = 24) \cite{Roundtree20191}.

\subsection{Analysis}

The mixed analysis is based on 56 participants from both evaluations. The first twelve decisions made per participant using each model were analyzed. The majority of the objective metrics were analyzed by SA level (overall ($SA_{O}$), perception ($SA_{1}$), comprehension ($SA_{2}$), and projection ($SA_{3}$)), decision difficulty (overall, easy, and hard), timing with respect to a SA probe question (15 seconds before asking, while being asked, or during response to a SA probe question), or per participant. Non-parametric statistical methods, including Mann-Whitney-Wilcoxon tests with one degree of freedom (DOF = 1) and Spearman Correlations, were calculated due to a lack of normality. The correlations were with respect to SA Probe Accuracy and Selection Success Rate. The Collective evaluation data was reanalyzed using the same methods. Secondary research question's hypotheses, associated metrics, results, and discussion are presented for each research question, Sections 5-8.  

\section{$R_{1}$: System Design Element Influence on Human Operator}

\begin{figure}[h]
\begin{center}
	\includegraphics[width=\textwidth]{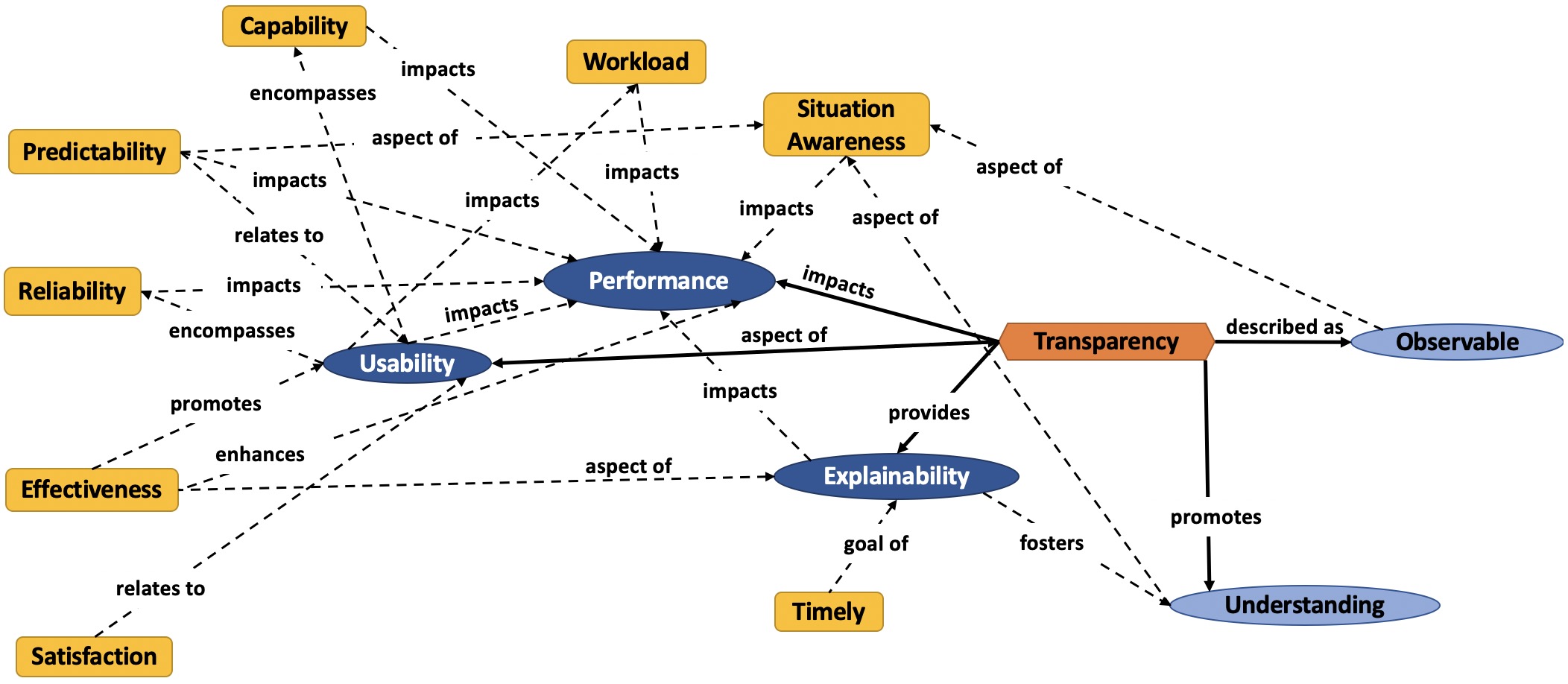}
	\caption{$R_{1}$ concept map of the assessed direct and indirect transparency factors.}
	\label{fig: Model Vis Concept Map R5}
	\end{center}
\end{figure}

Understanding \textit{how the model and visualization influenced the operator}, $R_{1}$, is necessary to determine if the transparency embedded into the system design aided operators with different capabilities. The associated objective dependent variables were (1) the operator's ability to influence the collective in order to choose the highest valued target, (2) SA, (3) visualization clutter, (4) the operator's spatial reasoning capability, and (5) the operator's working memory capacity. The specific direct and indirect transparency factors related to $R_{1}$ are identified in Figure \ref{fig: Model Vis Concept Map R5}. The relationship between the variables and the corresponding hypotheses, as well as the direct and indirect transparency factors, are identified in Table \ref{table:Impacts,Variables}. Additional relationships (not identified in Figure \ref{fig: Concept Map}) between the variable and transparency factors are identified due to correlation analyses.

\begin{table}[!t]
\centering
\caption{Interaction of system design elements influence on the human operator objective (obj) and subjective (subj) variables (vars), relationship to the hypotheses (H), as well as the associated direct and indirect transparency factors, as shown in Figure \ref{fig: Concept Map}.}
\label{table:Impacts,Variables}
\begin{tabular}{?l|c?c|c|c|c|c?c|c|c|c|c|c|c|c?}
\Cline{1pt}{3-15}
\multicolumn{1}{c}{} & & \multicolumn{13}{c?}{\textbf{Transparency Factors}} \\ \Cline{1pt}{3-15}
\multicolumn{1}{c}{} & & \multicolumn{5}{c?}{\textbf{Direct}} & \multicolumn{8}{c?}{\textbf{Indirect}} \\ \Cline{1pt}{3-15}
\multicolumn{1}{c}{} & & {\multirow[b]{6}{*}{\rotatebox{90}{\textbf{Explainability}}}} & {\multirow[b]{6}{*}{\rotatebox{90}{\textbf{Observable}}}} & {\multirow[b]{6}{*}{\rotatebox{90}{\textbf{Performance}}}} & {\multirow[b]{6}{*}{\rotatebox{90}{\textbf{Understanding}}}} & {\multirow[b]{6}{*}{\rotatebox{90}{\textbf{Usability}}}} & {\multirow[b]{6}{*}{\rotatebox{90}{\textbf{Capability}}}} & {\multirow[b]{6}{*}{\rotatebox{90}{\textbf{Effectiveness}}}} & {\multirow[b]{6}{*}{\rotatebox{90}{\textbf{Predictability}}}} & {\multirow[b]{6}{*}{\rotatebox{90}{\textbf{Reliability}}}} &  {\multirow[b]{6}{*}{\rotatebox{90}{\textbf{SA}}}} & {\multirow[b]{6}{*}{\rotatebox{90}{\textbf{Satisfaction}}}} & {\multirow[b]{6}{*}{\rotatebox{90}{\textbf{Timing}}}} &  {\multirow[b]{6}{*}{\rotatebox{90}{\textbf{Workload}}}} \\ 
\multicolumn{1}{c}{} & & & & & & & & & & & & & & \\ 
\multicolumn{1}{c}{} & & & & & & & & & & & & & & \\ 
\multicolumn{1}{c}{} & & & & & & & & & & & & & & \\ 
\multicolumn{1}{c}{} & & & & & & & & & & & & & & \\ \Cline{1pt}{1-2}
\multicolumn{1}{?c|}{\textbf{Obj Vars}} & {\textbf{H}} & & & & & & & & & & & & & \\ \Cline{1pt}{1-15}
{Target Value} & $H_{1}$ & & & & & {\checkmark} & & & & & & & & \\ \hline
{SA Probe Accuracy} & {$H_{1}$} & {\checkmark} & {\checkmark} & {\checkmark} & {\checkmark} & & & {\checkmark} & {\checkmark} & & {\checkmark} & & & \\ \hline
{Global Clutter} & {$H_{1}$} & {\checkmark} & {\checkmark} & {\checkmark} & {\checkmark} & {\checkmark} & & {\checkmark} & {\checkmark} & & {\checkmark} & & & \\ \hline
{Mental Rotations} & \multirow{2}{*}{$H_{2}$} & & & \multirow{2}{*}{\checkmark} & & & \multirow{2}{*}{\checkmark} & & & & & & & \\
{Assessment} & & & & & & & & & & & & & & \\ \hline
{Working Memory} & \multirow{2}{*}{$H_{2}$} & & & \multirow{2}{*}{\checkmark} & & & \multirow{2}{*}{\checkmark} & & & & & & & \\ 
{Capacity} & & & & & & & & & & & & & & \\ \Cline{1pt}{1-15}
\multicolumn{1}{?c}{\textbf{Subj Vars}} & \multicolumn{14}{c?}{\textbf{}} \\ \Cline{1pt}{1-15}
{Weekly Hours on a} & \multirow{2}{*}{$H_{2}$} & & & \multirow{2}{*}{\checkmark} & & & \multirow{2}{*}{\checkmark} & & & & & & & \\
{Desktop or Laptop} & & & & & & & & & & & & & & \\ \hline
{NASA-TLX} & $H_{1}$,$H_{3}$ & {\checkmark} & & {\checkmark} & & & {\checkmark} & & & & & {\checkmark} & {\checkmark} & {\checkmark} \\ \hline
{Post-Experiment} & {$H_{2}$} & & & & {\checkmark} & {\checkmark} & {\checkmark} & & & {\checkmark} & & & {\checkmark} & \\ \Cline{1pt}{1-15}
\end{tabular}
\end{table}

The hypotheses in this section and the subsequent result Sections 6-8 are phrased using the $M_{2}$ model and Collective visualization, because each individual system design element provided the best transparency in their respective evaluation \cite{Cody2020,Roundtree2020visual}. Operators may have performed differently depending on their individual differences. It was hypothesized ($H_{1}$) that operators using the $M_{2}$ model and Collective visualization will experience significantly higher SA and lower workload. SA represents an operator's ability to perceive and comprehend information in order to project future actions that must be taken in order to fulfill a task \citep{Endsley1995}. Usability influences the perception of information \citep{Roundtree2019} and will impact workload, which is the amount of stress an operator experiences in order to accomplish a task in a particular duration of time \cite{Wickens2004}. It was hypothesized ($H_{2}$) that operators with different individual capabilities will not perform significantly different using the $M_{2}$ model and Collective visualization. Ideal system design elements will enable operators with different capabilities to perceive, comprehend, and influence collectives relatively the same. The operator's attitude and sentiments towards a system, which is dependent on system usability, provides essential information related to the system's design \cite{Kizilcec2016}. Good designs promote higher operator satisfaction. It was hypothesized ($H_{3}$) that operators using the $M_{2}$ model and Collective visualization will experience significantly less frustration (i.e., higher satisfaction).

\subsection{Metrics and Results}
\label{section:R1 metrics}

\begin{table}[!b]
\begin{minipage}{0.5\linewidth}
\centering
\caption{Selected target value mean (SD) by decision difficulty (Dec Diff), where the maximum possible value was 100 and the minimum possible value was 67.}
\label{table:Impacts,Target Value}
\begin{tabular}{c|c|c|c|}
\cline{2-4}
 & \textbf{Dec Diff} & \textbf{IA} & \textbf{Collective} \\ \hline
\multicolumn{1}{|c|}{\multirow{3}{*}{$M_{2}$}} & Overall & 90.29 (7.11) & 92.05 (5.08) \\ \cline{2-4} 
\multicolumn{1}{|c|}{} & \cellcolor{gray1}Easy & \cellcolor{gray1}90.21 (7.29) & \cellcolor{gray1}92.09 (5.54) \\ \cline{2-4} 
\multicolumn{1}{|c|}{} & \cellcolor{gray2}Hard & \cellcolor{gray2}90.4 (6.88) & \cellcolor{gray2}92 (4.5) \\ \hline
\multicolumn{1}{|c|}{\multirow{3}{*}{$M_{3}$}} & Overall & 89.52 (8.05) & 92.22 (4.34) \\ \cline{2-4} 
\multicolumn{1}{|c|}{} & \cellcolor{gray1}Easy & \cellcolor{gray1}90.3 (7.31) & \cellcolor{gray1}91.73 (4.59) \\ \cline{2-4} 
\multicolumn{1}{|c|}{} & \cellcolor{gray2}Hard & \cellcolor{gray2}88.39 (8.93) & \cellcolor{gray2}92.92 (3.88) \\ \hline
\end{tabular}
\end{minipage} 
\hfill
\begin{minipage}{0.45\linewidth}
\centering
	\includegraphics[width=62mm]{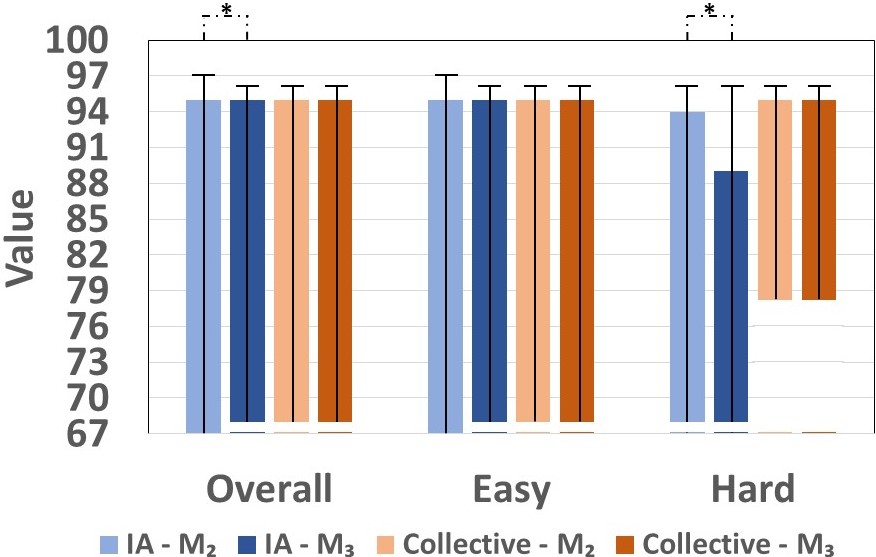}
	\captionof{figure}{Target value median (min/max) and Mann-Whitney-Wilcoxin test by decision difficulty with significance ($\rho$ $<$ 0.001 - ***, $\rho$ $<$ 0.01 - **, and $\rho$ $<$ 0.05 - *) between models.}
    \label{fig: Target Value}
\end{minipage}
\end{table}

Assessing variables, such as the selected target's value for each human-collective decision, is necessary in order to determine whether operators were able to perceive the target value correctly and influence the collectives positively. The objective of the human-collective team was to select the highest valued target for each decision from a range of target values (67 to 100). The selected target value is the average of all target's respective values that were selected by the human-collective teams during a trial. The mean (SD) for the selected target value per decision difficulty (i.e., overall, easy, and hard) are shown in Table \ref{table:Impacts,Target Value} \citep{Roundtree2020visual}. IA operators using the $M_{2}$ model chose higher valued targets compared to the $M_{3}$ model, regardless of the decision difficulty, while Collective operators using the $M_{3}$ model chose higher valued targets for overall and hard decisions. The target value median, min, max, and the Mann-Whitney-Wilcoxon significant effects between models for each model and visualization combination are shown in Figure \ref{fig: Target Value}. IA operators had significantly different selected target values between models for overall and easy decisions, while no differences were found for the Collective operators. Additional between visualizations Mann-Whitney-Wilcoxon tests identified moderate significant effects using the $M_{3}$ model for overall (n = 672, U = 63946, $\rho$ $<$ 0.01) and highly significant effects for hard decisions (n = 276, U = 12058, $\rho$ $<$ 0.001). Collective operators influenced the collective to choose higher valued targets compared to the IA operators.

The SA dependent variable was \textit{SA probe accuracy}, or the percentage of correctly answered SA probes questions \cite{Cody2018}. Each question corresponded to the three SA levels: perception, comprehension, and projection \cite{Endsley1995}. The five \textit{$SA_{1}$} questions determined the operator's ability to perceive information about the collectives and targets, such as ``What collectives are investigating Target 3?'' The operator's comprehension of information was determined by four \textit{$SA_{2}$} questions, such as ``Which Collective has achieved a majority support for Target 7?'' Three \textit{$SA_{3}$} questions were related to the operator estimating the collectives' future state, such as ``Will support for Target 1 decrease?'' The overall SA value, $SA_{O}$, was the percent of correctly answered SA probes. The SA probe accuracy mean (SD) are shown in Table \ref{table:Impacts,SA Probe Accuracy} \cite{Roundtree20191, Roundtree2020visual, Cody2020}. Operators from both evaluations using the $M_{2}$ model, when compared to $M_{3}$ model, had higher $SA_{3}$, while the IA operators had higher $SA_{2}$, and the Collective operators had higher $SA_{O}$. The SA probe accuracy median, min, max, and the Mann-Whitney-Wilcoxon significant effects between models for each model and visualization combination are presented in Figure \ref{fig: SA Probe}. Significant differences between models were found for IA operators answering $SA_{1}$ probe questions and for Collective operators answering $SA_{3}$ probe questions. Additional between visualizations Mann-Whitney-Wilcoxon tests (n = 56) identified highly significant effects using the $M_{2}$ model for $SA_{O}$ (U = 702, $\rho$ $<$ 0.001) and $SA_{1}$ (U = 714.5, $\rho$ $<$ 0.001); and moderately significant effects for $SA_{2}$ (U = 572.5, $\rho$ $<$ 0.01) and $SA_{3}$ (U = 554, $\rho$ $<$ 0.01). Highly significant effects between visualizations were found using the $M_{3}$ model for $SA_{O}$ (U = 657.5, $\rho$ $<$ 0.001), $SA_{2}$ (U = 648, $\rho$ $<$ 0.001), and $SA_{3}$ (U = 645.5, $\rho$ $<$ 0.001). A moderately significant effect between visualizations was found using the $M_{3}$ model for $SA_{1}$ (U = 564, $\rho$ $<$ 0.01). Operators using the Collective visualization had higher SA probe accuracy in general.

\begin{table}[h]
\begin{minipage}{0.5\linewidth}
\centering
\caption{SA probe accuracy (\%) mean (SD) by SA level.}
\label{table:Impacts,SA Probe Accuracy}
\begin{tabular}{c|c|c|c|}
\cline{2-4}
& \textbf{SA Level} & \textbf{IA} & \textbf{Collective)} \\ \hline
\multicolumn{1}{|c|}{\multirow{4}{*}{$M_{2}$}} & $SA_{O}$ & 65.3 (18.87) & 89.88 (10.96) \\ \cline{2-4} 
\multicolumn{1}{|c|}{} & \cellcolor{gray1}$SA_{1}$ & \cellcolor{gray1}58.57 (23.05) & \cellcolor{gray1}91.67 (11.11) \\ \cline{2-4} 
\multicolumn{1}{|c|}{} & \cellcolor{gray2}$SA_{2}$ & \cellcolor{gray2}72.32 (21.88) & \cellcolor{gray2}88.39 (14.6) \\ \cline{2-4} 
\multicolumn{1}{|c|}{} & \cellcolor{gray3}$SA_{3}$ & \cellcolor{gray3}65.48 (34.52) & \cellcolor{gray3}89.88 (20.46) \\ \hline
\multicolumn{1}{|c|}{\multirow{4}{*}{$M_{3}$}} & $SA_{O}$ & 68.15 (16.36) & 87.2 (10.75) \\ \cline{2-4} 
\multicolumn{1}{|c|}{} & \cellcolor{gray1}$SA_{1}$ & \cellcolor{gray1}80 (19.63) & \cellcolor{gray1}94.05 (13) \\ \cline{2-4} 
\multicolumn{1}{|c|}{} & \cellcolor{gray2}$SA_{2}$ & \cellcolor{gray2}65.18 (28.33) & \cellcolor{gray2}91.43 (12.68) \\ \cline{2-4} 
\multicolumn{1}{|c|}{} & \cellcolor{gray3}$SA_{3}$ & \cellcolor{gray3}52.38 (27.86) & \cellcolor{gray3}76.79 (16.57) \\ \hline
\end{tabular}
\end{minipage} 
\hfill
\begin{minipage}{0.45\linewidth}
\centering
	\includegraphics[width=62mm]{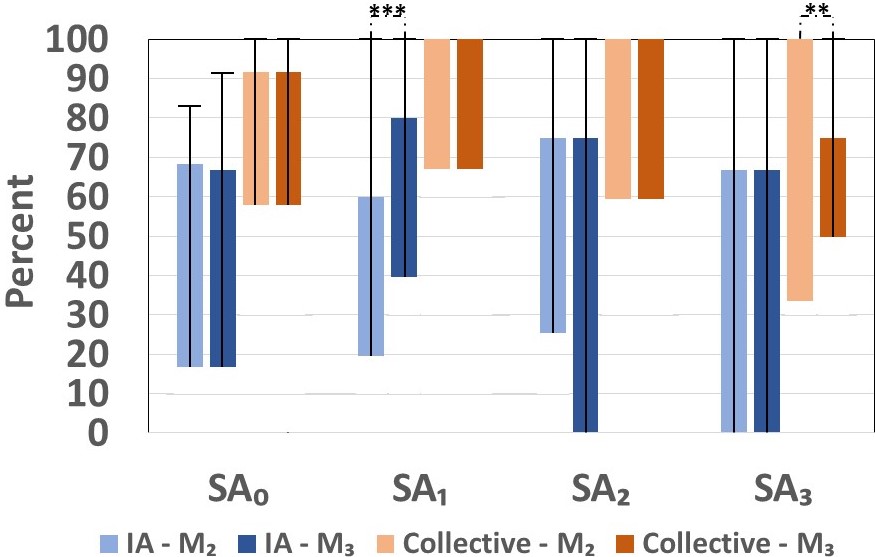}
	\captionof{figure}{SA probe accuracy median (min/max) and Mann-Whitney-Wilcoxin test by SA level with significance ($\rho$ $<$ 0.001 - ***, $\rho$ $<$ 0.01 - **, and $\rho$ $<$ 0.05 - *) between models.}
    \label{fig: SA Probe}
\end{minipage}
\end{table}

\begin{table}[!b]
\centering
\caption{Global clutter mean (SD) percentage 15 seconds before asking, while being asked, and during response to SA probe question by SA level.}
\label{table:Impacts,Global Clutter Percentage}
\begin{tabular}{c|c|c|c|c|}
\cline{2-5}
 & \textbf{Timing} & \textbf{SA Level} & \textbf{IA} & \textbf{Collective} \\ \hline
\multicolumn{1}{|c|}{\multirow{12}{*}{$M_{2}$}} & \multirow{4}{*}{Before} & $SA_{O}$ & 30.2 (3.06) & 31.37 (4.97) \\ \cline{3-5} 
\multicolumn{1}{|c|}{} & & \cellcolor{gray1}$SA_{1}$ & \cellcolor{gray1}29.88 (2.8) & \cellcolor{gray1}31.38 (5) \\ \cline{3-5} 
\multicolumn{1}{|c|}{} & & \cellcolor{gray2}$SA_{2}$ & \cellcolor{gray2}30.41 (3.05) & \cellcolor{gray2}31.25 (5.09) \\ \cline{3-5}
\multicolumn{1}{|c|}{} & & \cellcolor{gray3}$SA_{3}$ & \cellcolor{gray3}30.45 (3.45) & \cellcolor{gray3}31.56 (4.76) \\ \cline{2-5} 
\multicolumn{1}{|c|}{} & \multirow{4}{*}{Asking} & $SA_{O}$ & 30.25 (3.13) & 31.43 (5.13) \\ \cline{3-5} 
\multicolumn{1}{|c|}{} & & \cellcolor{gray1}$SA_{1}$ & \cellcolor{gray1}29.95 (2.91) & \cellcolor{gray1}31.24 (5.26) \\ \cline{3-5}
\multicolumn{1}{|c|}{} & & \cellcolor{gray2}$SA_{2}$ & \cellcolor{gray2}30.41 (3.12) & \cellcolor{gray2}31.52 (5.2) \\ \cline{3-5} 
\multicolumn{1}{|c|}{} & & \cellcolor{gray3}$SA_{3}$ & \cellcolor{gray3}30.52 (3.49) & \cellcolor{gray3}31.69 (4.78) \\ \cline{2-5} 
\multicolumn{1}{|c|}{} & \multirow{4}{*}{Responding} & $SA_{O}$ & 30.09 (3.02) & 31.41 (5.15) \\ \cline{3-5} 
\multicolumn{1}{|c|}{} & & \cellcolor{gray1}$SA_{1}$ & \cellcolor{gray1}29.83 (2.81) & \cellcolor{gray1}31.43 (5.43) \\ \cline{3-5} 
\multicolumn{1}{|c|}{} & & \cellcolor{gray2}$SA_{2}$ & \cellcolor{gray2}30.22 (3) & \cellcolor{gray2}31.34 (5.08) \\ \cline{3-5} 
\multicolumn{1}{|c|}{} & & \cellcolor{gray3}$SA_{3}$ & \cellcolor{gray3}30.37 (3.38) & \cellcolor{gray3}31.49 (4.66) \\ \hline
\multicolumn{1}{|c|}{\multirow{12}{*}{$M_{3}$}} & \multirow{4}{*}{Before} & $SA_{O}$ & 31.26 (3.41) & 31.76 (5.23) \\ \cline{3-5} 
\multicolumn{1}{|c|}{} & & \cellcolor{gray1}$SA_{1}$ & \cellcolor{gray1}31.2 (3.48) & \cellcolor{gray1}31.51 (5.05) \\ \cline{3-5} 
\multicolumn{1}{|c|}{} & & \cellcolor{gray2}$SA_{2}$ & \cellcolor{gray2}31.78 (3.4) & \cellcolor{gray2}32.11 (5.21) \\ \cline{3-5}
\multicolumn{1}{|c|}{} & & \cellcolor{gray3}$SA_{3}$ & \cellcolor{gray3}30.68 (3.24) & \cellcolor{gray3}31.51 (5.51) \\ \cline{2-5} 
\multicolumn{1}{|c|}{} & \multirow{4}{*}{Asking} & $SA_{O}$ & 31.49 (3.59) & 31.7 (5.23) \\ \cline{3-5} 
\multicolumn{1}{|c|}{} & & \cellcolor{gray1}$SA_{1}$ & \cellcolor{gray1}31.6 (3.74) & \cellcolor{gray1}31.15 (5.05) \\ \cline{3-5}
\multicolumn{1}{|c|}{} & & \cellcolor{gray2}$SA_{2}$ & \cellcolor{gray2}31.83 (3.54) & \cellcolor{gray2}32.33 (5.52) \\ \cline{3-5} 
\multicolumn{1}{|c|}{} & & \cellcolor{gray3}$SA_{3}$ & \cellcolor{gray3}30.82 (3.34) & \cellcolor{gray3}31.4 (4.9) \\ \cline{2-5} 
\multicolumn{1}{|c|}{} & \multirow{4}{*}{Responding} & $SA_{O}$ & 31.16 (3.36) & 31.7 (5.27) \\ \cline{3-5} 
\multicolumn{1}{|c|}{} & & \cellcolor{gray1}$SA_{1}$ & \cellcolor{gray1}31.25 (3.4) & \cellcolor{gray1}31.24 (5.12) \\ \cline{3-5} 
\multicolumn{1}{|c|}{} & & \cellcolor{gray2}$SA_{2}$ & \cellcolor{gray2}31.49 (3.41) & \cellcolor{gray2}32.25 (5.56) \\ \cline{3-5} 
\multicolumn{1}{|c|}{} & & \cellcolor{gray3}$SA_{3}$ & \cellcolor{gray3}30.56 (3.2) & \cellcolor{gray3}31.37 (4.93) \\ \hline
\end{tabular}
\end{table}

Global clutter percentages were analyzed for each SA probe question. Clutter is the area occupied by objects on a display, relative to the display's total area \cite{Wickens2004}. Clutter becomes an issue when presenting too much information in close proximity requires a longer search time \cite{Wickens2004} and negatively influences the operator's ability to perform a task. Area coverage was calculated as the number of pixels an item covered on the computer visualization. One meter for the IA visualization was approximately 1.97 pixels per meter. The Collective visualization computer display size was unknown; therefore, global clutter percentage calculations used the corresponding item and computer display (2073600 $pixels^{2}$) dimensions from the IA visualization. The \textit{global clutter percentage} variable was the percentage of area obstructed by all objects displayed on the computer displays, using Equation \ref{eq:Global Clutter}: 

\begin{equation}
Global Clutter (\%) = \left(\frac{ICA + GHA + GHTA + GTA + GAICE + GTIW + GCIW}{2073600}\right) \cdot 100, 
\label{eq:Global Clutter}
\end{equation} where ICA represents the static interface components areas (493414 $pixels^{2}$). GHA represents the area covered by Collectives I-IV (9856 $pixels^{2}$), which were visible throughout the trial. The area corresponding to highlighted targets (2350 $pixels^{2}$ per highlighted target), which have outlines and are in range of the selected collective, are represented as GHTA. Remaining targets that are not highlighted (1720 $pixels^{2}$ per target) are represented as GTA. GAICE is the area consumed by the 800 individual collective entities (51200 $pixels^{2}$), only considered for the IA visualization. The area corresponding to the number of target information pop-up windows (32922 $pixels^{2}$ per target information pop-up window) is represented as GTIW, and the corresponding collective information pop-up windows is represented as GCIW (25740 $pixels^{2}$ per collective information pop-up window). The potential clutter associated with the background map was not considered in the global clutter calculation for two reasons. First, the underlying map is identical for both visualizations, but differed slightly due to the computer screen size, making a between evaluation assessment unattainable. Second, the operators did not depend on the underlying map to complete the sequential best-of-\textit{n} task. The map provided an ecologically valid background, representative of the task in corresponding real-world dynamic environments. Future evaluations must consider how the background may influence clutter, especially if the map becomes dynamic, or the operator can zoom in and out on particular areas.

\def \figwidth {0.49}
\begin{figure*}[h]
    \captionsetup[subfigure]{justification=centering}
    \begin{subfigure}{\figwidth\linewidth}
		\centering
		\includegraphics[keepaspectratio,height=1.71in]{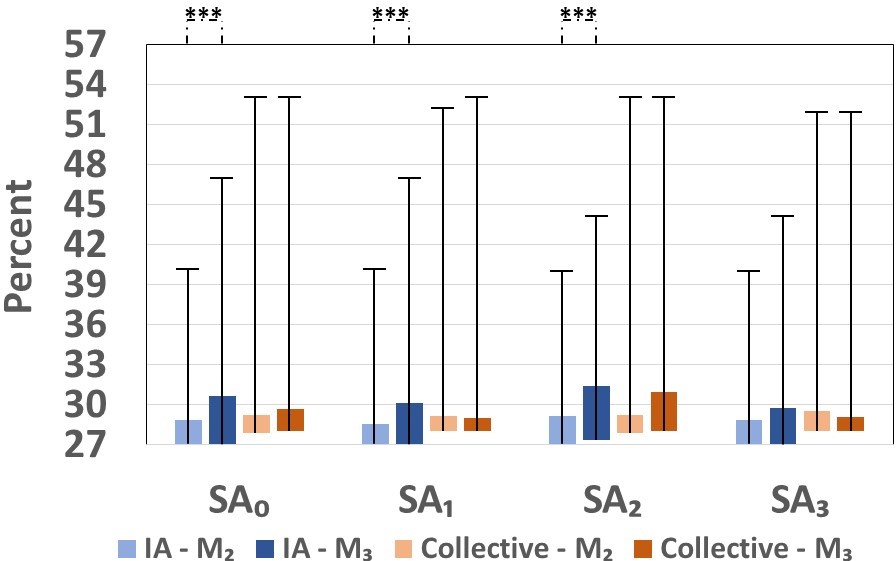}
		\captionsetup{width=\linewidth}
		\caption{15 seconds before asking a SA probe question.}
	\end{subfigure}
\begin{subfigure}{\figwidth\linewidth}
	\centering
	\includegraphics[keepaspectratio,height=1.71in]{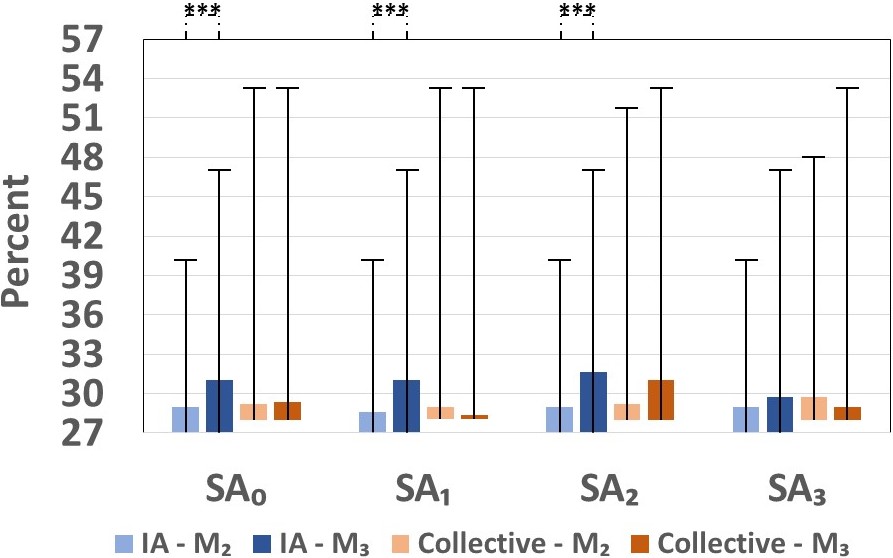}
	\captionsetup{width=\linewidth}
	\caption{While being asked a SA probe question.}
\end{subfigure}
\begin{subfigure}{\figwidth\linewidth}
	\centering
	\includegraphics[keepaspectratio,height=1.705in]{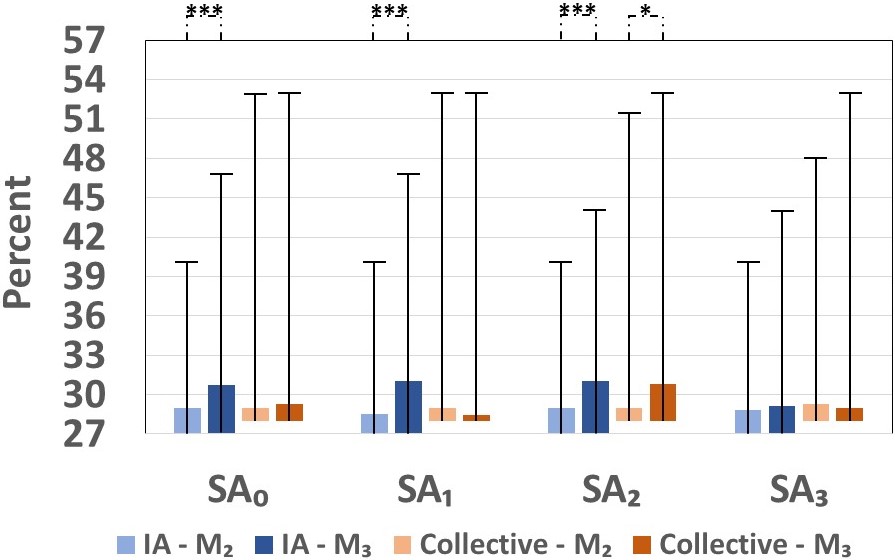}
	\captionsetup{width=\linewidth}
	\caption{During response to a SA probe question.}
\end{subfigure}
	\caption{Global clutter percentage median (min/max) and Mann-Whitney-Wilcoxin test by SA level between models a) 15 seconds before asking, b) while being asked, and c) during response to a SA probe question.}
	\label{fig: Global Clutter}
\end{figure*}

The global clutter mean (SD) percentage 15 seconds before asking, while being asked, and during response to a SA probe question are shown in Table \ref{table:Impacts,Global Clutter Percentage} \citep{Roundtree2020visual}. IA operators who used the $M_{2}$ model had lower global clutter percentages compared to when they used the $M_{3}$ model. Collective operators in general had lower global clutter percentages using the $M_{2}$ model. $SA_{3}$ at all timings and $SA_{1}$ while being asked a SA probe question were lower when Collective operators used the $M_{3}$ model. The global clutter percentage median, min, max, and the Mann-Whitney-Wilcoxon significant effects between models are shown in Figure \ref{fig: Global Clutter}. Significant differences between models were found for IA operators at all timings for $SA_{O}$, $SA_{1}$, and $SA_{2}$ probe questions, while a significant difference between models was identified for Collective operators during response to $SA_{2}$ probe questions. 

Additional between visualizations Mann-Whitney-Wilcoxin tests were conducted. All significant differences between visualizations occurred when using the $M_{2}$ model. A highly significant effect between visualizations was found when responding to a SA probe question for $SA_{O}$ (n = 670, U = 64442, $\rho$ $<$ 0.001). Moderate significant effects between visualizations were found for $SA_{O}$ 15 seconds before asking (U = 64188, $\rho$ $<$ 0.01) and while being asked a SA probe question (U = 63728, $\rho$ $<$ 0.01). Significant effects between visualizations were found 15 seconds before asking a SA probe question for $SA_{1}$ (n = 294, U = 12487, $\rho$ = 0.02) and $SA_{3}$ (n = 152, U = 3445.5, $\rho$ = 0.03); while being asked a SA probe question for $SA_{1}$ (U = 12301, $\rho$ = 0.03) and $SA_{3}$ (U = 3452, $\rho$ = 0.05); and during the response to a SA probe question for $SA_{1}$ (U = 12216, $\rho$ = 0.04). Correlations between the global clutter percentage and SA probe accuracy were only revealed when using the Collective visualization 15 seconds before asking a SA probe question. The Spearman correlation analysis revealed a moderate correlation using the $M_{3}$ model for $SA_{3}$ (r = 0.45, $\rho$ $<$ 0.001), and weak correlations when using the $M_{2}$ model for $SA_{1}$ (r = 0.16, $\rho$ = 0.05) and when using the $M_{3}$ model for $SA_{O}$ (r = 0.2, $\rho$ $<$ 0.001). The IA visualization had lower global clutter percentages in general compared to the Collective visualization. Collective operators using the $M_{3}$ model; however, had lower global clutter while being asked and during response to a $SA_{1}$ probe question.

There were no significant effects between visualizations for operator spatial reasoning, based on the Mental Rotations Assessment \cite{Vandenberg1978}. Correlations between the Mental Rotations Assessment and SA probe accuracy only existed for the IA visualization. The Spearman correlation analysis revealed weak correlations with the $M_{2}$ model for $SA_{O}$ (r = 0.17, $\rho$ $<$ 0.01), $SA_{1}$ (r = 0.18, $\rho$ = 0.03), and $SA_{2}$ (r = 0.27, $\rho$ $<$ 0.01). Weak correlations were revealed with the $M_{3}$ model for $SA_{O}$ (r = 0.15, $\rho$ $<$ 0.01), $SA_{1}$ (r = 0.19, $\rho$ = 0.03), and $SA_{2}$ (r = 0.18, $\rho$ = 0.05). A moderate correlation existed between Working Memory Capacity, which assessed operator higher-order cognitive task abilities \citep{Engle2002}, and SA probe accuracy for the IA visualization using the $M_{2}$ model for $SA_{3}$ (r = 0.45, $\rho$ $<$ 0.001). Weak correlations existed with the $M_{2}$ model for $SA_{O}$ (r = 0.23, $\rho$ $<$ 0.001) and $SA_{1}$ (r = 0.17, $\rho$ = 0.04), and when using the $M_{3}$ model for $SA_{O}$ (r = 0.14, $\rho$ = 0.01). The Mann-Whitney-Wilcoxon tests identified no significant effects between visualizations for the weekly hours spent using a desktop or laptop. Weak correlations were found between weekly hours using a desktop or laptop and SA probe accuracy when using the $M_{2}$ model for the IA visualization for $SA_{O}$ (r = 0.12, $\rho$ = 0.04) and $SA_{1}$ (r = 0.21, $\rho$ = 0.01), and when using the Collective visualization for $SA_{2}$ (r = 0.21, $\rho$ = 0.02). 

\begin{table}[h]
\begin{minipage}{0.5\linewidth}
\centering
\caption{NASA-TLX mean (SD).}
\label{table:Impacts,NASA TLX}
\begin{tabular}{c|c|c|c|}
\cline{2-4}
 & \textbf{TLX} & \textbf{IA} & \textbf{Collective} \\ \hline
\multicolumn{1}{|c|}{\multirow{7}{*}{$M_{2}$}} & Overall & 62.14 (14.81) & 57.06 (16.47) \\ \cline{2-4} 
\multicolumn{1}{|c|}{} & \cellcolor{gray1}Mental & \cellcolor{gray1}19.25 (8.8) & \cellcolor{gray1}23.58 (6.34) \\ \cline{2-4} 
\multicolumn{1}{|c|}{} & \cellcolor{gray2}Physical & \cellcolor{gray2}1.68 (3.32) & \cellcolor{gray2}0.46 (1.17) \\ \cline{2-4} 
\multicolumn{1}{|c|}{} & \cellcolor{gray3}Temporal & \cellcolor{gray3}11.75 (8.24) & \cellcolor{gray3}10.94 (7.67) \\ \cline{2-4} 
\multicolumn{1}{|c|}{} & \cellcolor{gray4}Performance & \cellcolor{gray4}10.69 (5.87) & \cellcolor{gray4}5.1 (4.7) \\ \cline{2-4} 
\multicolumn{1}{|c|}{} & \cellcolor{gray5}Effort & \cellcolor{gray5}11.35 (6.68) & \cellcolor{gray5}12.32 (6.36) \\ \cline{2-4} 
\multicolumn{1}{|c|}{} & \cellcolor{gray6}Frustration & \cellcolor{gray6}7.43 (8.36) & \cellcolor{gray6}4.65 (6.84) \\ \hline 
\multicolumn{1}{|c|}{\multirow{7}{*}{$M_{3}$}} & Overall & 60.38 (16.5) & 50.63 (17.56) \\ \cline{2-4} 
\multicolumn{1}{|c|}{} & \cellcolor{gray1}Mental & \cellcolor{gray1}18.32 (9.4) & \cellcolor{gray1}16.54 (9.19) \\ \cline{2-4} 
\multicolumn{1}{|c|}{} & \cellcolor{gray2}Physical & \cellcolor{gray2}6.11 (10.27) & \cellcolor{gray2}1.81 (6.01) \\ \cline{2-4} 
\multicolumn{1}{|c|}{} & \cellcolor{gray3}Temporal & \cellcolor{gray3}8.85 (7.3) & \cellcolor{gray3}7.49 (6.63) \\ \cline{2-4} 
\multicolumn{1}{|c|}{} & \cellcolor{gray4}Performance & \cellcolor{gray4}9.08 (6.7) & \cellcolor{gray4}5.15 (4.79) \\ \cline{2-4} 
\multicolumn{1}{|c|}{} & \cellcolor{gray5}Effort & \cellcolor{gray5}14.25 (8.06) & \cellcolor{gray5}12.5 (5.17) \\ \cline{2-4} 
\multicolumn{1}{|c|}{} & \cellcolor{gray6}Frustration & \cellcolor{gray6}3.77 (5.92) & \cellcolor{gray6}7.14 (8.31) \\ \hline
\end{tabular}
\end{minipage} 
\hfill
\begin{minipage}{0.45\linewidth}
\centering
	\includegraphics[width=62mm]{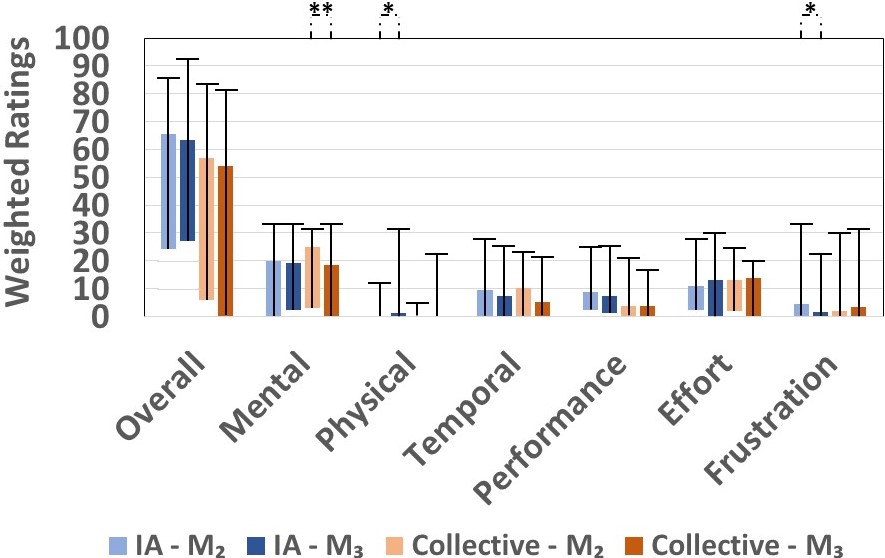}
	\captionof{figure}{NASA-TLX median (min/max) and Mann-Whitney-Wilcoxin test between models.}
    \label{fig: NASA TLX}
\end{minipage}
\end{table}

The NASA Task Load Index (\textit{NASA-TLX}) assessed the six workload subscales and the weighted overall workload \cite{Hart1988}. The mean (SD) for the NASA-TLX overall workload and imposed demands are presented in Table \ref{table:Impacts,NASA TLX} \citep{Roundtree2020visual, Cody2020}. IA operators using the $M_{2}$ model had lower physical demand and effort when compared to $M_{3}$, while those using the Collective visualization had lower physical demand, effort, and frustration when using the $M_{2}$ model. The NASA-TLX median, min, max, and the Mann-Whitney-Wilcoxon significant effects between models are presented in Figure \ref{fig: NASA TLX}. IA operators had significantly different rankings between models for physical demand and frustration, while mental demand was significantly different between models for Collective operators. Additional between visualizations Mann-Whitney-Wilcoxin tests (n = 56) identified a significant effect when using the $M_{2}$ model for mental demand (U = 515, $\rho$ = 0.04) and a highly significant effect for performance (U = 159.5, $\rho$ $<$ 0.001). Significant effects were found between visualizations using the $M_{3}$ model for overall workload (U = 266.5, $\rho$ = 0.04), performance (U = 242.5, $\rho$ = 0.01), and frustration (U = 511, $\rho$ = 0.05), as well as a highly significant effect for physical demand (U = 208, $\rho$ $<$ 0.001). The Collective visualization imposed a lower overall workload, had lower physical and temporal demands, and caused less frustration compared to the IA visualization.

The post-experiment questionnaire assessed the collective's \textit{responsiveness} to requests, the participants' \textit{ability} to choose the highest valued target, and their \textit{understanding} of the collective behavior, from best (1) to worst (2 for the IA evaluation and 3 for the Collective evaluation). The post-experiment questionnaire mean (SD) are shown in Table \ref{table:Impacts,PE} \cite{Cody2018}. The best collective responsiveness as well as operator ability and understanding occurred when IA operators used the $M_{2}$ model versus the $M_{3}$ model. Collective operators ranked the collective's responsiveness highest using the $M_{3}$ model, while operator ability and understanding were highest when using the $M_{2}$ model. The post-experiment questionnaire median, min, max, and the Mann-Whitney-Wilcoxon significant effects between models are presented in Figure \ref{fig: Post Experiment}. System responsiveness, operator ability, and understanding were ranked significantly different between models for IA operators, while Collective operators ranked system responsiveness and operator understanding significantly different. 

\begin{table}[h!]
\begin{minipage}{0.5\linewidth}
\centering
\caption{Post-experiment responsiveness, ability, and understanding model ranking mean (SD) (1-best, 2-worst for IA evaluation and 3-worst for Collective evaluation).}
\label{table:Impacts,PE}
\begin{tabular}{c|c|c|c|}
\cline{2-4}
& \textbf{Metric} & \textbf{IA} & \textbf{Collective} \\ \hline
\multicolumn{1}{|c|}{\multirow{3}{*}{$M_{2}$}} & Responsiveness & 1.64 (0.49) & 1.5 (0.51) \\ \cline{2-4} 
\multicolumn{1}{|c|}{} & \cellcolor{gray1}Ability & \cellcolor{gray1}1.86 (0.36) & \cellcolor{gray1}2 (1.02) \\ \cline{2-4} 
\multicolumn{1}{|c|}{} & \cellcolor{gray2}Understanding & \cellcolor{gray2}1.79 (0.42) & \cellcolor{gray2}2.5 (0.51) \\ \hline
\multicolumn{1}{|c|}{\multirow{3}{*}{$M_{3}$}} & Responsiveness & 1.36 (0.49) & 3 (0) \\ \cline{2-4}
\multicolumn{1}{|c|}{} & \cellcolor{gray1}Ability & \cellcolor{gray1}1.14 (0.36) & \cellcolor{gray1}2 (0) \\ \cline{2-4} 
\multicolumn{1}{|c|}{} & \cellcolor{gray2}Understanding & \cellcolor{gray2}1.21 (0.42) & \cellcolor{gray2}1 (0) \\ \hline
\end{tabular}
\end{minipage} 
\hfill
\begin{minipage}{0.45\linewidth}
\centering
	\includegraphics[width=62mm]{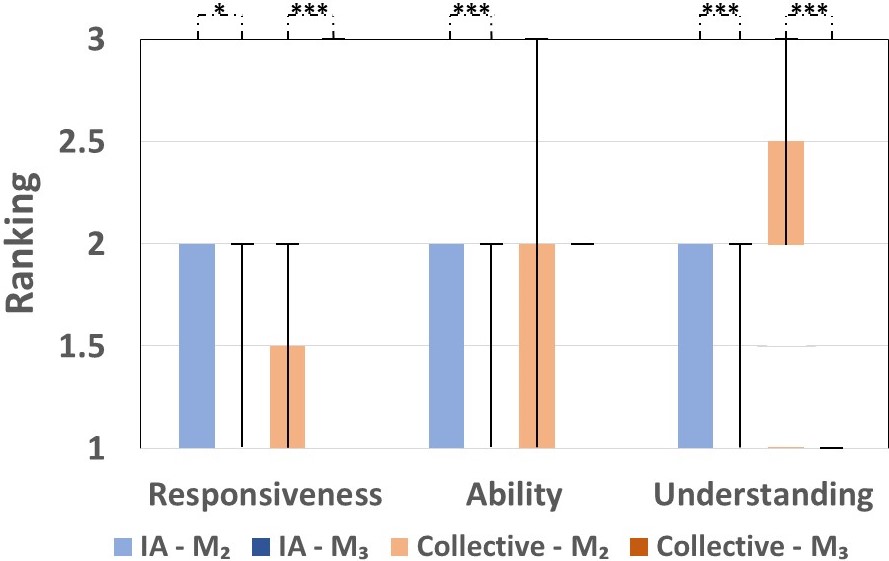}
	\captionof{figure}{Post-experiment responsiveness, ability, and understanding model ranking median (min/max) and Mann-Whitney-Wilcoxin test between models. The ranking was from 1-best to either 2-worst for the IA evaluation, or 3-worst for the Collective evaluation.}
    \label{fig: Post Experiment}
\end{minipage}
\end{table}

A summary of $R_{1}$'s results that show the hypotheses with associated significant results is provided in Table \ref{table:Impacts,Combined}. This summary table is intended to facilitate the discussion.

\begin{table}[h]
\centering
\caption{A synopsis of $R_{1}$'s hypotheses associated with significant results. The SA probe timings are all timings (All), 15 seconds Before asking (B), While being asked (W), and During response (D) to a SA probe question.}
\label{table:Impacts,Combined}
\begin{tabular}{?l|cc|cc|cc|c|c|c|c|c|c?}
\Cline{1pt}{1-10}
\multicolumn{1}{?c|}{\multirow{4}{*}{\textbf{Variable}}} & \multirow{3}{*}{\textbf{Sub-}} & \multicolumn{2}{?c?}{\textbf{Within}} & \multicolumn{2}{c?}{\textbf{Between}} & \multicolumn{4}{c?}{\multirow{2}{*}{\textbf{Correlation}}} \\
& \multirow{3}{*}{\textbf{Variable}} & \multicolumn{2}{?c?}{\textbf{Model}} & \multicolumn{2}{c?}{\textbf{Visualization}} & \multicolumn{4}{c?}{} \\ \Cline{1pt}{3-10}
& & \multicolumn{1}{?c|}{\multirow{2}{*}{IA}} & \multicolumn{1}{c?}{\multirow{2}{*}{Coll.}} & \multirow{2}{*}{$M_{2}$} & \multicolumn{1}{c?}{\multirow{2}{*}{$M_{3}$}} & \multicolumn{2}{c|}{IA} & \multicolumn{2}{c?}{Coll.} \\ \Cline{1pt}{7-10}
& & \multicolumn{1}{?c|}{} & \multicolumn{1}{c?}{} & & \multicolumn{1}{c?}{} & $M_{2}$ & $M_{3}$ & $M_{2}$ & \multicolumn{1}{c?}{$M_{3}$} \\ \Cline{1pt}{1-10}
\multirow{2}{*}{Target Value} & Overall & \multicolumn{1}{?c|}{{$H_{1}$}} & \multicolumn{1}{c?}{} & & \multicolumn{1}{c?}{$H_{1}$} & \multicolumn{4}{c?}{\multirow{6}{*}{-----------------}} \\ \cline{2-6}
& Hard & \multicolumn{1}{?c|}{{$H_{1}$}} & \multicolumn{1}{c?}{} & & \multicolumn{1}{c?}{$H_{1}$} & \multicolumn{1}{c}{} & \multicolumn{1}{c}{} & \multicolumn{1}{c}{} & \multicolumn{1}{c?}{} \\ \Cline{1pt}{1-6}
\multirow{4}{*}{SA Probe Accuracy} & $SA_{O}$ & \multicolumn{1}{?c|}{} & \multicolumn{1}{c?}{} & \bm{$H_{1}$} & \multicolumn{1}{c?}{$H_{1}$} & \multicolumn{1}{c}{} & \multicolumn{1}{c}{} & \multicolumn{1}{c}{} & \multicolumn{1}{c?}{} \\ \cline{2-6}
& $SA_{1}$ & \multicolumn{1}{?c|}{\color{gray7}\bm{$H_{1}$}} & \multicolumn{1}{c?}{} & \bm{$H_{1}$} & \multicolumn{1}{c?}{$H_{1}$} & \multicolumn{1}{c}{} & \multicolumn{1}{c}{} & \multicolumn{1}{c}{} & \multicolumn{1}{c?}{} \\ \cline{2-6}
& $SA_{2}$ & \multicolumn{1}{?c|}{} & \multicolumn{1}{c?}{} & \bm{$H_{1}$} & \multicolumn{1}{c?}{$H_{1}$} & \multicolumn{1}{c}{} & \multicolumn{1}{c}{} & \multicolumn{1}{c}{} & \multicolumn{1}{c?}{} \\ \cline{2-6}
& $SA_{3}$ & \multicolumn{1}{?c|}{} & \multicolumn{1}{c?}{} & \bm{$H_{1}$} & \multicolumn{1}{c?}{$H_{1}$} & \multicolumn{1}{c}{} & \multicolumn{1}{c}{} & \multicolumn{1}{c}{} & \multicolumn{1}{c?}{} \\ \Cline{1pt}{1-10}
\multirow{8}{*}{Global Clutter} & \multirow{2}{*}{$SA_{O}$} & \multicolumn{1}{?c|}{{$H_{1}$}} & \multicolumn{1}{c?}{} & {$H_{1}$} & \multicolumn{1}{c?}{} & & & & \multicolumn{1}{c?}{$H_{1}$} \\ 
& & \multicolumn{1}{?c|}{{-All}} & \multicolumn{1}{c?}{} & {-All} & \multicolumn{1}{c?}{} & & & & \multicolumn{1}{c?}{-B} \\ \cline{2-10}
& \multirow{2}{*}{$SA_{1}$} & \multicolumn{1}{?c|}{{$H_{1}$}} & \multicolumn{1}{c?}{} & {$H_{1}$} & \multicolumn{1}{c?}{} & & & $H_{1}$ & \multicolumn{1}{c?}{} \\ 
& & \multicolumn{1}{?c|}{{-All}} & \multicolumn{1}{c?}{} & {-All} & \multicolumn{1}{c?}{} & & & -B & \multicolumn{1}{c?}{} \\ \cline{2-10}
& \multirow{2}{*}{$SA_{2}$} & \multicolumn{1}{?c|}{{$H_{1}$}} & \multicolumn{1}{c?}{\color{gray7}\bm{$H_{1}$}} & {$H_{1}$} & \multicolumn{1}{c?}{} & & & & \multicolumn{1}{c?}{} \\ 
& & \multicolumn{1}{?c|}{{-All}} & \multicolumn{1}{c?}{\color{gray7}\bm{$-D$}} & {-B,W} & \multicolumn{1}{c?}{} & & & & \multicolumn{1}{c?}{} \\ \cline{2-10}
& \multirow{2}{*}{$SA_{3}$} & \multicolumn{1}{?c|}{} & \multicolumn{1}{c?}{} & & \multicolumn{1}{c?}{} & & & & \multicolumn{1}{c?}{$H_{1}$} \\ 
& & \multicolumn{1}{?c|}{} & \multicolumn{1}{c?}{} & & \multicolumn{1}{c?}{} & & & & \multicolumn{1}{c?}{-B} \\ \Cline{1pt}{1-10}
\multirow{2}{*}{Mental Rotations} & $SA_{O}$ & \multicolumn{4}{?c?}{\multirow{10}{*}{-----------------}} & \color{gray7}\bm{$H_{2}$} & \color{gray7}\bm{$H_{2}$} & & \multicolumn{1}{c?}{} \\ \cline{2-2} \cline{7-10}
\multirow{2}{*}{Assessment} & $SA_{1}$ & \multicolumn{1}{?c}{} & \multicolumn{1}{c}{} & \multicolumn{1}{c}{} & \multicolumn{1}{c?}{} & \color{gray7}\bm{$H_{2}$} & \color{gray7}\bm{$H_{2}$} & & \multicolumn{1}{c?}{} \\ \cline{2-2} \cline{7-10}
& $SA_{2}$ & \multicolumn{1}{?c}{} & \multicolumn{1}{c}{} & \multicolumn{1}{c}{} & \multicolumn{1}{c?}{} & & \color{gray7}\bm{$H_{2}$} & & \multicolumn{1}{c?}{} \\ \Cline{1pt}{1-2} \Cline{1pt}{7-10}
\multirow{2}{*}{Working Memory} & $SA_{O}$ & \multicolumn{1}{?c}{} & \multicolumn{1}{c}{} & \multicolumn{1}{c}{} & \multicolumn{1}{c?}{} & \color{gray7}\bm{$H_{2}$} & \color{gray7}\bm{$H_{2}$} & & \multicolumn{1}{c?}{} \\ \cline{2-2} \cline{7-10}
\multirow{2}{*}{Capacity} & $SA_{1}$ & \multicolumn{1}{?c}{} & \multicolumn{1}{c}{} & \multicolumn{1}{c}{} & \multicolumn{1}{c?}{} & \color{gray7}\bm{$H_{2}$} & & & \multicolumn{1}{c?}{} \\ \cline{2-2} \cline{7-10}
& $SA_{3}$ & \multicolumn{1}{?c}{} & \multicolumn{1}{c}{} & \multicolumn{1}{c}{} & \multicolumn{1}{c?}{} & \color{gray7}\bm{$H_{2}$} & & & \multicolumn{1}{c?}{} \\ \Cline{1pt}{1-2} \Cline{1pt}{7-10}
\multirow{2}{*}{Weekly Hours on} & {$SA_{O}$} & \multicolumn{1}{?c}{} & \multicolumn{1}{c}{} & \multicolumn{1}{c}{} & \multicolumn{1}{c?}{} & {\color{gray7}\bm{$H_{2}$}} & & & \multicolumn{1}{c?}{} \\ \cline{2-2} \cline{7-10}
\multirow{2}{*}{Desktop or Laptop} & $SA_{1}$ & \multicolumn{1}{?c}{} & \multicolumn{1}{c}{} & \multicolumn{1}{c}{} & \multicolumn{1}{c?}{} & \color{gray7}\bm{$H_{2}$} & & & \multicolumn{1}{c?}{} \\ \cline{2-2} \cline{7-10}
& $SA_{2}$ & \multicolumn{1}{?c}{} & \multicolumn{1}{c}{} & \multicolumn{1}{c}{} & \multicolumn{1}{c?}{} & & & \color{gray7}\bm{$H_{2}$} & \multicolumn{1}{c?}{} \\ \Cline{1pt}{1-10}
\multirow{6}{*}{NASA-TLX} & Overall & \multicolumn{1}{?c|}{} & \multicolumn{1}{c?}{} & & \multicolumn{1}{c?}{$H_{1}$} & \multicolumn{4}{c?}{\multirow{7}{*}{-----------------}} \\ \cline{2-6}
& Mental & \multicolumn{1}{?c|}{} & \multicolumn{1}{c?}{\color{gray7}\bm{$H_{1}$}} & \color{gray7}\bm{$H_{1}$} & \multicolumn{1}{c?}{} & \multicolumn{1}{c}{} & \multicolumn{1}{c}{} & \multicolumn{1}{c}{} & \multicolumn{1}{c?}{} \\ \cline{2-6}
& Physical & \multicolumn{1}{?c|}{$H_{1}$} & \multicolumn{1}{c?}{} & & \multicolumn{1}{c?}{$H_{1}$} & \multicolumn{1}{c}{} & \multicolumn{1}{c}{} & \multicolumn{1}{c}{} & \multicolumn{1}{c?}{} \\ \cline{2-6}
& Performance & \multicolumn{1}{?c|}{} & \multicolumn{1}{c?}{} & \bm{$H_{1}$} & \multicolumn{1}{c?}{$H_{1}$} & \multicolumn{1}{c}{} & \multicolumn{1}{c}{} & \multicolumn{1}{c}{} & \multicolumn{1}{c?}{} \\ \cline{2-6}
& \multirow{2}{*}{Frustration} & \multicolumn{1}{?c|}{\color{gray7}\bm{$H_{1},$}} & \multicolumn{1}{c?}{} & & \multicolumn{1}{c?}{\color{gray7}\bm{$H_{1},$}} & \multicolumn{1}{c}{} & \multicolumn{1}{c}{} & \multicolumn{1}{c}{} & \multicolumn{1}{c?}{} \\
& & \multicolumn{1}{?c|}{\color{gray7}\bm{$H_{3}$}} & \multicolumn{1}{c?}{} & & \multicolumn{1}{c?}{\color{gray7}\bm{$H_{3}$}} & \multicolumn{1}{c}{} & \multicolumn{1}{c}{} & \multicolumn{1}{c}{} & \multicolumn{1}{c?}{} \\ \Cline{1pt}{1-6}
Post-Experiment & {Ability} & \multicolumn{1}{?c|}{{$H_{1}$}} & \multicolumn{1}{c?}{} & \multicolumn{1}{c}{} & \multicolumn{1}{c}{} & \multicolumn{1}{c}{} & \multicolumn{1}{c}{} & \multicolumn{1}{c}{} & \multicolumn{1}{c?}{} \\ \Cline{1pt}{1-10}
\end{tabular}
\end{table}

\subsection{Discussion}

Relationships to the transparency factors provided in Table \ref{table:Impacts,Variables} are emphasized using italics. The analysis of how the model and visualization influenced operators with different individual \textit{capabilities} suggests that the $M_{2}$ model promoted transparency as \textit{effectively} as the $M_{3}$ model, while the Collective visualization promoted better transparency compared to the IA visualization. $H_{1}$, which hypothesized that operators using the $M_{2}$ model and Collective visualization will experience significantly higher \textit{SA} and lower \textit{workload}, was not supported. \textit{SA performance} (i.e., accuracy) varied across the SA levels depending on the model and \textit{workload} varied across the \textit{workload} subscales depending on the model and visualization. The $M_{2}$ model was \textit{effective} at enabling operators to more accurately \textit{predict} future collective behaviors, while the $M_{3}$ model enabled better \textit{observability} of the collectives' behaviors. Better \textit{predictability} may have occurred, because the $M_{2}$ model aligned with the operators expectations: that the model was designed to choose the highest value target. \textit{Predictability} of future collective states may have also improved due to the visualization. Favoring entities in the IA visualization created streamlines between hubs and targets, which may have directed the operator's attention to particular targets. The $M_{3}$ model may have promoted better perception of the collectives' behaviors, because the operator was required to direct those behaviors in order to achieve the task. Operator \textit{workload} was alleviated by the $M_{2}$ model by requiring less operator \textit{capabilities}, such as physical demand and effort, as well as promoting higher \textit{performance}, which was expected since operator influence was not required in order to make decisions. The $M_{3}$ model alleviated operator \textit{workload} by also requiring less operator \textit{capabilities}, such as mental demands, and improving \textit{satisfaction} (i.e., less frustration) by mitigating temporal (i.e., \textit{timing}) demands.  More operator control of the decision-making process, such making decision quickly or more slowly, may have contributed to these lower \textit{workload} subscales. 

Transparency embedded into the Collective visualization partially supported $H_{1}$, because it promoted higher \textit{SA performance} via the color-coded icons and outlines, state information identified on the collective icon, information provided in the pop-up windows, as well as feedback provided in the Collective Assignments and System Messages areas. Collective operators encountered more clutter; however, due to the long duration of time the target information pop-up windows were visible. The increased clutter has both positive and negative implications for transparency. Clutter, from an \textit{usability} perspective, is not ideal if operators are unable to perform their tasks \textit{effectively}. The Collective operators, who had higher clutter were able to answer more \textit{SA} probe questions accurately, which suggests that the operators were not hindered by the clutter and \textit{performed} better. The dependence on the visible target information pop-up windows may have been caused by the type of \textit{SA} probe questions asked. Thirteen of twenty-four \textit{SA} probe questions relied on information provided in the target information-pop up windows. An example question, such as ``What collectives are investigating Target 3?", required using the target information pop-up window, if Target 3 was in range of multiple collectives. The operator was able to identify which collectives were within range of a particular target by left-clicking on the respective collective; however, target information pop-up windows were needed in order to see the numeric collective support values from a specific collective, or multiple collectives. Experimental design modifications can ensure a more even distribution of \textit{SA} probe questions that rely on other information, such as the icons, system messages, or collective assignments versus information pop-up windows. Target icon design modifications that indicate which collectives support a particular target may improve \textit{explainability}, \textit{reliability}, and increase the reliance on the target icon instead of the information pop-up window. 

The Collective visualization partially supported $H_{1}$ by requiring less operator \textit{capabilities}, such as physical demands, and improving \textit{satisfaction} (i.e., less frustration) by mitigating temporal (i.e., \textit{timing}) demands. Not visualizing entities may have reduced operator stress, because the rate of a collective's state change was not easily perceived. The need or desire to influence collective behaviors may not have been as apparent, which attributed to lower physical demand and frustration. Higher operator mental demand when using the $M_{2}$ model and Collective visualization may have occurred if collective behaviors, or state changes, were not \textit{observable} and required more interactions to deduce what was happening, such as accessing information pop-up windows. 

$H_{2}$, which hypothesized that operators with different individual \textit{capabilities} did not \textit{perform} significantly different using the $M_{2}$ model and the Collective visualization, was partially supported. Individuals with different spatial reasoning and working memory capacity \textit{capabilities performed} relatively the same. Operators who had a higher level of computer knowledge; however, had a better \textit{understanding} of the collective behaviors. This finding was anticipated considering the computer simulation environment. Further investigations are needed to identify what particular aspects of computer knowledge attribute to better \textit{understanding}. 

Collective operators using the $M_{2}$ model were more \textit{satisfied} (i.e., less frustration), which supported $H_{3}$. Dis\textit{satisfaction} transpires when the system is not transparent and prohibits the operator from \textit{understanding} what is currently happening, or the interface appears visually noisy due to clutter \cite{Preece2007}. A more autonomous model, one with more decision-making \textit{capabilities}, and an abstract collective visualization may mitigate dis\textit{satisfaction}. More metrics, such as the Questionnaire for User Interface Satisfaction \cite{Chin1988}, are needed to properly assess how the transparency embedded in the models and visualizations influence operator \textit{satisfaction}.

The transparency embedded in the Collective visualization in general supported operators with individual differences better than the IA visualization. The $M_{2}$ model; however, did not support all operators. More computer experience, for example, aided operator \textit{SA performance}. Mitigating the need for operators to have particular \textit{capability} levels is desired in order to design \textit{effective} human-collective systems. Higher \textit{SA performance} also varied between the models, which suggests system design changes must be considered in order to improve the perception, comprehension, and projection of future collective behaviors when using the $M_{2}$ model. \textit{Usability} considerations need to identify the ideal amount of operator influence in the decision-making process in order to alleviate \textit{workload} (e.g., mental demand) and promote better \textit{SA}.

\section{$R_{2}$: System Design Element Promotion of Operator Comprehension}

\begin{figure}[!b]
\begin{center}
	\includegraphics[width=\textwidth]{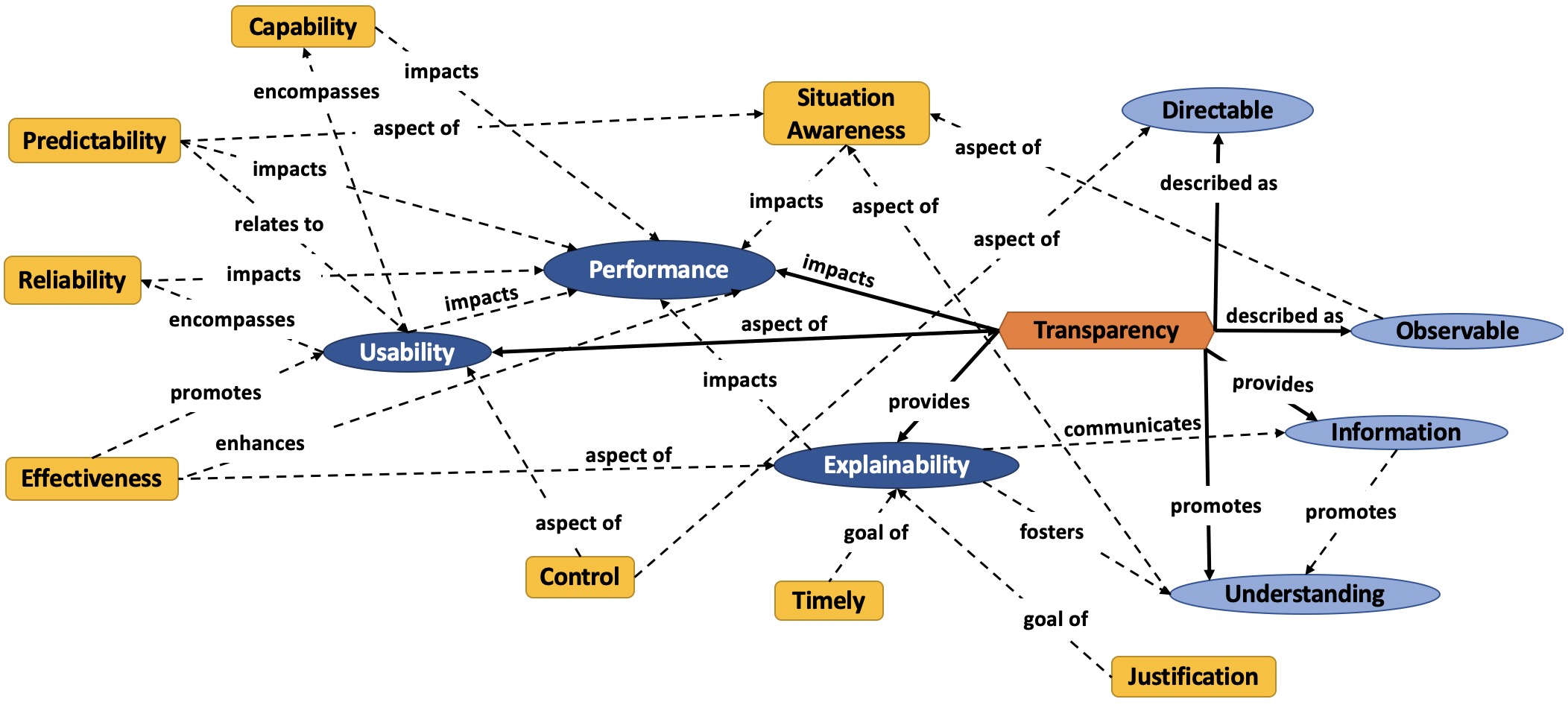}
	\caption{$R_{2}$ concept map of the assessed direct and indirect transparency factors.}
	\label{fig: Model Vis Concept Map R6}
	\end{center}
\end{figure}

The explainability direct transparency factor was explored in $R_{2}$, which was interested in determining whether the transparency embedded in \textit{the model and visualization promoted operator comprehension}. Perception and comprehension of the presented information are necessary to inform future operator actions. The associated objective dependent variables were (1) SA, (2) collective and target left- or right-clicks, (3) collective and target observations, (4) interventions, (5) the percentage of times the highest value target was abandoned, and (6) whether the information pop-up window was open when a target was abandoned. The specific direct and indirect transparency factors related to $R_{2}$ are identified in Figure \ref{fig: Model Vis Concept Map R6}. The relationship between the variables and the corresponding hypotheses, as well as the direct and indirect transparency factors, are identified in Table \ref{table:Comprehension,Variables}. Relationships between the variable and the direct or indirect transparency factors that are not shown in Figure \ref{fig: Concept Map}, were identified after conducting correlation analyses.

\begin{table}[h]
\centering
\caption{Interaction of system design elements promotion of human operator comprehension objective (obj) and subjective (subj) variables (vars), relationship to the hypotheses (H), as well as the associated direct and indirect transparency factors, are presented in Figure \ref{fig: Concept Map}.}
\label{table:Comprehension,Variables}
\begin{tabular}{?l|c?c|c|c|c|c|c|c?c|c|c|c|c|c|c|c?}
\Cline{1pt}{3-17}
\multicolumn{1}{c}{} & & \multicolumn{15}{c?}{\textbf{Transparency Factors}} \\ \Cline{1pt}{3-17}
\multicolumn{1}{c}{} & & \multicolumn{7}{c?}{\textbf{Direct}} & \multicolumn{8}{c?}{\textbf{Indirect}} \\ \Cline{1pt}{3-17}
\multicolumn{1}{c}{} & & {\multirow[b]{6}{*}{\rotatebox{90}{\textbf{Directable}}}} &  {\multirow[b]{6}{*}{\rotatebox{90}{\textbf{Explainability}}}} & {\multirow[b]{6}{*}{\rotatebox{90}{\textbf{Information}}}} & {\multirow[b]{6}{*}{\rotatebox{90}{\textbf{Observable}}}} & {\multirow[b]{6}{*}{\rotatebox{90}{\textbf{Performance}}}} &  {\multirow[b]{6}{*}{\rotatebox{90}{\textbf{Understanding}}}} & {\multirow[b]{6}{*}{\rotatebox{90}{\textbf{Usability}}}} & {\multirow[b]{6}{*}{\rotatebox{90}{\textbf{Capability}}}} & {\multirow[b]{6}{*}{\rotatebox{90}{\textbf{Control}}}} & {\multirow[b]{6}{*}{\rotatebox{90}{\textbf{Effectiveness}}}} & {\multirow[b]{6}{*}{\rotatebox{90}{\textbf{Justification}}}} & {\multirow[b]{6}{*}{\rotatebox{90}{\textbf{Predictability}}}}& {\multirow[b]{6}{*}{\rotatebox{90}{\textbf{Reliability}}}} & {\multirow[b]{6}{*}{\rotatebox{90}{\textbf{SA}}}} & {\multirow[b]{6}{*}{\rotatebox{90}{\textbf{Timing}}}} \\
\multicolumn{1}{c}{} & & & & & & & & & & & & & & & & \\
\multicolumn{1}{c}{} & & & & & & & & & & & & & & & & \\
\multicolumn{1}{c}{} & & & & & & & & & & & & & & & & \\
\multicolumn{1}{c}{} & & & & & & & & & & & & & & & & \\ \Cline{1pt}{1-2}
\multicolumn{1}{?c|}{\textbf{Obj Vars}} & {\textbf{H}} & & & & & & & & & & & & & & & \\ \Cline{1pt}{1-17}
{SA Probe} & \multirow{2}{*}{$H_{4}$} & & & & \multirow{2}{*}{\checkmark} & \multirow{2}{*}{\checkmark} & \multirow{2}{*}{\checkmark} & & & & \multirow{2}{*}{\checkmark} & & \multirow{2}{*}{\checkmark} & & \multirow{2}{*}{\checkmark} & \\
{Accuracy} & & & & & & & & & & & & & & & & \\ \hline
{Collective Left-} & \multirow{2}{*}{$H_{5}$} & & \multirow{2}{*}{\checkmark} & & & & & \multirow{2}{*}{\checkmark} & & & \multirow{2}{*}{\checkmark} & \multirow{2}{*}{\checkmark} & & & & \\ 
{Clicks} & & & & & & & & & & & & & & & & \\ \hline
{Target Right-} & \multirow{3}{*}{$H_{5}$} & & \multirow{3}{*}{\checkmark} & & & & & \multirow{3}{*}{\checkmark} & & & \multirow{3}{*}{\checkmark} & \multirow{3}{*}{\checkmark} & & & & \\ 
{Clicks by SA} & & & & & & & & & & & & & & & & \\
{Level} & & & & & & & & & & & & & & & & \\ \hline
{Collective} & \multirow{2}{*}{$H_{5}$} & & \multirow{2}{*}{\checkmark} & \multirow{2}{*}{\checkmark} & & & & \multirow{2}{*}{\checkmark} & & & \multirow{2}{*}{\checkmark} & \multirow{2}{*}{\checkmark} & & & & \\ 
{Observations} & & & & & & & & & & & & & & & & \\ \hline
{Target} & \multirow{2}{*}{$H_{4}$} & & \multirow{2}{*}{\checkmark} & & & & \multirow{2}{*}{\checkmark} & \multirow{2}{*}{\checkmark} & & & & & & & & \\
{Observations} & & & & & & & & & & & & & & & & \\ \hline
{Collective Right-} & \multirow{2}{*}{$H_{5}$} & & \multirow{2}{*}{\checkmark} & \multirow{2}{*}{\checkmark} & & & & \multirow{2}{*}{\checkmark} & & & \multirow{2}{*}{\checkmark} & \multirow{2}{*}{\checkmark} & & & & \\ 
{Clicks} & & & & & & & & & & & & & & & & \\ \hline
{Target Right-} & \multirow{3}{*}{$H_{5}$} & & \multirow{3}{*}{\checkmark} & \multirow{3}{*}{\checkmark} & & & & \multirow{3}{*}{\checkmark} & & & \multirow{3}{*}{\checkmark} & \multirow{3}{*}{\checkmark} & & & & \\ 
{Clicks per} & & & & & & & & & & & & & & & & \\
{Decision} & & & & & & & & & & & & & & & & \\ \hline
{Interventions} & {$H_{4}$} & {\checkmark} & {\checkmark} & & & & {\checkmark} & {\checkmark} & & {\checkmark} & {\checkmark} & & {\checkmark} & {\checkmark} & & \\ \hline
{Highest Value} & {$H_{4}$,} & \multirow{2}{*}{\checkmark} & \multirow{2}{*}{\checkmark} & & & & \multirow{2}{*}{\checkmark} & \multirow{2}{*}{\checkmark} & & \multirow{2}{*}{\checkmark} & \multirow{2}{*}{\checkmark} & \multirow{2}{*}{\checkmark} & & & & \\ 
{Target Abandon} & {$H_{5}$} & & & & & & & & & & & & & & & \\ \hline
{Abandon Target} & \multirow{3}{*}{$H_{5}$} & & \multirow{3}{*}{\checkmark} & & & & & \multirow{3}{*}{\checkmark} & & & \multirow{3}{*}{\checkmark} & \multirow{3}{*}{\checkmark} & & & & \\ 
{Information} & & & & & & & & & & & & & & & & \\
{Window Open} & & & & & & & & & & & & & & & & \\ \Cline{1pt}{1-17}
\multicolumn{1}{?c}{\textbf{Subj Vars}} & \multicolumn{16}{c?}{\textbf{}} \\ \Cline{1pt}{1-17}
{Post-Trial} & \multirow{3}{*}{$H_{4}$} & & \multirow{3}{*}{\checkmark} & & & \multirow{3}{*}{\checkmark} & \multirow{3}{*}{\checkmark} & & \multirow{3}{*}{\checkmark} & & & & & & & \\
{Performance and} & & & & & & & & & & & & & & & & \\ 
{Understanding} & & & & & & & & & & & & & & & & \\ \hline
{Post-Experiment} & {$H_{4}$} & & & & & & {\checkmark} & {\checkmark} & {\checkmark} & & & & & {\checkmark} & & {\checkmark} \\ \Cline{1pt}{1-17}
\end{tabular}
\end{table}

Models designed to aid operators to fulfill a best-of-\textit{n} decision making task can help mitigate workload by reducing repetitive interactions, ensuring task progress in case an operator becomes distracted, and allowing more time to establish situational awareness and understanding. Display principles, associated with perceptual operations, mental models, as well as human attention and memory \cite{Wickens2004}, may also improve understanding by providing information that is legible, clear, concise, organized, easily accessible, and consistent. Providing information, such as the collective state, on the collective's hub, rather than using all of the individual collective entities is more clear, concise, organized, and consistent. It was hypothesized ($H_{4}$) that operators will have a better understanding of the $M_{2}$ model and information provided by the Collective visualization. Appropriate expectations of the model's capabilities and contributions towards a goal, as well as providing information redundantly via icons, colors, messages, and the collective and target information pop-up windows can aid operator comprehension and justify their future actions. It was hypothesized ($H_{5}$) that operators using the $M_{2}$ model and the Collective visualization were able to accurately justify their actions. An ideal system will enable operators to perceive and comprehend information that is explainable, which will support effective human-collective interactions. 

\subsection{Metrics and Results}
\label{sec: R2 metrics}

\begin{table}[!b]
\centering
\caption{Collective left-clicks mean (SD) 15 seconds before asking, while being asked, and during response to SA probe question by SA level.}
\label{table:Comprehension,Collective Left-Clicks}
\begin{tabular}{c|c|c|c|c|}
\cline{2-5}
 & \textbf{Timing} & \textbf{SA Level} & \textbf{IA} & \textbf{Collective} \\ \hline
\multicolumn{1}{|c|}{\multirow{12}{*}{$M_{2}$}} & \multirow{4}{*}{Before} & $SA_{O}$ & 1.64 (1.84) & 1.95 (1.57) \\ \cline{3-5} 
\multicolumn{1}{|c|}{} & & \cellcolor{gray1}$SA_{1}$ & \cellcolor{gray1}1.53 (1.75) & \cellcolor{gray1}1.88 (1.47) \\ \cline{3-5} 
\multicolumn{1}{|c|}{} & & \cellcolor{gray2}$SA_{2}$ & \cellcolor{gray2}1.78 (1.9) & \cellcolor{gray2}2.13 (1.68) \\ \cline{3-5}
\multicolumn{1}{|c|}{} & & \cellcolor{gray3}$SA_{3}$ & \cellcolor{gray3}1.65 (1.92) & \cellcolor{gray3}1.83 (1.61) \\ \cline{2-5} 
\multicolumn{1}{|c|}{} & \multirow{4}{*}{Asking} & $SA_{O}$ & 0.49 (0.76) & 0.69 (0.88) \\ \cline{3-5} 
\multicolumn{1}{|c|}{} & & \cellcolor{gray1}$SA_{1}$ & \cellcolor{gray1}0.3 (0.6) & \cellcolor{gray1}0.51 (0.79) \\ \cline{3-5}
\multicolumn{1}{|c|}{} & & \cellcolor{gray2}$SA_{2}$ & \cellcolor{gray2}0.42 (0.77) & \cellcolor{gray2}0.91 (0.89) \\ \cline{3-5}
\multicolumn{1}{|c|}{} & & \cellcolor{gray3}$SA_{3}$ & \cellcolor{gray3}0.33 (0.61) & \cellcolor{gray3}0.73 (0.96) \\ \cline{2-5} 
\multicolumn{1}{|c|}{} & \multirow{4}{*}{Responding} & $SA_{O}$ & 1.68 (1.79) & 1.52 (1.21) \\ \cline{3-5} 
\multicolumn{1}{|c|}{} & & \cellcolor{gray1}$SA_{1}$ & \cellcolor{gray1}1.14 (1.46) & \cellcolor{gray1}1.32 (1.02) \\ \cline{3-5} 
\multicolumn{1}{|c|}{} & & \cellcolor{gray2}$SA_{2}$ & \cellcolor{gray2}1.46 (1.8) & \cellcolor{gray2}1.57 (1.21) \\ \cline{3-5}
\multicolumn{1}{|c|}{} & & \cellcolor{gray3}$SA_{3}$ & \cellcolor{gray3}1.53 (1.98) & \cellcolor{gray3}1.89 (1.48) \\ \hline
\multicolumn{1}{|c|}{\multirow{12}{*}{$M_{3}$}} & \multirow{4}{*}{Before} & $SA_{O}$ & 2.48 (2.28) & 2.58 (1.76) \\ \cline{3-5} 
\multicolumn{1}{|c|}{} & & \cellcolor{gray1}$SA_{1}$ & \cellcolor{gray1}2.42 (2.15) & \cellcolor{gray1}2.27 (1.73) \\ \cline{3-5} 
\multicolumn{1}{|c|}{} & & \cellcolor{gray2}$SA_{2}$ & \cellcolor{gray2}2.45 (2.33) & \cellcolor{gray2}2.71 (1.7) \\ \cline{3-5}
\multicolumn{1}{|c|}{} & & \cellcolor{gray3}$SA_{3}$ & \cellcolor{gray3}2.61 (2.43) & \cellcolor{gray3}2.79 (1.85) \\ \cline{2-5} 
\multicolumn{1}{|c|}{} & \multirow{4}{*}{Asking} & $SA_{O}$ & 0.63 (0.82) & 0.85 (0.83) \\ \cline{3-5} 
\multicolumn{1}{|c|}{} & & \cellcolor{gray1}$SA_{1}$ & \cellcolor{gray1}0.46 (0.65) & \cellcolor{gray1}0.88 (0.87) \\ \cline{3-5}
\multicolumn{1}{|c|}{} & & \cellcolor{gray2}$SA_{2}$ & \cellcolor{gray2}0.91 (0.89) & \cellcolor{gray2}0.83 (0.8) \\ \cline{3-5}
\multicolumn{1}{|c|}{} & & \cellcolor{gray3}$SA_{3}$ & \cellcolor{gray3}0.52 (0.88) & \cellcolor{gray3}0.85 (0.86) \\ \cline{2-5} 
\multicolumn{1}{|c|}{} & \multirow{4}{*}{Responding} & $SA_{O}$ & 2.02 (1.81) & 1.97 (1.39) \\ \cline{3-5} 
\multicolumn{1}{|c|}{} & & \cellcolor{gray1}$SA_{1}$ & \cellcolor{gray1}1.78 (1.7) & \cellcolor{gray1}1.83 (1.25) \\ \cline{3-5} 
\multicolumn{1}{|c|}{} & & \cellcolor{gray2}$SA_{2}$ & \cellcolor{gray2}2.21 (1.78) & \cellcolor{gray2}2.04 (1.43) \\ \cline{3-5}
\multicolumn{1}{|c|}{} & & \cellcolor{gray3}$SA_{3}$ & \cellcolor{gray3}2.16 (1.98) & \cellcolor{gray3}2.05 (1.5) \\ \hline
\end{tabular}
\end{table}

The operator had access to supplementary information that was not displayed continually, such as different colored target borders that identified which targets were in range and had been abandoned, or information pop-up windows that provided collective state and target support information, in order to aid comprehension ($SA_{2}$) of collective behavior and inform particular actions. The results of \textit{SA probe accuracy}, which is the percentage of correctly answered SA probes questions used to assess the operator's SA during a trial, identified that IA and Collective operators using the $M_{2}$ model had higher $SA_{3}$ compared to the $M_{3}$ model, while the IA operators had higher $SA_{2}$ and the Collective operators had higher $SA_{O}$. Operators using the Collective visualization had higher SA probe accuracy, regardless of the SA level, compared to the IA visualization. Further details regarding the statistical tests were provided in the Metrics and Results Section \ref{section:R1 metrics}. 

\def \figwidth {0.49}
\begin{figure*}[!b]
    \captionsetup[subfigure]{justification=centering}
    \begin{subfigure}{\figwidth\linewidth}
		\centering
		\includegraphics[keepaspectratio,height=1.71in]{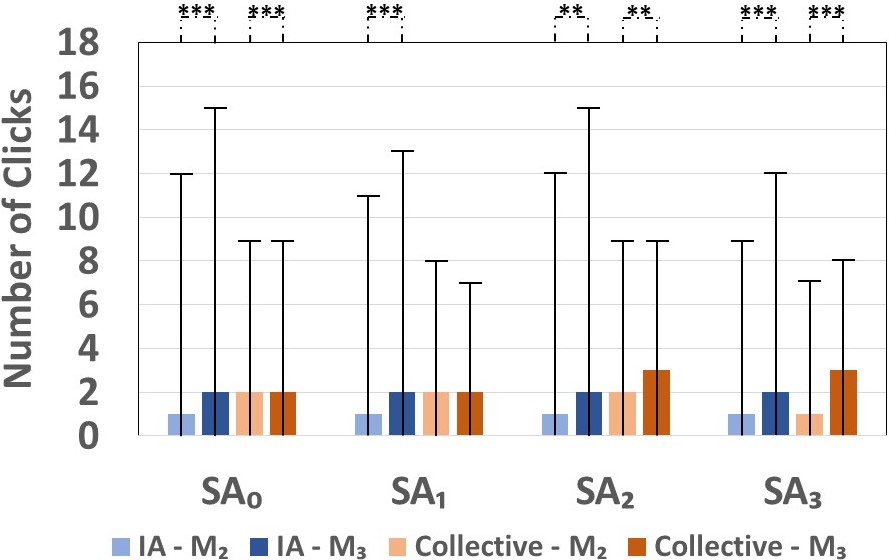}
		\captionsetup{width=\linewidth}
		\caption{15 seconds before asking a SA probe question.}
	\end{subfigure}
\begin{subfigure}{\figwidth\linewidth}
	\centering
	\includegraphics[keepaspectratio,height=1.71in]{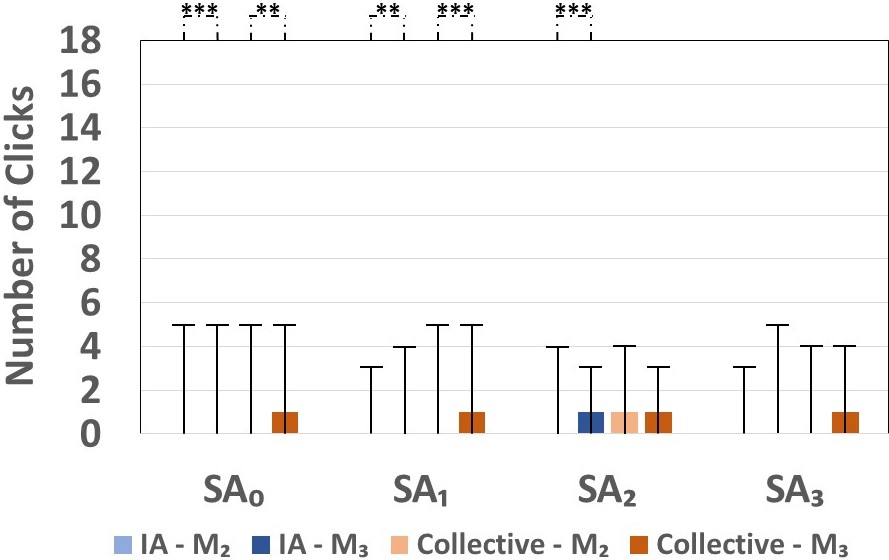}
	\captionsetup{width=\linewidth}
	\caption{While being asked a SA probe question.}
\end{subfigure}
\begin{subfigure}{\figwidth\linewidth}
	\centering
	\includegraphics[keepaspectratio,height=1.705in]{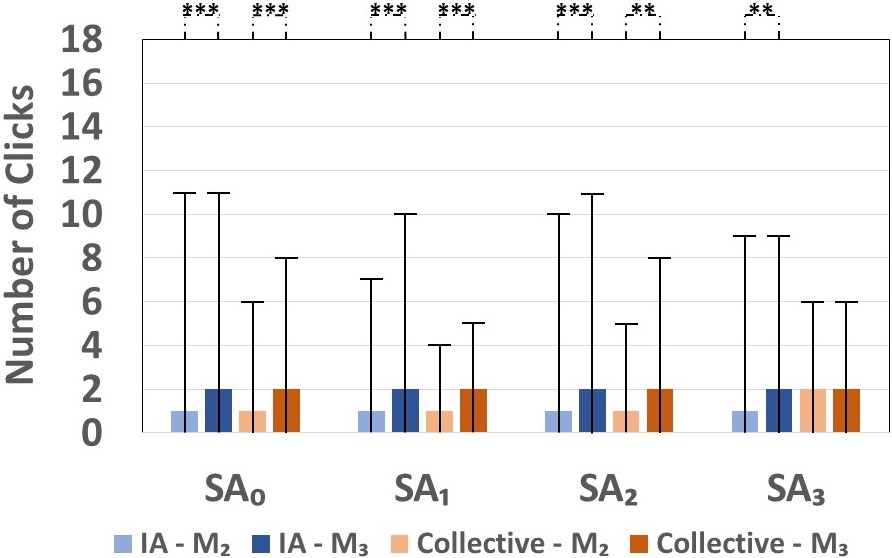}
	\captionsetup{width=\linewidth}
	\caption{During response to a SA probe question.}
\end{subfigure}
	\caption{Collective left-clicks median (min/max) and Mann-Whitney-Wilcoxin test by SA level between models a) 15 seconds before asking, b) while being asked, and c) during response to a SA probe question.}
	\label{fig: Collective Left-Clicks}
\end{figure*}

\textit{Collective left-clicks} identified all targets that were within range of a collective and was the first click required to issue a command. The number of collective left-clicks mean (SD) 15 seconds before asking, while being asked, and during response to a SA probe question are presented in Table \ref{table:Comprehension,Collective Left-Clicks} \citep{Roundtree2020visual}. The $M_{2}$ model in general had fewer collective left-clicks compared to the $M_{3}$ model. Collective operators using the $M_{3}$ model while being asked a SA probe question had fewer collective left-clicks for $SA_{2}$. The number of collective left-clicks median, min, max, and the Mann-Whitney-Wilcoxon significant effects between models are presented in Figure \ref{fig: Collective Left-Clicks}. IA operators had significantly different collective left-clicks between models for $SA_{O}$, $SA_{1}$, and $SA_{2}$ at all timings, as well as $SA_{3}$ 15 seconds before asking and during response to a SA probe question. Significantly different collective left-clicks between models were identified 15 seconds before asking a SA probe question for $SA_{O}$, $SA_{2}$, and $SA_{3}$, while being asked a SA probe question for $SA_{O}$ and $SA_{1}$, and during response to a SA probe question for $SA_{O}$, $SA_{1}$, and $SA_{2}$.

\begin{table}[!b]
\centering
\caption{Target right-clicks mean (SD) 15 seconds before asking, while being asked, and during response to SA probe question by SA level.}
\label{table:Comprehension,Target Right-Clicks}
\begin{tabular}{c|c|c|c|c|}
\cline{2-5}
 & \textbf{Timing} & \textbf{SA Level} & \textbf{IA} & \textbf{Collective} \\ \hline
\multicolumn{1}{|c|}{\multirow{12}{*}{$M_{2}$}} & \multirow{4}{*}{Before} & $SA_{O}$ & 1.68 (2.38) & 1.52 (2.41) \\ \cline{3-5} 
\multicolumn{1}{|c|}{} & & \cellcolor{gray1}$SA_{1}$ & \cellcolor{gray1}1.92 (2.62) & \cellcolor{gray1}1.79 (2.71) \\ \cline{3-5} 
\multicolumn{1}{|c|}{} & & \cellcolor{gray2}$SA_{2}$ & \cellcolor{gray2}1.28 (1.91) & \cellcolor{gray2}1.17 (1.94) \\ \cline{3-5}
\multicolumn{1}{|c|}{} & & \cellcolor{gray3}$SA_{3}$ & \cellcolor{gray3}1.8 (2.5) & \cellcolor{gray3}1.49 (2.35) \\ \cline{2-5} 
\multicolumn{1}{|c|}{} & \multirow{4}{*}{Asking} & $SA_{O}$ & 0.37 (0.79) & 0.5 (1) \\ \cline{3-5} 
\multicolumn{1}{|c|}{} & & \cellcolor{gray1}$SA_{1}$ & \cellcolor{gray1}0.44 (0.74) & \cellcolor{gray1}0.49 (0.86) \\ \cline{3-5}
\multicolumn{1}{|c|}{} & & \cellcolor{gray2}$SA_{2}$ & \cellcolor{gray2}0.31 (0.67) & \cellcolor{gray2}0.55 (1.31) \\ \cline{3-5} 
\multicolumn{1}{|c|}{} & & \cellcolor{gray3}$SA_{3}$ & \cellcolor{gray3}0.37 (0.75) & \cellcolor{gray3}0.44 (0.69) \\ \cline{2-5} 
\multicolumn{1}{|c|}{} & \multirow{4}{*}{Responding} & $SA_{O}$ & 1.07 (1.77) & 0.99 (1.7) \\ \cline{3-5} 
\multicolumn{1}{|c|}{} & & \cellcolor{gray1}$SA_{1}$ & \cellcolor{gray1}1.11 (1.69) & \cellcolor{gray1}1.01 (1.74) \\ \cline{3-5} 
\multicolumn{1}{|c|}{} & & \cellcolor{gray2}$SA_{2}$ & \cellcolor{gray2}1.1 (1.75) & \cellcolor{gray2}0.84 (1.44) \\ \cline{3-5} 
\multicolumn{1}{|c|}{} & & \cellcolor{gray3}$SA_{3}$ & \cellcolor{gray3}1.68 (2.24) & \cellcolor{gray3}1.21 (1.98) \\ \hline
\multicolumn{1}{|c|}{\multirow{12}{*}{$M_{3}$}} & \multirow{4}{*}{Before} & $SA_{O}$ & 1.04 (1.68) & 1.17 (2) \\ \cline{3-5} 
\multicolumn{1}{|c|}{} & & \cellcolor{gray1}$SA_{1}$ & \cellcolor{gray1}1.34 (1.54) & \cellcolor{gray1}1.44 (2.44) \\ \cline{3-5} 
\multicolumn{1}{|c|}{} & & \cellcolor{gray2}$SA_{2}$ & \cellcolor{gray2}0.79 (1.53) & \cellcolor{gray2}0.96 (1.63) \\ \cline{3-5}
\multicolumn{1}{|c|}{} & & \cellcolor{gray3}$SA_{3}$ & \cellcolor{gray3}0.88 (2.03) & \cellcolor{gray3}1.15 (1.88) \\ \cline{2-5} 
\multicolumn{1}{|c|}{} & \multirow{4}{*}{Asking} & $SA_{O}$ & 0.36 (0.86) & 0.42 (0.95) \\ \cline{3-5} 
\multicolumn{1}{|c|}{} & & \cellcolor{gray1}$SA_{1}$ & \cellcolor{gray1}0.42 (0.82) & \cellcolor{gray1}0.38 (0.96) \\ \cline{3-5}
\multicolumn{1}{|c|}{} & & \cellcolor{gray2}$SA_{2}$ & \cellcolor{gray2}0.29 (0.79) & \cellcolor{gray2}0.44 (0.98) \\ \cline{3-5} 
\multicolumn{1}{|c|}{} & & \cellcolor{gray3}$SA_{3}$ & \cellcolor{gray3}0.35 (1) & \cellcolor{gray3}0.45 (0.9) \\ \cline{2-5} 
\multicolumn{1}{|c|}{} & \multirow{4}{*}{Responding} & $SA_{O}$ & 0.89 (1.64) & 0.72 (1.3) \\ \cline{3-5} 
\multicolumn{1}{|c|}{} & & \cellcolor{gray1}$SA_{1}$ & \cellcolor{gray1}0.91 (1.69) & \cellcolor{gray1}0.67 (1.49) \\ \cline{3-5} 
\multicolumn{1}{|c|}{} & & \cellcolor{gray2}$SA_{2}$ & \cellcolor{gray2}0.8 (1.68) & \cellcolor{gray2}0.72 (1.17) \\ \cline{3-5} 
\multicolumn{1}{|c|}{} & & \cellcolor{gray3}$SA_{3}$ & \cellcolor{gray3}0.98 (1.52) & \cellcolor{gray3}0.8 (1.24) \\ \hline
\end{tabular}
\end{table}

Additional between visualizations Mann-Whitney-Wilcoxon tests identified highly significant effects when using the $M_{2}$ model 15 seconds before asking a SA probe question for $SA_{O}$ (n = 664, U = 64213, $\rho$ $<$ 0.001), a moderate significant effect for $SA_{1}$ (n = 290, U = 12534, $\rho$ $<$ 0.01), and a significant effect for $SA_{2}$ (n = 223, U = 7210.5, $\rho$ = 0.04). Highly significant effects between visualizations when using the $M_{2}$ model while being asked a SA probe question were found for $SA_{O}$ (U = 67670, $\rho$ $<$ 0.001), and $SA_{2}$ (U = 8317 $\rho$ $<$ 0.001), as were moderately significant effects for $SA_{1}$ (U = 12043, $\rho$ $<$ 0.01), and $SA_{3}$ (n = 151, U = 3472, $\rho$ $<$ 0.01). A highly significant effect between visualizations when using the $M_{2}$ model during response to a SA probe question was found for $SA_{O}$ (U = 64710, $\rho$ $<$ 0.001), a moderate significant effect for $SA_{1}$ (U = 12414, $\rho$ $<$ 0.01), and a significant effect for $SA_{3}$ (U = 3489, $\rho$ = 0.01). A significant effect between visualizations when using the $M_{3}$ model 15 seconds before asking a SA probe question was found for $SA_{O}$ (n = 665, U = 60696, $\rho$ = 0.03). Highly significant effects between visualizations when using the $M_{3}$ model while being asked a SA probe question were found for $SA_{O}$ (U = 64376, $\rho$ $<$ 0.001), $SA_{1}$ (n = 251, U = 9959.5, $\rho$ $<$ 0.001), as well as a moderate significant effect for $SA_{3}$ (n = 162, U = 4114, $\rho$ $<$ 0.01). Correlations between the collective left-clicks and SA probe accuracy were only revealed when using the $M_{3}$ model. The Spearman correlation analysis revealed weak correlations for the IA visualization for $SA_{3}$ 15 seconds before asking (r = -0.26, $\rho$ = 0.02) and while being asked a SA probe question (r = -0.33, $\rho$ $<$ 0.01). Weak correlations were also revealed for the Collective visualization while being asked a SA probe question for $SA_{O}$ (r = 0.13, $\rho$ = 0.02) and $SA_{1}$ (r = 0.22, $\rho$ = 0.02). The IA visualization had fewer collective left-clicks in general compared to the Collective visualization. Collective operators who used the $M_{2}$ model during response to a SA probe question had fewer left-clicks for $SA_{O}$, and when using the $M_{3}$ model for all SA levels. 

\def \figwidth {0.49}
\begin{figure*}[!b]
    \captionsetup[subfigure]{justification=centering}
    \begin{subfigure}{\figwidth\linewidth}
		\centering
		\includegraphics[keepaspectratio,height=1.71in]{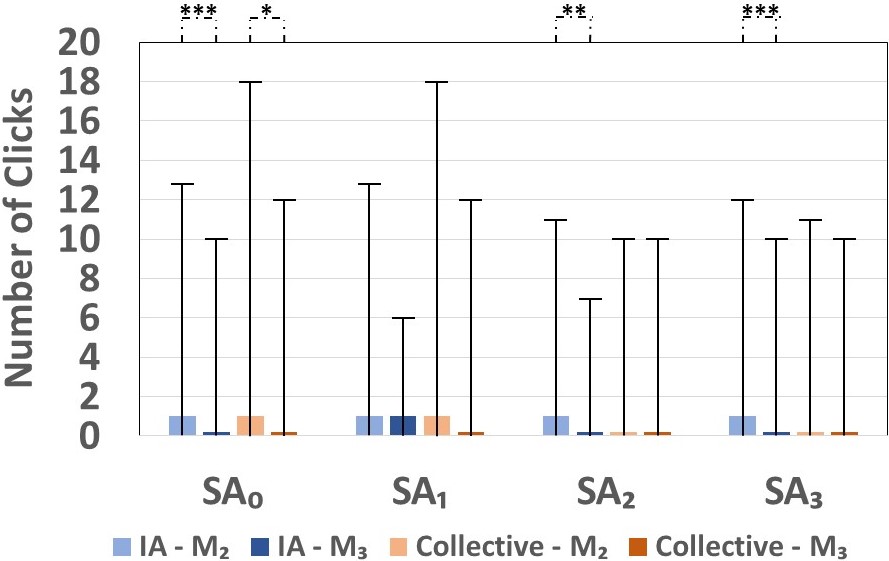}
		\captionsetup{width=\linewidth}
		\caption{15 seconds before asking a SA probe question.}
	\end{subfigure}
\begin{subfigure}{\figwidth\linewidth}
	\centering
	\includegraphics[keepaspectratio,height=1.71in]{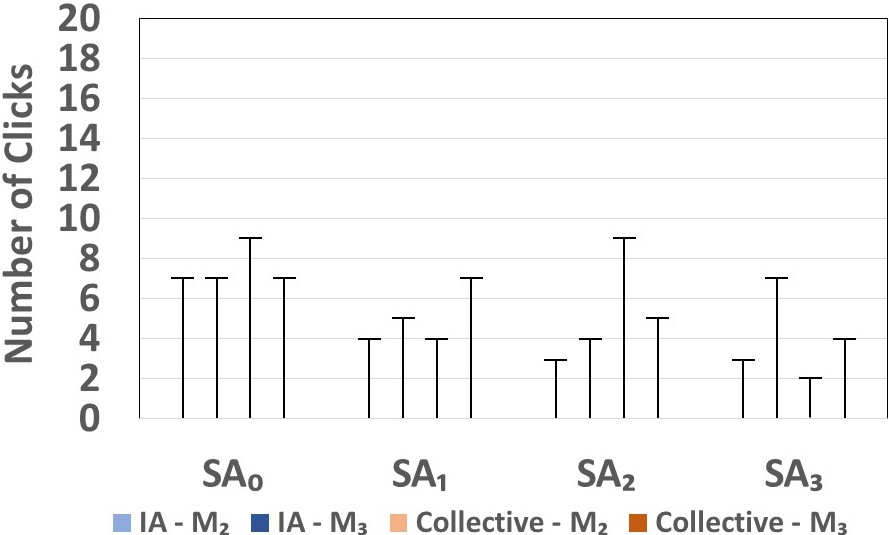}
	\captionsetup{width=\linewidth}
	\caption{While being asked a SA probe question.}
\end{subfigure}
\begin{subfigure}{\figwidth\linewidth}
	\centering
	\includegraphics[keepaspectratio,height=1.705in]{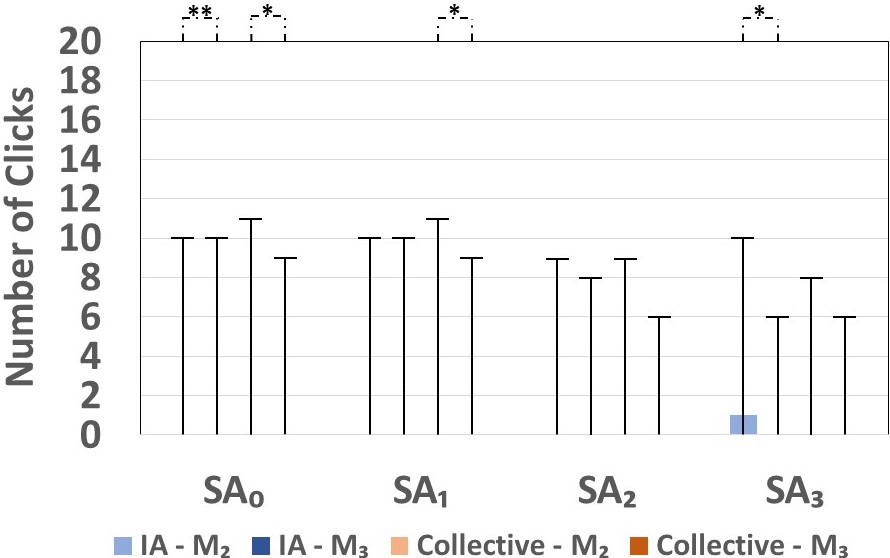}
	\captionsetup{width=\linewidth}
	\caption{During response to a SA probe question.}
\end{subfigure}
	\caption{Target right-clicks median (min/max) and Mann-Whitney-Wilcoxin test by SA level between models a) 15 seconds before asking, b) while being asked, and c) during response to a SA probe question.}
	\label{fig: Target Right-Clicks}
\end{figure*}

\textit{Target right-clicks} allowed the operator to access target information pop-up windows that provided each collective's percentage of support for a respective target. Operators may have used the support information to justify issuing commands. The number of target right-clicks mean (SD) 15 seconds before asking, while being asked, and during response to a SA probe question are shown in Table \ref{table:Comprehension,Target Right-Clicks} \citep{Roundtree2020visual}. The $M_{2}$ model in general had fewer target right-clicks for both visualizations. Collective operators who used the $M_{3}$ model while being asked a SA probe question had fewer target right-clicks for $SA_{3}$. The number of target right-clicks median, min, max, and the Mann-Whitney-Wilcoxon significant effects between models are shown in Figure \ref{fig: Target Right-Clicks}. IA operators had significantly different collective left-clicks between models 15 seconds before asking a SA probe question for $SA_{O}$, $SA_{2}$, and $SA_{3}$, as well as during response to a SA probe question for $SA_{O}$ and $SA_{3}$. Significantly different collective left-clicks between models were found 15 seconds before asking a SA probe question for $SA_{O}$ and during response to a SA probe question for $SA_{O}$ and $SA_{1}$. No significant effects between visualizations were found. The Collective visualization using the $M_{2}$ model had fewer target right-clicks for all SA levels, 15 seconds before asking and during response to a SA probe question compared to the IA visualization. Fewer target right-clicks, 15 seconds before asking and while being asked a SA probe question, occurred when IA operators used the $M_{3}$ model compared to the Collective visualization. The Spearman correlation analysis revealed weak correlations between the number of target right-clicks and SA probe accuracy for the IA visualization using the $M_{2}$ model 15 seconds before asking a SA probe question for $SA_{O}$ (r = 0.17, $\rho$ $<$ 0.01) and $SA_{2}$ (r = 0.37, $\rho$ $<$ 0.001). Weak correlations were found for the IA visualization using the $M_{3}$ model 15 seconds before asking a SA probe question for $SA_{O}$ (r = 0.11, $\rho$ = 0.04), $SA_{1}$ (r = 0.2, $\rho$ = 0.02), and for the Collective visualization for $SA_{1}$ 15 seconds before asking (r = -0.24, $\rho$ = 0.01) and while being asked a SA probe question (r = -0.21, $\rho$ = 0.03). 

\textit{Collective observations} were the subset of collective left-clicks that only identified targets within range of a collective (i.e., white borders indicated that the individual collective entities were investigating the target, while yellow indicated no investigation) and whether the targets had been abandoned (i.e., red borders). The percentage of collective left-clicks associated with collective observations mean (SD) by decision difficulty are shown in Table \ref{table:Comprehension,Collective Observations} \citep{Cody2020}. IA operators using the $M_{3}$ model had fewer collective observations compared to the $M_{2}$ model, while Collective operators had fewer collective observations when using the $M_{2}$ model. The collective observations median, min, max, and the Mann-Whitney-Wilcoxon significant effects between models are presented in Figure \ref{fig: Collective Observations}. IA operators had significantly different collective observations between models for all decision difficulties, while Collective operators had significantly different collective observations between models for easy decisions. Additional between visualizations Mann-Whitney-Wilcoxon tests identified a moderate significant effect when using the $M_{2}$ model for overall (n = 672, U = 61152, $\rho$ $<$ 0.01) and a significant effect for easy decisions (n = 374, U = 19008, $\rho$ = 0.05). Highly significant effects between visualizations when using the $M_{3}$ model were found for overall (U = 73920, $\rho$ $<$ 0.001), easy (n = 396, U = 25587, $\rho$ $<$ 0.001), and hard decisions (n = 276, U = 12520, $\rho$ $<$ 0.001). The IA visualization had fewer collective observations compared to the Collective visualization.

\begin{table}[h]
\begin{minipage}{0.5\linewidth}
\centering
\caption{Collective observations (\%) mean (SD) by decision difficulty (Dec Diff).}
\label{table:Comprehension,Collective Observations}
\begin{tabular}{c|c|c|c|}
\cline{2-4}
 & \textbf{Dec Diff} & \textbf{IA} & \textbf{Collective} \\ \hline
\multicolumn{1}{|c|}{\multirow{3}{*}{$M_{2}$}} & Overall & 77.68 (41.7) & 86.01 (34.74) \\ \cline{2-4} 
\multicolumn{1}{|c|}{} & \cellcolor{gray1}Easy & \cellcolor{gray1}71.13 (45.43) & \cellcolor{gray1}80 (40.11) \\ \cline{2-4} 
\multicolumn{1}{|c|}{} & \cellcolor{gray2}Hard & \cellcolor{gray2}86.62 (34.16) & \cellcolor{gray2}92.95 (25.68) \\ \hline
\multicolumn{1}{|c|}{\multirow{3}{*}{$M_{3}$}} & Overall & 59.23 (49.21) & 90.18 (29.8) \\ \cline{2-4} 
\multicolumn{1}{|c|}{} & \cellcolor{gray1}Easy & \cellcolor{gray1}57.79 (49.51) & \cellcolor{gray1}88.32 (32.19) \\ \cline{2-4} 
\multicolumn{1}{|c|}{} & \cellcolor{gray2}Hard & \cellcolor{gray2}61.31 (48.88) & \cellcolor{gray2}92.81 (25.93) \\ \hline
\end{tabular}
\end{minipage} 
\hfill
\begin{minipage}{0.45\linewidth}
\centering
	\includegraphics[width=62mm]{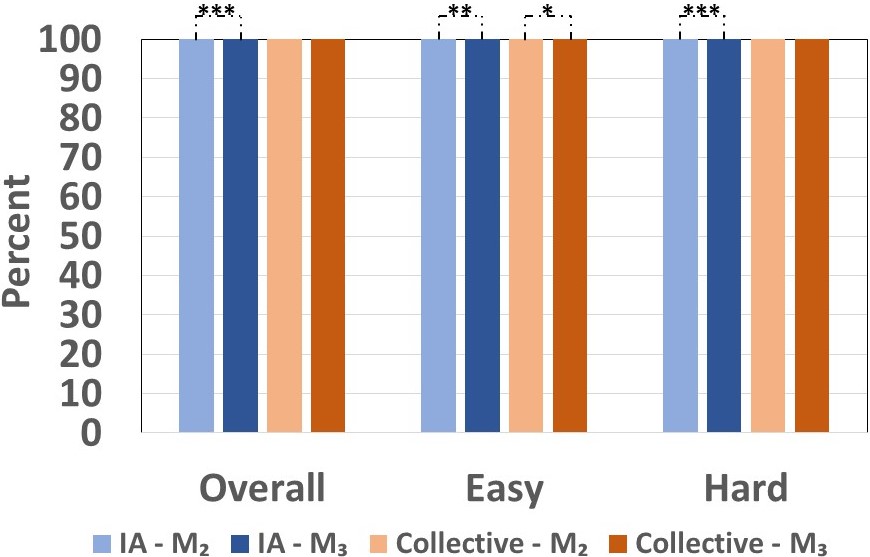}
	\captionof{figure}{Collective observations median (min/max) and Mann-Whitney-Wilcoxin test by decision difficulty between models.}
    \label{fig: Collective Observations}
\end{minipage}
\end{table}

\textit{Target observations} represent the subset of target left-clicks not associated with issuing a command. The percentage mean (SD) for target left-clicks that were target observations by decision difficulty are shown in Table \ref{table:Comprehension,Target Observations} \citep{Cody2020}. The $M_{2}$ model and Collective visualization had fewer target observations, regardless of decision difficulty. The target observations median, min, max, and the Mann-Whitney-Wilcoxon significant effects between models are presented in Figure \ref{fig: Target Observations}. IA operators had significantly different target observations between models for overall decisions, while Collective operators had significantly different target observations between models for all decision difficulties. Additional between visualizations Mann-Whitney-Wilcoxon tests identified highly significant effects when using the $M_{2}$ model for overall (n = 672, U = 35280, $\rho$ $<$ 0.001), easy (n = 374, U = 10886, $\rho$ $<$ 0.001), and hard decisions (n = 298, U = 6910, $\rho$ $<$ 0.001). Highly significant effects between visualizations when using the $M_{3}$ model were also found for overall (U = 41664, $\rho$ $<$ 0.001), easy (n = 396, U = 15053, $\rho$ $<$ 0.001), and hard decisions (n = 276, U = 6615, $\rho$ $<$ 0.001).

\begin{table}[h]
\begin{minipage}{0.5\linewidth}
\centering
\caption{Target observations (\%) mean (SD) by decision difficulty (Dec Diff).}
\label{table:Comprehension,Target Observations}
\begin{tabular}{c|c|c|c|}
\cline{2-4}
& \textbf{Dec Diff} & \textbf{IA} & \textbf{Collective} \\ \hline
\multicolumn{1}{|c|}{\multirow{3}{*}{$M_{2}$}} & Overall & 60.12 (49.04) & 22.62 (41.9) \\ \cline{2-4} 
\multicolumn{1}{|c|}{} & \cellcolor{gray1}Easy & \cellcolor{gray1}58.76 (49.35) & \cellcolor{gray1}21.11 (40.92) \\ \cline{2-4} 
\multicolumn{1}{|c|}{} & \cellcolor{gray2}Hard & \cellcolor{gray2}61.97 (48.72) & \cellcolor{gray2}24.36 (43.06) \\ \hline
\multicolumn{1}{|c|}{\multirow{3}{*}{$M_{3}$}} & Overall & 67.26 (47) & 41.07 (49.27) \\ \cline{2-4} 
\multicolumn{1}{|c|}{} & \cellcolor{gray1}Easy & \cellcolor{gray1}64.32 (48.03) & \cellcolor{gray1}41.12 (49.33) \\ \cline{2-4} 
\multicolumn{1}{|c|}{} & \cellcolor{gray2}Hard & \cellcolor{gray2}71.53 (45.29) & \cellcolor{gray2}41.01 (49.36) \\ \hline
\end{tabular}
\end{minipage} 
\hfill
\begin{minipage}{0.45\linewidth}
\centering
	\includegraphics[width=62mm]{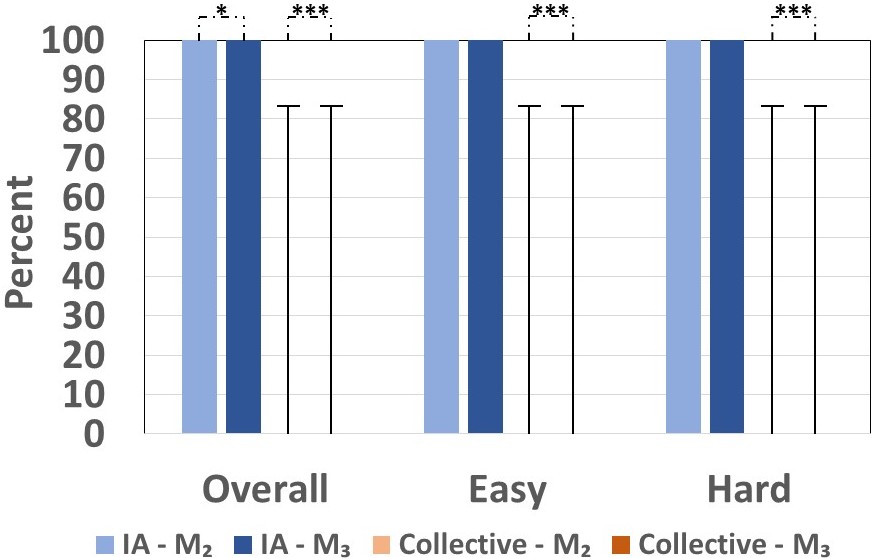}
	\captionof{figure}{Target observations median (min/max) and Mann-Whitney-Wilcoxin test by decision difficulty between models.}
    \label{fig: Target Observations}
\end{minipage}
\end{table}

\textit{Collective right-clicks} allowed the operator to open or close collective information pop-up windows, which provided the number of individual collective entities in each decision-making state. Operators may have used the information to justify issuing commands. The number of collective right-clicks per decision was only assessed for the IA evaluation, because the Collective evaluation did not record which particular collective pop-up window was opened or closed. The number of collective right-clicks mean (SD) per decision difficulty are presented in Table \ref{table:Comprehension,Collective Right Clicks Per Decision}. The $M_{3}$ model had fewer collective right-clicks compared to the $M_{2}$ model, regardless of decision difficulty. The collective right-clicks median, min, max, and the Mann-Whitney-Wilcoxon significant effects between models are presented in Figure \ref{fig: Collective Right-Clicks}. Significantly different collective right-clicks between models were found for overall and hard decisions.

\begin{table}[h]
\begin{minipage}{0.5\linewidth}
\centering
\caption{Collective right-clicks per decision mean (SD) by decision difficulty (Dec Diff).}
\label{table:Comprehension,Collective Right Clicks Per Decision}
\begin{tabular}{c|c|c|}
\cline{2-3}
& \textbf{Dec Diff} & \textbf{IA} \\ \hline
\multicolumn{1}{|c|}{\multirow{3}{*}{$M_{2}$}} & Overall & 1.55 (2.25) \\ \cline{2-3} 
\multicolumn{1}{|c|}{} & \cellcolor{gray1}Easy & \cellcolor{gray1}1.16 (1.87) \\ \cline{2-3} 
\multicolumn{1}{|c|}{} & \cellcolor{gray2}Hard & \cellcolor{gray2}2.09 (2.6) \\ \hline
\multicolumn{1}{|c|}{\multirow{3}{*}{$M_{3}$}} & Overall & 0.88 (2.2) \\ \cline{2-3} 
\multicolumn{1}{|c|}{} & \cellcolor{gray1}Easy & \cellcolor{gray1}0.87 (2.54) \\ \cline{2-3} 
\multicolumn{1}{|c|}{} & \cellcolor{gray2}Hard & \cellcolor{gray2}0.89 (1.57) \\ \hline
\end{tabular}
\end{minipage} 
\hfill
\begin{minipage}{0.45\linewidth}
\centering
	\includegraphics[width=62mm]{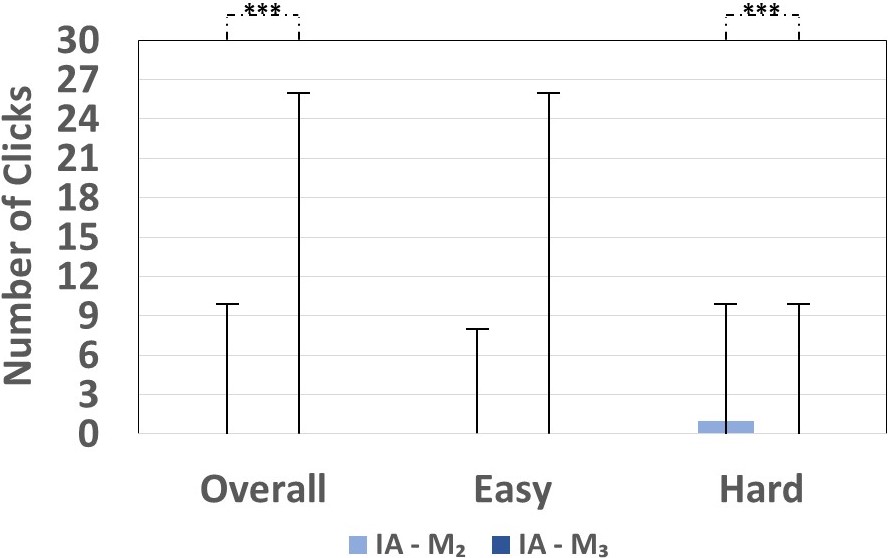}
	\captionof{figure}{Collective right-clicks median (min/max) and Mann-Whitney-Wilcoxin test by decision difficulty between models.}
    \label{fig: Collective Right-Clicks}
\end{minipage}
\end{table}

\textit{Target right-clicks} allowed the operator to open or close target information pop-up windows, which provided the percentage of support each collective had for a respective target. The target support information may have also been used to justify issuing commands, such as increasing or decreasing support from particular collectives. The mean (SD) for the number of target right-clicks by decision difficulty are presented in Table \ref{table:Comprehension,Target Right Clicks Per Decision}. IA operators using the $M_{2}$ model had fewer target right-clicks, while Collective operators had fewer target right-clicks using the $M_{3}$ model. The target right-clicks median, min, max, and the Mann-Whitney-Wilcoxon significant effects between models are presented in Figure \ref{fig: Target Right-Clicks DD}. IA operators had significantly different target right-clicks between models for easy decisions, while no differences were found for the Collective operators. The Collective visualization had fewer target right-clicks compared to the IA visualization; however, no significant effects between visualizations were found.

\begin{table}[h]
\begin{minipage}{0.5\linewidth}
\centering
\caption{Target right-clicks per decision mean (SD) by decision difficulty (Dec Diff).}
\label{table:Comprehension,Target Right Clicks Per Decision}
\begin{tabular}{c|c|c|c|}
\cline{2-4}
& \textbf{Dec Diff} & \textbf{IA} & \textbf{Collective} \\ \hline
\multicolumn{1}{|c|}{\multirow{3}{*}{$M_{2}$}} & Overall & 3.54 (4.18) & 3.09 (3.56) \\ \cline{2-4} 
\multicolumn{1}{|c|}{} & \cellcolor{gray1}Easy & \cellcolor{gray1}2.64 (3.14) & \cellcolor{gray1}2.61 (2.87) \\ \cline{2-4} 
\multicolumn{1}{|c|}{} & \cellcolor{gray2}Hard & \cellcolor{gray2}4.77 (5.03) & \cellcolor{gray2}3.64 (4.17) \\ \hline
\multicolumn{1}{|c|}{\multirow{3}{*}{$M_{3}$}} & Overall & 3.75 (5.38) & 3.04 (3.49) \\ \cline{2-4} 
\multicolumn{1}{|c|}{} & \cellcolor{gray1}Easy & \cellcolor{gray1}3.8 (5.82) & \cellcolor{gray1}2.95 (3.46) \\ \cline{2-4} 
\multicolumn{1}{|c|}{} & \cellcolor{gray2}Hard & \cellcolor{gray2}3.67 (4.69) & \cellcolor{gray2}3.15 (3.54) \\ \hline
\end{tabular}
\end{minipage} 
\hfill
\begin{minipage}{0.45\linewidth}
\centering
	\includegraphics[width=62mm]{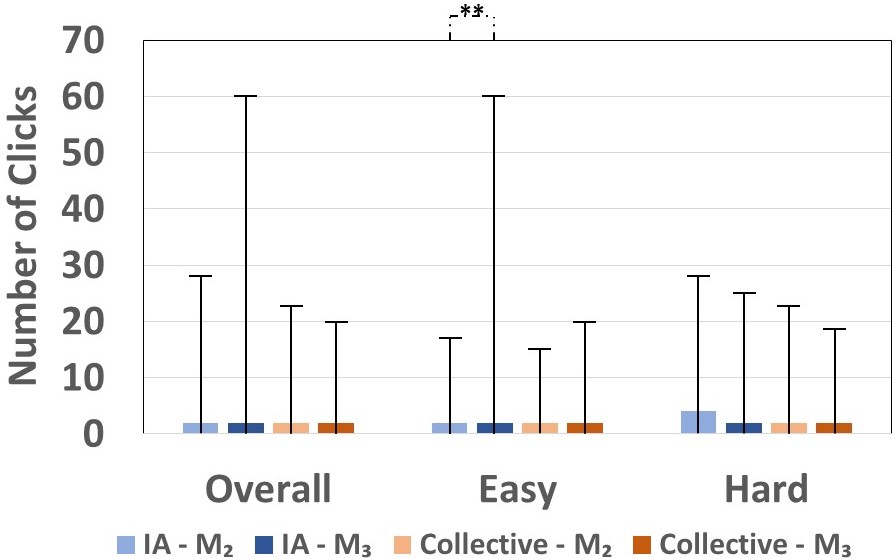}
	\captionof{figure}{Target right-clicks median (min/max) and Mann-Whitney-Wilcoxin test by decision difficulty between models.}
    \label{fig: Target Right-Clicks DD}
\end{minipage}
\end{table}

\textit{Interventions} occurred when the operator abandoned a target with greater than 10\% collective support. Abandoning low-value targets was a desired intervention. Interventions were assessed per participant, due to the inability to associate an intervention to a decision, and the descriptive statistics are shown in Table \ref{table:Comprehension,Interventions Per Operator} \citep{Cody2020}. The $M_{2}$ model and IA visualization had fewer interventions. The Mann-Whitney-Wilcoxon tests found a significant effect between models for the IA visualization (n = 56, U = 270.5, $\rho$ = 0.04). No significant effects between visualizations were found.

\begin{table}[h]
\centering
\caption{Interventions (abandoned targets with 10\% support) per participant descriptive statistics.}
\label{table:Comprehension,Interventions Per Operator}
\begin{tabular}{c|c|c|c|}
\cline{2-4}
 & \textbf{Model} & \textbf{Mean (SD)} & \textbf{Median (Min/Max)} \\ \hline
\multicolumn{1}{|c|}{\multirow{2}{*}{IA}} & {$M_{2}$} & 1.5 (2.03) & 0.5 (0/7) \\ \cline{2-4}
\multicolumn{1}{|c|}{} & \cellcolor{gray1}{$M_{3}$} & \cellcolor{gray1}3.75 (4.27 & \cellcolor{gray1}3 (0/17) \\ \hline
\multicolumn{1}{|c|}{\multirow{2}{*}{Collective}} & {$M_{2}$} & 2.21 (1.99) & 1.5 (0/7) \\ \cline{2-4}
\multicolumn{1}{|c|}{} & \cellcolor{gray1}{$M_{3}$} & \cellcolor{gray1}5 (5.11) & \cellcolor{gray1}3.5 (0/18) \\ \hline
\end{tabular}
\end{table}

The abandon command discontinued a collective's investigation of a particular target. Ideally lower valued targets were abandoned, since the objective was to aid each collective in selecting and moving to the highest valued target. The percentage of times the \textit{highest value target was abandoned} per participant mean (SD) are presented in Table \ref{table:Comprehension,Highest Value Target Abandoned} \citep{Roundtree2020visual}. Operators using the $M_{3}$ model abandoned the highest value target less frequently compared to the $M_{2}$ model. The highest value target abandoned median, min, max, and the Mann-Whitney-Wilcoxon significant effects between models are presented in Figure \ref{fig: High Value Target Abandoned}. IA operators had significantly different highest value target abandoned percentages between models for easy decisions, while Collective operators had significant differences between models for overall decisions. Operators using the IA visualization abandoned the highest value target less frequently compared to those using the Collective visualization; however, no significant effects were found between the visualizations.

\begin{table}[h]
\begin{minipage}{0.5\linewidth}
\centering
\caption{Highest value target abandoned (\%) mean (SD) per participant by decision difficulty (Dec Diff).}
\label{table:Comprehension,Highest Value Target Abandoned}
\begin{tabular}{c|c|c|c|}
\cline{2-4}
& \textbf{Dec Diff} & \textbf{IA} & \textbf{Collective} \\ \hline
\multicolumn{1}{|c|}{\multirow{3}{*}{$M_{2}$}} & Overall & 32.36 (29.53) & 43.6 (31.94) \\ \cline{2-4} 
\multicolumn{1}{|c|}{} & \cellcolor{gray1}Easy & \cellcolor{gray1}31.2 (27.17) & \cellcolor{gray1}33.25 (35.96) \\ \cline{2-4} 
\multicolumn{1}{|c|}{} & \cellcolor{gray2}Hard & \cellcolor{gray2}42.1 (40.53) & \cellcolor{gray2}48.72 (36.85) \\ \hline
\multicolumn{1}{|c|}{\multirow{3}{*}{$M_{3}$}} & Overall & 18.56 (18.38) & 21.04 (21.19) \\ \cline{2-4} 
\multicolumn{1}{|c|}{} & \cellcolor{gray1}Easy & \cellcolor{gray1}11.35 (20.82) & \cellcolor{gray1}11.64 (14.26) \\ \cline{2-4} 
\multicolumn{1}{|c|}{} & \cellcolor{gray2}Hard & \cellcolor{gray2}22.47 (11.89) & \cellcolor{gray2}28.09 (22.39) \\ \hline
\end{tabular}
\end{minipage} 
\hfill
\begin{minipage}{0.45\linewidth}
\centering
	\includegraphics[width=62mm]{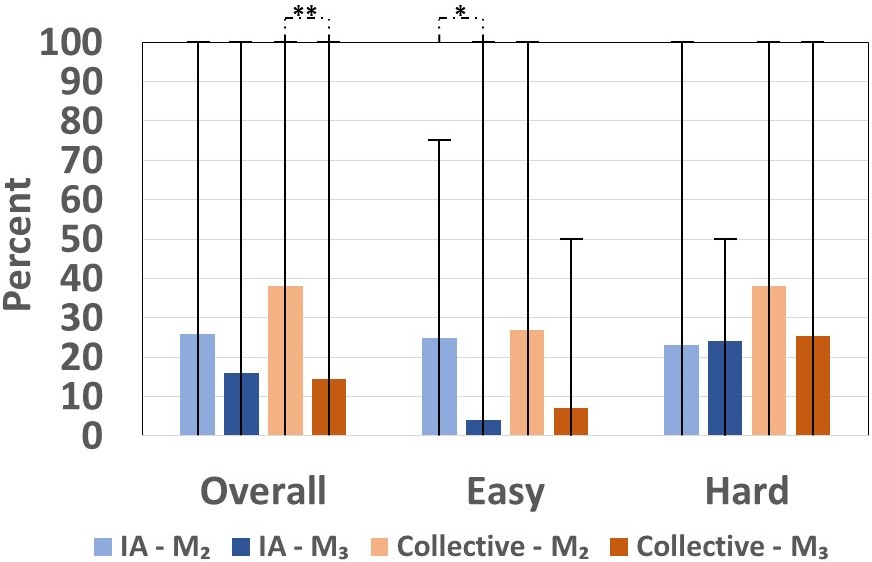}
	\captionof{figure}{Highest value target abandoned median (min/max) and Mann-Whitney-Wilcoxin test by decision difficulty between models.}
    \label{fig: High Value Target Abandoned}
\end{minipage}
\end{table}

The percentage of times an \textit{abandoned target information pop-up window was open} per participant was evaluated and the mean (SD) are presented in Table \ref{table:Comprehension,Info Window Open} \citep{Roundtree2020visual}. The operator may have used the support information in order to justify abandoning a target. Operators using the $M_{3}$ model had fewer abandoned target information pop-up windows open compared to the $M_{2}$ model. The abandoned target information pop-up window open median, min, max, and the Mann-Whitney-Wilcoxon significant effects between models are presented in Figure \ref{fig: High Value Target Window Open}, but no significant effects between models were found. Additional between visualizations Mann-Whitney-Wilcoxon tests identified significant effects when using the $M_{3}$ model for overall (n = 49, U = 414.5, $\rho$ = 0.02) and easy decisions (n = 45, U = 352, $\rho$ = 0.02). Fewer abandoned target information pop-up windows were open when using the IA visualization compared to the Collective visualization.

\begin{table}[h]
\begin{minipage}{0.5\linewidth}
\centering
\caption{Abandoned target information pop-up window open (\%) mean (SD) per participant by decision difficulty (Dec Diff).}
\label{table:Comprehension,Info Window Open}
\begin{tabular}{c|c|c|c|}
\cline{2-4}
& \textbf{Dec Diff} & \textbf{IA} & \textbf{Collective} \\ \hline
\multicolumn{1}{|c|}{\multirow{3}{*}{$M_{2}$}} & Overall & 23.86 (31.43) & 33.8 (34.9) \\ \cline{2-4} 
\multicolumn{1}{|c|}{} & \cellcolor{gray1}Easy & \cellcolor{gray1}22.2 (30.95) & \cellcolor{gray1}30.7 (37.85) \\ \cline{2-4} 
\multicolumn{1}{|c|}{} & \cellcolor{gray2}Hard & \cellcolor{gray2}28.7 (37.89) & \cellcolor{gray2}36.08 (40.87) \\ \hline
\multicolumn{1}{|c|}{\multirow{3}{*}{$M_{3}$}} & Overall & 8.48 (15.6) & 26.96 (35.48) \\ \cline{2-4} 
\multicolumn{1}{|c|}{} & \cellcolor{gray1}Easy & \cellcolor{gray1}9.17 (16.13) & \cellcolor{gray1}28.18 (34.02) \\ \cline{2-4} 
\multicolumn{1}{|c|}{} & \cellcolor{gray2}Hard & \cellcolor{gray2}8.65 (18.5) & \cellcolor{gray2}25.18 (38.69) \\ \hline
\end{tabular}
\end{minipage} 
\hfill
\begin{minipage}{0.45\linewidth}
\centering
	\includegraphics[width=62mm]{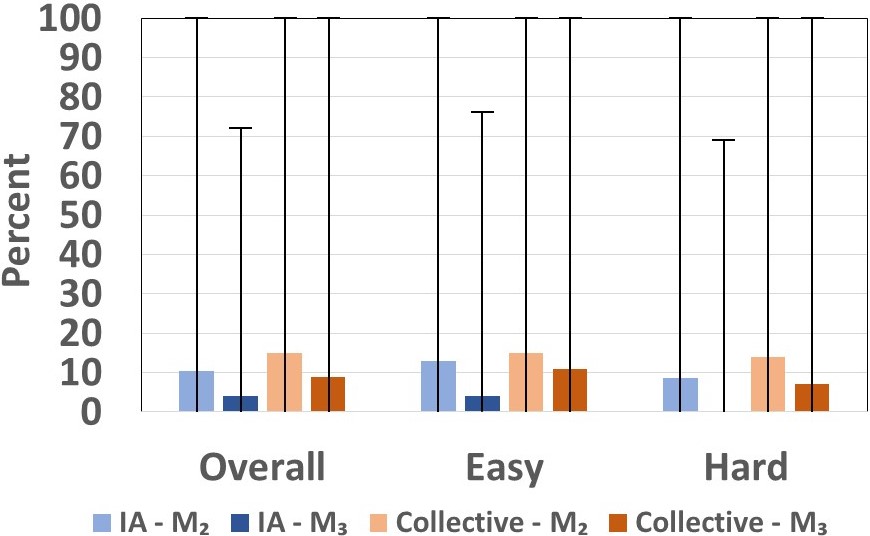}
	\captionof{figure}{Abandoned target information pop-up window open median (min/max) and Mann-Whitney-Wilcoxin test by decision difficulty between models.}
    \label{fig: High Value Target Window Open}
\end{minipage}
\end{table}

The post-trial questionnaire assessed the participants' \textit{understanding of collective behavior}, never (1) to always (7), and their \textit{ability} to choose the best target per decision, never (1) to always (7). The post-trial questionnaire mean (SD) are shown in Table \ref{table:Performance,PT Performance} \cite{Cody2018, Roundtree2020visual}. The performance and understanding rankings were higher for Collective operators using the $M_{3}$ model. The post-trial performance and understanding median, min, max, and the Mann-Whitney-Wilcoxon significant effects between models are presented in Figure \ref{fig: Post Trial}. IA operators ranked understanding significantly different between models. Additional between visualizations Mann-Whitney-Wilcoxon tests identified a significant effect for understanding using the $M_{2}$ model (n = 56, U = 513, $\rho$ = 0.04).

\begin{table}[h]
\begin{minipage}{0.5\linewidth}
\centering
\caption{Post-trial performance and understanding model ranking mean (SD) (1-low, 7-high).}
\label{table:Performance,PT Performance}
\begin{tabular}{c|c|c|c|}
\cline{2-4}
& \textbf{Metric} & \textbf{IA} & \textbf{Collective} \\ \hline
\multicolumn{1}{|c|}{\multirow{2}{*}{$M_{2}$}} & Performance & 5.25 (1.69) & 5.54 (1.29) \\ \cline{2-4} 
\multicolumn{1}{|c|}{} & \cellcolor{gray1}Understanding & \cellcolor{gray1}4.89 (1.75) & \cellcolor{gray1}5.82 (1.16) \\ \hline
\multicolumn{1}{|c|}{\multirow{2}{*}{$M_{3}$}} & Performance & 5.57 (1.43) & 5.75 (1.43) \\ \cline{2-4} 
\multicolumn{1}{|c|}{} & \cellcolor{gray1}Understanding & \cellcolor{gray1}5.93 (1.02) & \cellcolor{gray1}5.93 (1.46) \\ \hline
\end{tabular}
\end{minipage} 
\hfill
\begin{minipage}{0.45\linewidth}
\centering
	\includegraphics[width=62mm]{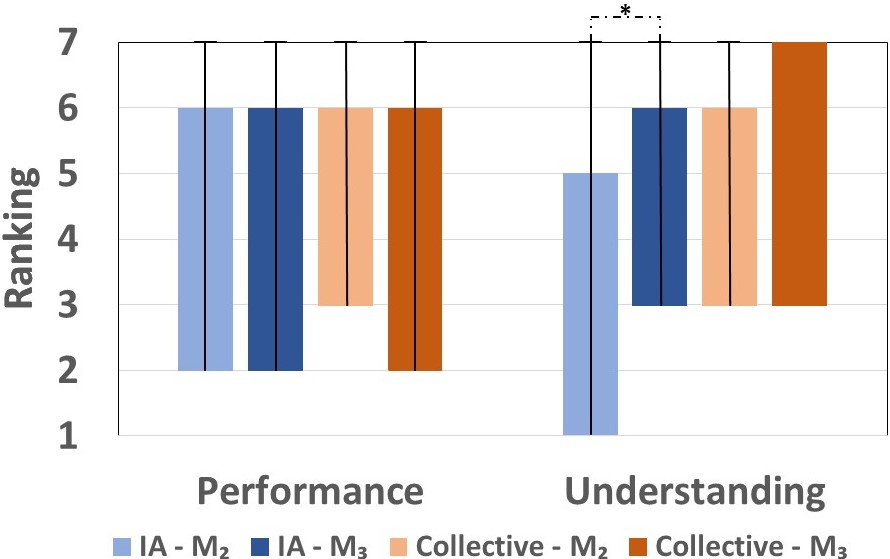}
	\captionof{figure}{Post-trial performance and understanding model ranking median (min/max) and Mann-Whitney-Wilcoxin test between models.}
    \label{fig: Post Trial}
\end{minipage}
\end{table}

The post-experiment questionnaire assessed the collective's \textit{responsiveness} to requests, the participants' \textit{ability} to choose the highest valued target, and their \textit{understanding} of the collective behavior. IA operators who used the $M_{2}$ model had the best collective responsiveness, operator ability, and understanding versus the $M_{3}$ model. Collective operators ranked the collective's responsiveness highest using the $M_{3}$ model, while operator ability and understanding were highest using the $M_{2}$ model. Details regarding the statistical tests were provided in the Metrics and Results Section \ref{section:R1 metrics}.

A summary of $R_{2}$'s results that show the hypotheses with associated significant results is provided in Table \ref{table:Comprehension,Combined}. This summary table is intended to facilitate the discussion.

\begin{table}[h]
\centering
\caption{A synopsis of $R_{2}$'s hypotheses associated with significant results. The SA probe timings are all timings (All), 15 seconds Before asking (B), While being asked (W), and During response (D) to a SA probe question.}
\label{table:Comprehension,Combined}
\begin{tabular}{?l|cc|cc|cc|c|c|c|c|c|c?}
\Cline{1pt}{1-10}
\multicolumn{1}{?c|}{\multirow{4}{*}{\textbf{Variable}}} & & \multicolumn{2}{?c?}{\textbf{Within}} & \multicolumn{2}{c?}{\textbf{Between}} & \multicolumn{4}{c?}{\multirow{2}{*}{\textbf{Correlation}}} \\
& \textbf{Sub-} & \multicolumn{2}{?c?}{\textbf{Model}} & \multicolumn{2}{c?}{\textbf{Visualization}} & \multicolumn{4}{c?}{} \\ \Cline{1pt}{3-10}
& \textbf{Variable} & \multicolumn{1}{?c|}{\multirow{2}{*}{IA}} & \multicolumn{1}{c?}{\multirow{2}{*}{Coll.}} & \multirow{2}{*}{$M_{2}$} & \multicolumn{1}{c?}{\multirow{2}{*}{$M_{3}$}} & \multicolumn{2}{c|}{IA} & \multicolumn{2}{c?}{Coll.} \\ \Cline{1pt}{7-10}
& & \multicolumn{1}{?c|}{} & \multicolumn{1}{c?}{} & & \multicolumn{1}{c?}{} & $M_{2}$ & $M_{3}$ & $M_{2}$ & \multicolumn{1}{c?}{$M_{3}$} \\ 
\Cline{1pt}{1-10}
\multirow{4}{*}{SA Probe Accuracy} & $SA_{O}$ & \multicolumn{1}{?c|}{} & \multicolumn{1}{c?}{} & \bm{$H_{4}$} & \multicolumn{1}{c?}{$H_{4}$} &
\multicolumn{4}{c?}{\multirow{4}{*}{-----------------}} \\ \cline{2-6}
& $SA_{1}$ & \multicolumn{1}{?c|}{\color{gray7}\bm{$H_{4}$}} & \multicolumn{1}{c?}{} & \bm{$H_{4}$} & \multicolumn{1}{c?}{$H_{4}$} & \multicolumn{1}{c}{} & \multicolumn{1}{c}{} & \multicolumn{1}{c}{} & \multicolumn{1}{c?}{} \\ \cline{2-6}
& $SA_{2}$ & \multicolumn{1}{?c|}{} & \multicolumn{1}{c?}{} & \bm{$H_{4}$} & \multicolumn{1}{c?}{$H_{4}$} & \multicolumn{1}{c}{} & \multicolumn{1}{c}{} & \multicolumn{1}{c}{} & \multicolumn{1}{c?}{} \\ \cline{2-6}
& $SA_{3}$ & \multicolumn{1}{?c|}{} & \multicolumn{1}{c?}{} & \bm{$H_{4}$} & \multicolumn{1}{c?}{$H_{4}$} & \multicolumn{1}{c}{} & \multicolumn{1}{c}{} & \multicolumn{1}{c}{} & \multicolumn{1}{c?}{} \\ \Cline{1pt}{1-10}
\multirow{8}{*}{Collective Left-Clicks} & \multirow{2}{*}{$SA_{O}$} & \multicolumn{1}{?c|}{{$H_{5}$}} & \multicolumn{1}{c?}{\bm{$H_{5}$}} & {$H_{5}$} & \multicolumn{1}{c?}{$H_{5}$} & & & & \multicolumn{1}{c?}{{$H_{5}$}} \\
& & \multicolumn{1}{?c|}{-All} & \multicolumn{1}{c?}{\bm{$-All$}} & {-All} & \multicolumn{1}{c?}{-B,W} & & & & \multicolumn{1}{c?}{-W} \\ \cline{2-10}
& \multirow{2}{*}{$SA_{1}$} & \multicolumn{1}{?c|}{{$H_{5}$}} & \multicolumn{1}{c?}{{$H_{5}$}} & {$H_{5}$} & \multicolumn{1}{c?}{{$H_{5}$}} & & & & \multicolumn{1}{c?}{{$H_{5}$}} \\ 
& & \multicolumn{1}{?c|}{-All} & \multicolumn{1}{c?}{-W,D} & {-All} & \multicolumn{1}{c?}{-W} & & & & \multicolumn{1}{c?}{-W} \\ \cline{2-10}
& \multirow{2}{*}{$SA_{2}$} & \multicolumn{1}{?c|}{{$H_{5}$}} & \multicolumn{1}{c?}{$H_{5}$} & $H_{5}$, & \multicolumn{1}{c?}{} & & & & \multicolumn{1}{c?}{} \\
& & \multicolumn{1}{?c|}{-All} & \multicolumn{1}{c?}{-B,D} & -B,W & \multicolumn{1}{c?}{} & & & & \multicolumn{1}{c?}{} \\ \cline{2-10}
& \multirow{2}{*}{$SA_{3}$} & \multicolumn{1}{?c|}{{$H_{5}$}} & \multicolumn{1}{c?}{{$H_{5}$}} & {$H_{5}$} & \multicolumn{1}{c?}{{$H_{5}$}} & & {$H_{5}$} & & \multicolumn{1}{c?}{} \\
& & \multicolumn{1}{?c|}{-B,D} & \multicolumn{1}{c?}{-B} & -W,D & \multicolumn{1}{c?}{-W} & & -B,W & & \multicolumn{1}{c?}{} \\ \Cline{1pt}{1-10}

\end{tabular}
\end{table}

\begin{table}[!t]
\centering
\begin{tabular}{?l|cc|cc|cc|c|c|c|c|c|c?}
\Cline{1pt}{1-10}
\multicolumn{1}{?c|}{\multirow{4}{*}{\textbf{Variable}}} & \multicolumn{1}{c}{\multirow{3}{*}{\textbf{Sub-}}} & \multicolumn{2}{?c?}{\textbf{Within}} & \multicolumn{2}{c?}{\textbf{Between}} & \multicolumn{4}{c?}{\multirow{2}{*}{\textbf{Correlation}}} \\
& \multicolumn{1}{c}{\multirow{3}{*}{\textbf{Variable}}} & \multicolumn{2}{?c?}{\textbf{Model}} & \multicolumn{2}{c?}{\textbf{Visualization}} & \multicolumn{4}{c?}{} \\ \Cline{1pt}{3-10}
& & \multicolumn{1}{?c|}{\multirow{2}{*}{IA}} & \multicolumn{1}{c?}{\multirow{2}{*}{Coll.}} & \multirow{2}{*}{$M_{2}$} & \multicolumn{1}{c?}{\multirow{2}{*}{$M_{3}$}} & \multicolumn{2}{c|}{IA} & \multicolumn{2}{c?}{Coll.} \\ \Cline{1pt}{7-10}
& & \multicolumn{1}{?c|}{} & \multicolumn{1}{c?}{} & & \multicolumn{1}{c?}{} & $M_{2}$ & $M_{3}$ & $M_{2}$ & \multicolumn{1}{c?}{$M_{3}$} \\ \Cline{1pt}{1-10}
\multirow{7}{*}{Target Right-Clicks} & \multirow{2}{*}{$SA_{O}$} & \multicolumn{1}{?c|}{$H_{5}$} & \multicolumn{1}{c?}{$H_{5}$} & & \multicolumn{1}{c?}{} & {$H_{5}$} & {$H_{5}$} & & \multicolumn{1}{c?}{{$H_{5}$}} \\ 
\multirow{7}{*}{by SA Level} & & \multicolumn{1}{?c|}{-B,D} & \multicolumn{1}{c?}{-B,D} & & \multicolumn{1}{c?}{} & -B & -B & & \multicolumn{1}{c?}{-B} \\ \cline{2-10}
& \multirow{2}{*}{$SA_{1}$} & \multicolumn{1}{?c|}{\multirow{2}{*}{}} & \multicolumn{1}{c?}{{$H_{5}$}} & & \multicolumn{1}{c?}{\multirow{2}{*}{}} & & {$H_{5}$} & & \multicolumn{1}{c?}{$H_{5}$} \\ 
& & \multicolumn{1}{?c|}{} & \multicolumn{1}{c?}{-D} & & \multicolumn{1}{c?}{} & & -B & & \multicolumn{1}{c?}{-B,W} \\ \cline{2-10}
& \multirow{2}{*}{$SA_{2}$} & \multicolumn{1}{?c|}{$H_{5}$} & \multicolumn{1}{c?}{} & & \multicolumn{1}{c?}{} & $H_{5}$ & & & \multicolumn{1}{c?}{} \\
& & \multicolumn{1}{?c|}{-B} & \multicolumn{1}{c?}{} & & \multicolumn{1}{c?}{} & -B & & & \multicolumn{1}{c?}{} \\ \cline{2-10}
& \multirow{2}{*}{$SA_{3}$} & \multicolumn{1}{?c|}{$H_{5}$} & \multicolumn{1}{c?}{} & & \multicolumn{1}{c?}{} & & & & \multicolumn{1}{c?}{} \\
& & \multicolumn{1}{?c|}{-B,D} & \multicolumn{1}{c?}{} & & \multicolumn{1}{c?}{} & & & & \multicolumn{1}{c?}{} \\ \Cline{1pt}{1-10}
\multirow{2}{*}{Collective} & Overall & \multicolumn{1}{?c|}{\color{gray7}\bm{$H_{5}$}} & \multicolumn{1}{c?}{} & \color{gray7}\bm{$H_{5}$} & \multicolumn{1}{c?}{\color{gray7}\bm{$H_{5}$}} &
\multicolumn{4}{c?}{\multirow{20}{*}{-----------------}} \\ \cline{2-6}
\multirow{2}{*}{Observations} & Easy & \multicolumn{1}{?c|}{\color{gray7}\bm{$H_{5}$}} & \multicolumn{1}{c?}{{$H_{5}$}} & \color{gray7}\bm{$H_{5}$} & \multicolumn{1}{c?}{\color{gray7}\bm{$H_{5}$}} & \multicolumn{1}{c}{} & \multicolumn{1}{c}{} & \multicolumn{1}{c}{} & \multicolumn{1}{c?}{} \\ \cline{2-6}
& Hard & \multicolumn{1}{?c|}{\color{gray7}\bm{$H_{5}$}} & \multicolumn{1}{c?}{} & & \multicolumn{1}{c?}{\color{gray7}\bm{$H_{5}$}} & \multicolumn{1}{c}{} & \multicolumn{1}{c}{} & \multicolumn{1}{c}{} & \multicolumn{1}{c?}{} \\ \Cline{1pt}{1-6}
\multirow{3}{*}{Target Observations} & Overall & \multicolumn{1}{?c|}{{$H_{4}$}} & \multicolumn{1}{c?}{\bm{$H_{4}$}} & \bm{$H_{4}$} & \multicolumn{1}{c?}{$H_{4}$} & \multicolumn{1}{c}{} & \multicolumn{1}{c}{} & \multicolumn{1}{c}{} & \multicolumn{1}{c?}{} \\ \cline{2-6}
& Easy & \multicolumn{1}{?c|}{} & \multicolumn{1}{c?}{\bm{$H_{4}$}} & \bm{$H_{4}$} & \multicolumn{1}{c?}{$H_{4}$} & \multicolumn{1}{c}{} & \multicolumn{1}{c}{} & \multicolumn{1}{c}{} & \multicolumn{1}{c?}{} \\ \cline{2-6}
& Hard & \multicolumn{1}{?c|}{} & \multicolumn{1}{c?}{\bm{$H_{4}$}} & \bm{$H_{4}$} & \multicolumn{1}{c?}{$H_{4}$} & \multicolumn{1}{c}{} & \multicolumn{1}{c}{} & \multicolumn{1}{c}{} & \multicolumn{1}{c?}{} \\ \Cline{1pt}{1-6}
{Collective Right-} & Overall & \multicolumn{1}{?c|}{\color{gray7}\bm{$H_{5}$}} & \multicolumn{1}{c}{} & \multicolumn{1}{c}{} & \multicolumn{1}{c}{} & \multicolumn{1}{c}{} & \multicolumn{1}{c}{} & \multicolumn{1}{c}{} & \multicolumn{1}{c?}{} \\ \cline{2-3}
Clicks & Hard & \multicolumn{1}{?c|}{\color{gray7}\bm{$H_{5}$}} & \multicolumn{1}{c}{} & \multicolumn{1}{c}{} & \multicolumn{1}{c}{} & \multicolumn{1}{c}{} & \multicolumn{1}{c}{} & \multicolumn{1}{c}{} & \multicolumn{1}{c?}{} \\ \Cline{1pt}{1-6}
Target Right-Clicks & \multirow{2}{*}{Easy} & \multicolumn{1}{?c|}{\multirow{2}{*}{$H_{5}$}} & \multicolumn{1}{c?}{} & & \multicolumn{1}{c?}{} & \multicolumn{1}{c}{} & \multicolumn{1}{c}{} & \multicolumn{1}{c}{} & \multicolumn{1}{c?}{} \\
per Decision & & \multicolumn{1}{?c|}{} & \multicolumn{1}{c?}{} & & \multicolumn{1}{c?}{} & \multicolumn{1}{c}{} & \multicolumn{1}{c}{} & \multicolumn{1}{c}{} & \multicolumn{1}{c?}{} \\ \Cline{1pt}{1-6}
Interventions & & \multicolumn{1}{?c|}{{$H_{4}$}} & \multicolumn{1}{c?}{{}} & \multicolumn{1}{c}{} & \multicolumn{1}{c}{{}} & \multicolumn{1}{c}{} & \multicolumn{1}{c}{} & \multicolumn{1}{c}{} & \multicolumn{1}{c?}{} \\ \Cline{1pt}{1-6}
\multirow{3}{*}{Highest Value} & \multirow{2}{*}{Overall} & \multicolumn{1}{?c|}{{}} & \multicolumn{1}{c?}{\color{gray7}\bm{$H_{4},$}} & & \multicolumn{1}{c?}{{}} & \multicolumn{1}{c}{} & \multicolumn{1}{c}{} & \multicolumn{1}{c}{} & \multicolumn{1}{c?}{} \\ 
\multirow{3}{*}{Target Abandoned} & & \multicolumn{1}{?c|}{{}} & \multicolumn{1}{c?}{\color{gray7}\bm{$H_{5}$}} & & \multicolumn{1}{c?}{{}} & \multicolumn{1}{c}{} & \multicolumn{1}{c}{} & \multicolumn{1}{c}{} & \multicolumn{1}{c?}{} \\ \cline{2-6}
& \multirow{2}{*}{Easy} & \multicolumn{1}{?c|}{\color{gray7}\bm{$H_{4},$}} & \multicolumn{1}{c?}{{}} & & \multicolumn{1}{c?}{{}} & \multicolumn{1}{c}{} & \multicolumn{1}{c}{} & \multicolumn{1}{c}{} & \multicolumn{1}{c?}{} \\
& & \multicolumn{1}{?c|}{\color{gray7}\bm{$H_{5}$}} & \multicolumn{1}{c?}{{}} & & \multicolumn{1}{c?}{{}} & \multicolumn{1}{c}{} & \multicolumn{1}{c}{} & \multicolumn{1}{c}{} & \multicolumn{1}{c?}{} \\ \Cline{1pt}{1-6}
Abandoned Target & \multirow{2}{*}{Overall} & \multicolumn{1}{?c|}{{}} & \multicolumn{1}{c?}{{}} & & \multicolumn{1}{c?}{\multirow{2}{*}{\color{gray7}\bm{$H_{5}$}}} & \multicolumn{1}{c}{} & \multicolumn{1}{c}{} & \multicolumn{1}{c}{} & \multicolumn{1}{c?}{} \\ 
Information & & \multicolumn{1}{?c|}{{}} & \multicolumn{1}{c?}{{}} & & \multicolumn{1}{c?}{{}} & \multicolumn{1}{c}{} & \multicolumn{1}{c}{} & \multicolumn{1}{c}{} & \multicolumn{1}{c?}{} \\ \cline{2-6}
Window Open & {Easy} & \multicolumn{1}{?c|}{{}} & \multicolumn{1}{c?}{{}} & {} & \multicolumn{1}{c?}{\color{gray7}\bm{$H_{5}$}} & \multicolumn{1}{c}{} & \multicolumn{1}{c}{} & \multicolumn{1}{c}{} & \multicolumn{1}{c?}{} \\ \Cline{1pt}{1-6}
{Post-Trial} & Understanding & \multicolumn{1}{?c|}{\color{gray7}\bm{$H_{4}$}} & \multicolumn{1}{c?}{{}} & {$H_{4}$} & \multicolumn{1}{c?}{{}} & \multicolumn{1}{c}{} & \multicolumn{1}{c}{} & \multicolumn{1}{c}{} & \multicolumn{1}{c?}{} \\ \Cline{1pt}{1-6}
{Post-Experiment} & {Understanding} & \multicolumn{1}{?c|}{{\color{gray7}\bm{$H_{4}$}}} & \multicolumn{1}{c?}{{\color{gray7}\bm{$H_{4}$}}} & \multicolumn{1}{c}{{}} & \multicolumn{1}{c}{{}} & \multicolumn{1}{c}{} & \multicolumn{1}{c}{} & \multicolumn{1}{c}{} & \multicolumn{1}{c?}{} \\ \Cline{1pt}{1-10}
\end{tabular}
\end{table}

\subsection{Discussion}

The analysis of how the model and visualization promoted operator comprehension (i.e., the operator's \textit{capability} of \textit{understanding}) suggests that the $M_{3}$ model promoted transparency more \textit{effectively} than the $M_{2}$ model, while both visualizations had their respective advantages and disadvantages. Operators using the $M_{2}$ model had fewer undesired interactions, such as target observations (i.e., extra clicks that did not contribute to the task) and interventions. Fewer undesired interactions may have occurred, because the $M_{2}$ model was designed to fulfill the best-of-\textit{n} decision-making task with or without operator influence, which \textit{effectively} balanced \textit{control} between the collectives and operator, whereas the $M_{3}$ model relied on operator influence (\textit{directability}) in order to make a decision. More undesirable interactions, such as target observations, resulted in better task \textit{performance} for operators using the $M_{3}$ model, which suggests that some interactions deemed undesirable for one model may be advantageous for another. Target observations may have occurred due to poor interface and visualization \textit{usability}. Operators who issued commands first selected the desired command, then selected the desired collective and target, and clicked on the commit button to complete a request. Reissuing the same command required re-selecting the target and clicking on the commit button. More target observations may have occurred if operators forgot to re-select the target when reissuing the same commands. Design improvements, such as leaving the target selected, may help decrease target observations.

$H_{4}$, which hypothesized that operators will have a better \textit{understanding} of the $M_{2}$ model, was not supported, because operators using the $M_{2}$ model abandoned the highest value target more frequently. The operators may have become overloaded supervising the four collectives simultaneously, especially if they were distracted by the secondary task and were momentarily out-of-the-loop. The interface's 10 (Hz) update rate (i.e., \textit{timing}) may have negatively impacted the operator's \textit{capability} to \textit{understand} what the collectives were doing and planned (e.g., \textit{predictability}) to do. Introducing \textit{timing} delays to the display may afford operators more \textit{time} to \textit{understanding} the current situation; however, task completion will be prolonged, which is undesired in missions that require fast system responses. Providing \textit{predictive} collective behaviors instead of \textit{timing} delays may help mitigate the \textit{time} required for an operator to reenter back into-the-loop.

The highest value target was abandoned more frequently when using the Collective visualization. The target value may not have been \textit{observable} enough (i.e., salient) to distinguish it from other potential targets, which did not support $H_{4}$. Further investigations are required to determine if the target value must use the entire collective hub icon area, similar to the IA visualization, in order to be more recognizable, and to establish what levels of obscurity are needed in order to ensure that target values are \textit{reliably} distinguishable from one another. Making distinctions clearer, such as using integers compared to letters, to identify collectives versus targets, may improve visualization \textit{explainability} and mitigate mistakes when operators confused the roman numeral identifiers with the integer identifiers. IA operators experienced this mistake frequently, which may have contributed to lowering their \textit{understanding}. Ensuring that identifiers are unique and distinct will improve the \textit{effectiveness} of the \textit{SA} probe questions. 

The use of target borders (collective observations), \textit{information} pop-up windows (target right-clicks), and target value, were assessed to determine if operators used this \textit{information} to \textit{justify} actions \textit{reliably} (i.e., accurately). Collective operators using the $M_{2}$ model made better decisions with fewer collective observations and more target-right clicks. \textit{Understanding} which collectives supported targets, by seeing numerical percentages, was more valuable compared to outlines indicating which targets were within a collective's range. $H_{5}$, which hypothesized that operators using the $M_{2}$ model and the Collective visualization were able to \textit{justify} actions accurately, was not supported. Collective operators who issued more collective left-clicks while being asked a SA probe question had better perception when using the $M_{3}$ model. IA operators who issued more target right-clicks 15 seconds before asking a SA probe question had better comprehension when using the $M_{2}$ model. The interactions of both operators were accurate and \textit{justified}; however, the model and visualization combination did not support the hypothesis. Collective left-clicks can improve perception of targets in range of a particular collective and are attributed with issued commands, which require perception, comprehension, and projection. Target right-clicks provide more \textit{information} about collective support for a particular target, which may improve \textit{understanding}. 

Lower \textit{SA performance} may have occurred if operators were in the middle of an interaction when the \textit{SA} probe question was posed, while higher \textit{SA performance} may have occurred because the operators anticipated when a \textit{SA} probe question was going to be asked and took preventative actions, such as opening or closing \textit{information} windows. Operators using target \textit{information} pop-up windows to verify that a target was abandoned by a collective may have been confused if the reported target support was greater than zero. There were instances during the trial when a few individual entities became lost, as the collective hub transitioned to a new location, and they did not move with the hub. The lost entities may have continued to explore a now abandoned target, because they never received the abandon target message, which occurred inside of the hub. The operators, as a result, may have reissued additional abandon commands in an attempt to reduce the collective support to zero, although only one abandon command was needed. Strategies improving \textit{explainability}, such as reporting zero percent support when an abandon command is issued and identifying how many individual entities have been lost, may help mitigate erroneous repeated abandon command behavior and improve \textit{understanding}. IA operators may have also experienced confusion if they saw individual collective entities still travelling to an abandoned target. Not displaying lost entities after a specific period of \textit{time} once a collective hub has moved to a new location may also reduce the number of reissued abandon commands. Further analysis using eye-tracking technology may provide more \textit{reliable} metrics to determine operator comprehension by identifying exactly where an operator is focusing their attention.

The transparency embedded in the $M_{2}$ model and Collective visualization combination did not support the operator's \textit{capability} to \textit{understand} (i.e., comprehension) the collectives' behaviors the best. The $M_{3}$ model provided better operator comprehension, because operators were more involved in the decision-making process. More interactions, even if some were undesirable, contributed to better \textit{understanding} and task \textit{performance}. Strategies to increase operator involvement, without taking complete \textit{control} over the decision-making process, when using the $M_{2}$ model must be considered to improve it's \textit{effectiveness}. Design improvements, such as increasing \textit{explainability} by identifying how many individual entities became lost during a hub transition to a new location, can help mitigate abandoning the highest value target, which occurred most frequently for Collective operators using the $M_{2}$ model. \textit{Understanding} why particular interactions occurred for specific model and visualization combinations, and what aspects contributed to those interactions, can help aid designers to improving the transparency embedded in the $M_{2}$ model and Collective visualization.

\section{$R_{3}$: System Design Element Usability}

\begin{figure}[!b]
\begin{center}
	\includegraphics[width=\textwidth]{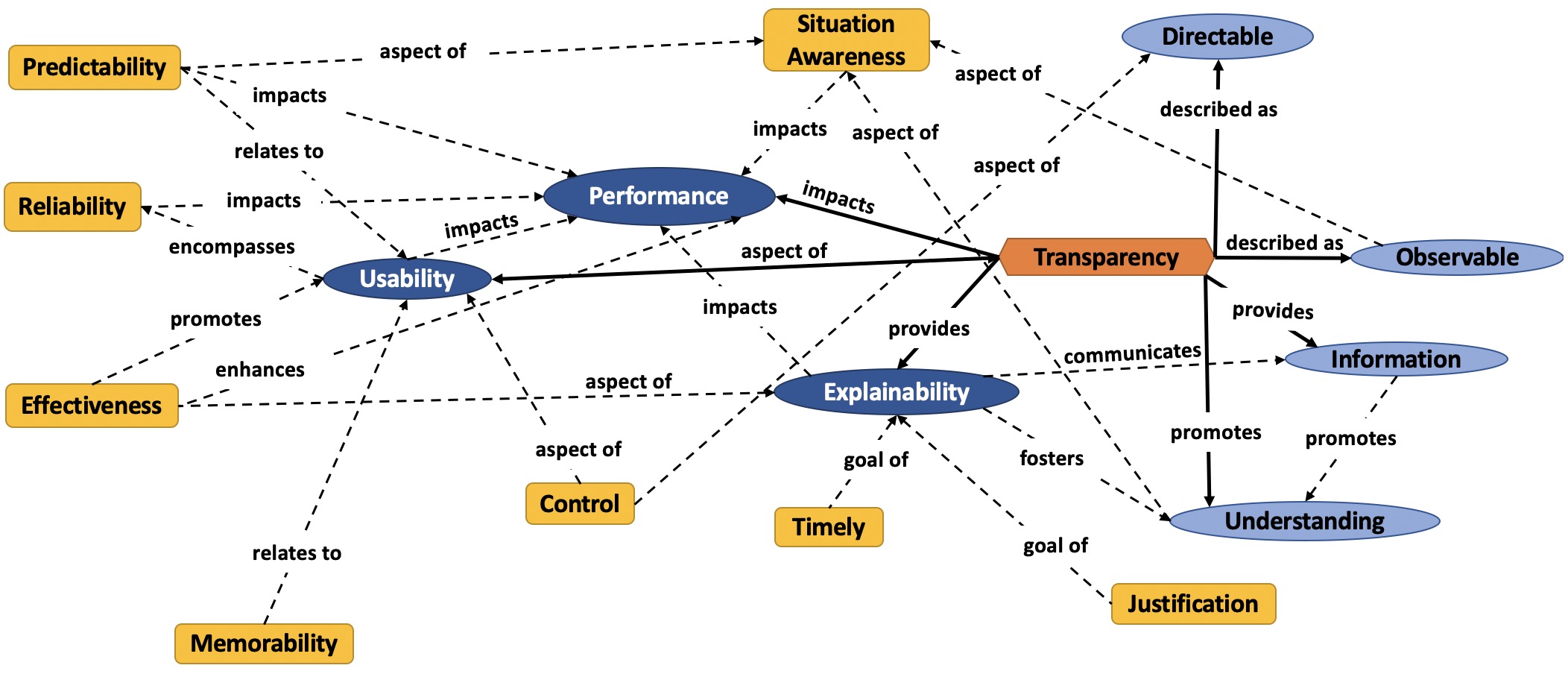}
	\caption{$R_{3}$ concept map of the assessed direct and indirect transparency factors.}
	\label{fig: Model Vis Concept Map R7}
	\end{center}
\end{figure}

Understanding \textit{which model and visualization promoted better usability}, $R_{3}$, is necessary to determine which system design elements promote effective transparency in human-collective systems. The associated objective dependent variables were (1) visualization clutter, (2) Euclidean distance, (3) whether an operator was in the middle of an action and completed that action when asked a SA probe question, (4) issued commands, (5) collective and target right-clicks, (6) metrics associated with abandoned targets, (7) the time between the committed state and an issued decide command, and (8) metrics associated with information pop-up windows. The specific direct and indirect transparency factors related to $R_{3}$ are identified in Figure \ref{fig: Model Vis Concept Map R7}. The relationship between the variables and the corresponding hypotheses, as well as the direct and indirect transparency factors are identified in Table \ref{table:Usability,Variables}. Additional relationships that are not shown in Figure \ref{fig: Concept Map}, between the variable and the direct or indirect transparency factors are provided after conducting correlation analyses.

\begin{table}[h]
\centering
\caption{Interaction of system design elements usability objective (obj) and subjective (subj) variables (vars), relationship to the hypotheses (H), as well as the associated direct and indirect transparency factors, are presented in Figure \ref{fig: Concept Map}.}
\label{table:Usability,Variables}
\begin{tabular}{?l|c?c|c|c|c|c|c|c?c|c|c|c|c|c|c|c?} \Cline{1pt}{3-17}
\multicolumn{1}{c}{} & & \multicolumn{15}{c?}{\textbf{Transparency Factors}} \\ \Cline{1pt}{3-17}
\multicolumn{1}{c}{} & & \multicolumn{7}{c?}{\textbf{Direct}} & \multicolumn{8}{c?}{\textbf{Indirect}} \\ \Cline{1pt}{3-17}
\multicolumn{1}{c}{} & & {\multirow[b]{6}{*}{\rotatebox{90}{\textbf{Directable}}}} & {\multirow[b]{6}{*}{\rotatebox{90}{\textbf{Explainability}}}} & {\multirow[b]{6}{*}{\rotatebox{90}{\textbf{Information}}}} & {\multirow[b]{6}{*}{\rotatebox{90}{\textbf{Observable}}}} & {\multirow[b]{6}{*}{\rotatebox{90}{\textbf{Performance}}}} & {\multirow[b]{6}{*}{\rotatebox{90}{\textbf{Understanding}}}} & {\multirow[b]{6}{*}{\rotatebox{90}{\textbf{Usability}}}} & {\multirow[b]{6}{*}{\rotatebox{90}{\textbf{Control}}}} & {\multirow[b]{6}{*}{\rotatebox{90}{\textbf{Effectiveness}}}} &  {\multirow[b]{6}{*}{\rotatebox{90}{\textbf{Justification}}}} & {\multirow[b]{6}{*}{\rotatebox{90}{\textbf{Memorability}}}} &  {\multirow[b]{6}{*}{\rotatebox{90}{\textbf{Predictability}}}} & {\multirow[b]{6}{*}{\rotatebox{90}{\textbf{Reliability}}}} & {\multirow[b]{6}{*}{\rotatebox{90}{\textbf{SA}}}} & {\multirow[b]{6}{*}{\rotatebox{90}{\textbf{Timing}}}} \\
\multicolumn{1}{c}{} & & & & & & & & & & & & & & & & \\ 
\multicolumn{1}{c}{} & & & & & & & & & & & & & & & & \\
\multicolumn{1}{c}{} & & & & & & & & & & & & & & & & \\
\multicolumn{1}{c}{} & & & & & & & & & & & & & & & & \\
\Cline{1pt}{1-2}
\multicolumn{1}{?c|}{\textbf{Obj Vars}} & {\textbf{H}} & & & & & & & & & & & & & & & \\ \Cline{1pt}{1-17}
{Global Clutter} & {$H_{6}$} & & & & {\checkmark} & {\checkmark} & {\checkmark} & {\checkmark} & & {\checkmark} & & & {\checkmark} & & {\checkmark} & \\ \hline
{Euclidean} & {\multirow{5}{*}{$H_{7}$}} & & & & & & & {\multirow{5}{*}{\checkmark}} & & {\multirow{5}{*}{\checkmark}} & & & & & & \\
{Distance Between} & & & & & & & & & & & & & & & & \\ 
{SA Probe} & & & & & & & & & & & & & & & & \\ 
{Interaction and} & & & & & & & & & & & & & & & & \\
{Clicks} & & & & & & & & & & & & & & & & \\ \hline
{Middle of Action} & \multirow{2}{*}{$H_{7}$} & \multirow{2}{*}{\checkmark} & & & \multirow{2}{*}{\checkmark} & & \multirow{2}{*}{\checkmark} & \multirow{2}{*}{\checkmark} & \multirow{2}{*}{\checkmark} & \multirow{2}{*}{\checkmark} & & & \multirow{2}{*}{\checkmark} & & \multirow{2}{*}{\checkmark} & \multirow{2}{*}{\checkmark} \\ 
{During SA Probe} & & & & & & & & & & & & & & & & \\ \hline
{Completed} & \multirow{2}{*}{$H_{6}$,} & \multirow{3}{*}{\checkmark} & \multirow{3}{*}{\checkmark} & & \multirow{3}{*}{\checkmark} & & \multirow{3}{*}{\checkmark} & \multirow{3}{*}{\checkmark} & \multirow{3}{*}{\checkmark} & \multirow{3}{*}{\checkmark} & & \multirow{3}{*}{\checkmark} & \multirow{3}{*}{\checkmark} & & \multirow{3}{*}{\checkmark} & \\
{Interrupted SA} & \multirow{2}{*}{$H_{7}$} & & & & & & & & & & & & & & & \\ 
{Probe Action} & \multirow{2}{*}{} & & & & & & & & & & & & & & & \\ \hline
{Investigate} & \multirow{2}{*}{$H_{7}$} & \multirow{2}{*}{\checkmark} & & & & \multirow{2}{*}{\checkmark} & & \multirow{2}{*}{\checkmark} & \multirow{2}{*}{\checkmark} & \multirow{2}{*}{\checkmark} & & & & & & \\
{Command} & & & & & & & & & & & & & & & & \\ \hline
{Abandon} & \multirow{2}{*}{$H_{7}$} & \multirow{2}{*}{\checkmark} & & & & \multirow{2}{*}{\checkmark} & & \multirow{2}{*}{\checkmark} & \multirow{2}{*}{\checkmark} & \multirow{2}{*}{\checkmark} & & & & & & \\
{Command} & & & & & & & & & & & & & & & & \\ \hline
{Decide} & {$H_{6}$,} & \multirow{2}{*}{\checkmark} & & & & \multirow{2}{*}{\checkmark} & & \multirow{2}{*}{\checkmark} & \multirow{2}{*}{\checkmark} & \multirow{2}{*}{\checkmark} & & & \multirow{2}{*}{\checkmark} & & & \\ 
Command & {$H_{7}$} & & & & & & & & & & & & & & & \\ \hline
{Collective Right-} & \multirow{2}{*}{$H_{7}$} & & & \multirow{2}{*}{\checkmark} & & \multirow{2}{*}{\checkmark} & & \multirow{2}{*}{\checkmark} & & \multirow{2}{*}{\checkmark} & & & & & & \\
{Clicks} & & & & & & & & & & & & & & & & \\ \hline
{Target Right-} & \multirow{3}{*}{$H_{7}$} & & & \multirow{3}{*}{\checkmark} & & \multirow{3}{*}{\checkmark} & & \multirow{3}{*}{\checkmark} & & \multirow{3}{*}{\checkmark} & & & & & & \\
{Clicks per} & & & & & & & & & & & & & & & & \\
{Decision} & & & & & & & & & & & & & & & & \\ \hline
{Highest Value} & \multirow{2}{*}{$H_{6}$} & & \multirow{2}{*}{\checkmark} & & & & \multirow{2}{*}{\checkmark} & \multirow{2}{*}{\checkmark} & & \multirow{2}{*}{\checkmark} & \multirow{2}{*}{\checkmark} & & & & & \\ 
{Target Abandon} & & & & & & & & & & & & & & & & \\ \hline
{Abandon Target} & \multirow{3}{*}{$H_{7}$} & & \multirow{3}{*}{\checkmark} & & & & & \multirow{3}{*}{\checkmark} & & \multirow{3}{*}{\checkmark} & \multirow{3}{*}{\checkmark} & & & & & \\
{Information} & & & & & & & & & & & & & & & & \\
{Window Open} & & & & & & & & & & & & & & & & \\ \hline
\end{tabular}
\end{table}

\begin{table}[!t]
\centering
\begin{tabular}{?l|c?c|c|c|c|c|c|c?c|c|c|c|c|c|c|c?} \Cline{1pt}{3-17}
\multicolumn{1}{c}{} & & \multicolumn{15}{c?}{\textbf{Transparency Factors}} \\ \Cline{1pt}{3-17}
\multicolumn{1}{c}{} & & \multicolumn{7}{c?}{\textbf{Direct}} & \multicolumn{8}{c?}{\textbf{Indirect}} \\ \Cline{1pt}{3-17}
\multicolumn{1}{c}{} & & {\multirow[b]{6}{*}{\rotatebox{90}{\textbf{Directable}}}} & {\multirow[b]{6}{*}{\rotatebox{90}{\textbf{Explainability}}}} & {\multirow[b]{6}{*}{\rotatebox{90}{\textbf{Information}}}} & {\multirow[b]{6}{*}{\rotatebox{90}{\textbf{Observable}}}} & {\multirow[b]{6}{*}{\rotatebox{90}{\textbf{Performance}}}} & {\multirow[b]{6}{*}{\rotatebox{90}{\textbf{Understanding}}}} & {\multirow[b]{6}{*}{\rotatebox{90}{\textbf{Usability}}}} & {\multirow[b]{6}{*}{\rotatebox{90}{\textbf{Control}}}} & {\multirow[b]{6}{*}{\rotatebox{90}{\textbf{Effectiveness}}}} &  {\multirow[b]{6}{*}{\rotatebox{90}{\textbf{Justification}}}} & {\multirow[b]{6}{*}{\rotatebox{90}{\textbf{Memorability}}}} &  {\multirow[b]{6}{*}{\rotatebox{90}{\textbf{Predictability}}}} & {\multirow[b]{6}{*}{\rotatebox{90}{\textbf{Reliability}}}} & {\multirow[b]{6}{*}{\rotatebox{90}{\textbf{SA}}}} & {\multirow[b]{6}{*}{\rotatebox{90}{\textbf{Timing}}}} \\
\multicolumn{1}{c}{} & & & & & & & & & & & & & & & & \\ 
\multicolumn{1}{c}{} & & & & & & & & & & & & & & & & \\
\multicolumn{1}{c}{} & & & & & & & & & & & & & & & & \\
\multicolumn{1}{c}{} & & & & & & & & & & & & & & & & \\
\Cline{1pt}{1-2}
\multicolumn{1}{?c|}{\textbf{Obj Vars}} & {\textbf{H}} & & & & & & & & & & & & & & & \\ \Cline{1pt}{1-17}
{Abandon Request} & \multirow{3}{*}{$H_{6}$} & & \multirow{3}{*}{\checkmark} & & & & \multirow{3}{*}{\checkmark} & \multirow{3}{*}{\checkmark} & & \multirow{3}{*}{\checkmark} & & & & & & \\ 
{Exceeded} & & & & & & & & & & & & & & & & \\
{Abandon Target} & & & & & & & & & & & & & & & & \\ \hline
{Time Commit} & {\multirow{3}{*}{$H_{6}$}} & & & & {\multirow{3}{*}{\checkmark}} & & {\multirow{3}{*}{\checkmark}} & {\multirow{3}{*}{\checkmark}} & & {\multirow{3}{*}{\checkmark}} & & & {\multirow{3}{*}{\checkmark}} & & {\multirow{3}{*}{\checkmark}} & {\multirow{3}{*}{\checkmark}} \\
{State and Issued} & & & & & & & & & & & & & & & & \\
{Decide Command} & & & & & & & & & & & & & & & & \\ \hline
{Frequency of} & \multirow{3}{*}{$H_{6}$,} & & \multirow{4}{*}{\checkmark} & \multirow{4}{*}{\checkmark} & & & & \multirow{4}{*}{\checkmark} & & \multirow{4}{*}{\checkmark} & & & & & & \\ 
{Accessed Target} & \multirow{3}{*}{$H_{7}$} & & & & & & & & & & & & & & & \\ 
{Information} & & & & & & & & & & & & & & & & \\
{Window} & & & & & & & & & & & & & & & & \\ \hline
{Time Target} & \multirow{3}{*}{$H_{6}$} & & & & & & & \multirow{3}{*}{\checkmark} & & \multirow{3}{*}{\checkmark} & & & & \multirow{3}{*}{\checkmark} & & \multirow{3}{*}{\checkmark} \\
{Information} & & & & & & & & & & & & & & & & \\
{Window Open} & & & & & & & & & & & & & & & & \\ \hline
{Time Decision} & \multirow{4}{*}{$H_{6}$} & & & & & & & \multirow{4}{*}{\checkmark} & & \multirow{4}{*}{\checkmark} & & & & & & \multirow{4}{*}{\checkmark} \\
{Collective} & & & & & & & & & & & & & & & & \\ 
{Information} & & & & & & & & & & & & & & & & \\ 
{Window Open} & & & & & & & & & & & & & & & & \\ \hline
{Time Decision} & \multirow{6}{*}{$H_{6}$} & & & & & & & \multirow{6}{*}{\checkmark} & & \multirow{6}{*}{\checkmark} & & & & & & \multirow{6}{*}{\checkmark} \\
{Target} & & & & & & & & & & & & & & & & \\ 
{Information} & & & & & & & & & & & & & & & & \\ 
{Window Open} & & & & & & & & & & & & & & & & \\ \Cline{1pt}{1-17}
\multicolumn{1}{?c}{\textbf{Subj Vars}} & \multicolumn{16}{c?}{\textbf{}} \\ \Cline{1pt}{1-17}
{Post-Trial} & \multirow{3}{*}{$H_{6}$} & & & & & & & \multirow{3}{*}{\checkmark} & & \multirow{3}{*}{\checkmark} & & & & \multirow{3}{*}{\checkmark} & & \\ 
{Command} & & & & & & & & & & & & & & & & \\ 
{Effectiveness} & & & & & & & & & & & & & & & & \\ \hline
{Post-Experiment} & {$H_{6}$} & & & & & & {\checkmark} & {\checkmark} & & & & & & {\checkmark} & & {\checkmark} \\ \Cline{1pt}{1-17}
\end{tabular}
\end{table}

The goal of usability is to design systems that are effective, efficient, safe to use, easy to learn, and are memorable \cite{Preece2007}. Good usability is necessary to ensure operators can perceive and understand the information presented on a visualization, and to promote effective interactions. It was hypothesized ($H_{6}$) that the $M_{2}$ model and Collective visualization will promote better usability by being more predictable and explainable. Providing information that is explainable may aid operator comprehension, while predictable information may expedite operator actions. An ideal system will not require constant operator interaction to perform well; therefore, it was hypothesized ($H_{7}$) that operators using the $M_{2}$ model and Collective visualization will require fewer interactions.

\subsection{Metrics and Results}
\label{sec: R3 metrics}

System features were available to the operators in order to aid task completion. The IA visualization had lower \textit{global clutter percentages}, which was the percentage of visualization area obstructed by all displayed objects. IA operators using the $M_{2}$ model had lower global clutter percentages compared to the $M_{3}$ model. Collective operators in general had lower global clutter percentages using the $M_{2}$ model. The IA visualization had lower global clutter percentages in general compared to the Collective visualization. The statistical test details were provided in Section \ref{section:R1 metrics}.

The Euclidean \textit{distance (pixels) between the SA probe interest and where the operator was interacting} with the visualization indicated where operators focused their attention, since no eye-tracker was used. Euclidean distance can be used to assess the effectiveness of the object placements on the display. Larger distances are not ideal, because more time \cite{Gillan1992} and effort is required to locate and interact with the object. The first requirement of calculating the Euclidean distance was to determine what the collective, or target of interest was in a SA probe question. Target 3 is the target of interest for the following question: ``What collectives are investigating Target 3?'' The second requirement was to determine where the operator was interacting with the system (i.e., clicking on the interface), which was recorded for both evaluations. The Euclidean distance between SA probe interest and clicks mean (SD) 15 seconds before asking, while being asked, and during response to a SA probe question are presented in Table \ref{table:Usability,Dist} \citep{Roundtree2020visual}. Operators from both visualizations when using the $M_{2}$ model in general had shorter Euclidean distances compared to the $M_{3}$ model. Shorter Euclidean distances however, occurred at all timings for $SA_{3}$ and 15 seconds before asking and during response to a SA probe question for $SA_{1}$ when the Collective operators used the $M_{3}$ model. 

\begin{table}[h]
\centering
\caption{Euclidean distance between SA probe interest and clicks mean (SD) 15 seconds before asking, while being asked, and during response to SA probe question by SA level.}
\label{table:Usability,Dist}
\begin{tabular}{c|c|c|c|c|}
\cline{2-5}
 & \textbf{Timing} & \textbf{SA Level} & \textbf{IA} & \textbf{Collective} \\ \hline
\multicolumn{1}{|c|}{\multirow{12}{*}{$M_{2}$}} & \multirow{4}{*}{Before} & $SA_{O}$ & 767.1 (262.5) & 820.7 (255.67) \\ \cline{3-5} 
\multicolumn{1}{|c|}{} & & \cellcolor{gray1}$SA_{1}$ & \cellcolor{gray1}759.5 (251.64) & \cellcolor{gray1}825.6 (264.1) \\ \cline{3-5} 
\multicolumn{1}{|c|}{} & & \cellcolor{gray2}$SA_{2}$ & \cellcolor{gray2}768.9 (282.07) & \cellcolor{gray2}812.9 (234.94) \\ \cline{3-5}
\multicolumn{1}{|c|}{} & & \cellcolor{gray3}$SA_{3}$ & \cellcolor{gray3}783.4 (262.89) & \cellcolor{gray3}821.6 (271.03) \\ \cline{2-5} 
\multicolumn{1}{|c|}{} & \multirow{4}{*}{Asking} & $SA_{O}$ & 758.44 (291.48) & 851.4 (293.91) \\ \cline{3-5} 
\multicolumn{1}{|c|}{} & & \cellcolor{gray1}$SA_{1}$ & \cellcolor{gray1}754.4 (284.65) & \cellcolor{gray1}845.5 (282.53) \\ \cline{3-5}
\multicolumn{1}{|c|}{} & & \cellcolor{gray2}$SA_{2}$ & \cellcolor{gray2}768.4 (316.09) & \cellcolor{gray2}879.5 (299.93) \\ \cline{3-5} 
\multicolumn{1}{|c|}{} & & \cellcolor{gray3}$SA_{3}$ & \cellcolor{gray3}753.7 (275.04) & \cellcolor{gray3}823.5 (314.47) \\ \cline{2-5} 
\multicolumn{1}{|c|}{} & \multirow{4}{*}{Responding} & $SA_{O}$ & 764.24 (298.84) & 827.7 (273.83) \\ \cline{3-5} 
\multicolumn{1}{|c|}{} & & \cellcolor{gray1}$SA_{1}$ & \cellcolor{gray1}760.9 (297.14) & \cellcolor{gray1}827.9 (279.21) \\ \cline{3-5} 
\multicolumn{1}{|c|}{} & & \cellcolor{gray2}$SA_{2}$ & \cellcolor{gray2}774.6 (319.08) & \cellcolor{gray2}845.2 (275.55) \\ \cline{3-5} 
\multicolumn{1}{|c|}{} & & \cellcolor{gray3}$SA_{3}$ & \cellcolor{gray3}757.71 (278.14) & \cellcolor{gray3}799.7 (261.1) \\ \hline
\multicolumn{1}{|c|}{\multirow{12}{*}{$M_{3}$}} & \multirow{4}{*}{Before} & $SA_{O}$ & 868.3 (239.4) & 845.1 (258.07) \\ \cline{3-5} 
\multicolumn{1}{|c|}{} & & \cellcolor{gray1}$SA_{1}$ & \cellcolor{gray1}814.4 (225.39) & \cellcolor{gray1}789.9 (261.93) \\ \cline{3-5} 
\multicolumn{1}{|c|}{} & & \cellcolor{gray2}$SA_{2}$ & \cellcolor{gray2}925.2 (243.31) & \cellcolor{gray2}896.9 (241.67) \\ \cline{3-5}
\multicolumn{1}{|c|}{} & & \cellcolor{gray3}$SA_{3}$ & \cellcolor{gray3}907.8 (238.96) & \cellcolor{gray3}805.9 (277.15) \\ \cline{2-5} 
\multicolumn{1}{|c|}{} & \multirow{4}{*}{Asking} & $SA_{O}$ & 862.3 (254.68) & 860.22 (266.91) \\ \cline{3-5} 
\multicolumn{1}{|c|}{} & & \cellcolor{gray1}$SA_{1}$ & \cellcolor{gray1}808.4 (250.01) & \cellcolor{gray1}846.4 (272.23) \\ \cline{3-5}
\multicolumn{1}{|c|}{} & & \cellcolor{gray2}$SA_{2}$ & \cellcolor{gray2}931.6 (249.36) & \cellcolor{gray2}933.5 (252.66) \\ \cline{3-5} 
\multicolumn{1}{|c|}{} & & \cellcolor{gray3}$SA_{3}$ & \cellcolor{gray3}865.7 (241.6) & \cellcolor{gray3}759 (248.65) \\ \cline{2-5} 
\multicolumn{1}{|c|}{} & \multirow{4}{*}{Responding} & $SA_{O}$ & 865 (262.27) & 837 (263.75) \\ \cline{3-5} 
\multicolumn{1}{|c|}{} & & \cellcolor{gray1}$SA_{1}$ & \cellcolor{gray1}816.7 (254.45) & \cellcolor{gray1}802.7 (270.79) \\ \cline{3-5} 
\multicolumn{1}{|c|}{} & & \cellcolor{gray2}$SA_{2}$ & \cellcolor{gray2}928.3 (264.62) & \cellcolor{gray2}901.6 (238.64) \\ \cline{3-5} 
\multicolumn{1}{|c|}{} & & \cellcolor{gray3}$SA_{3}$ & \cellcolor{gray3}860.4 (248.47) & \cellcolor{gray3}755.2 (274.87) \\ \hline
\end{tabular}
\end{table}

\def \figwidth {0.49}
\begin{figure*}[h]
    \captionsetup[subfigure]{justification=centering}
    \begin{subfigure}{\figwidth\linewidth}
		\centering
		\includegraphics[keepaspectratio,height=1.71in]{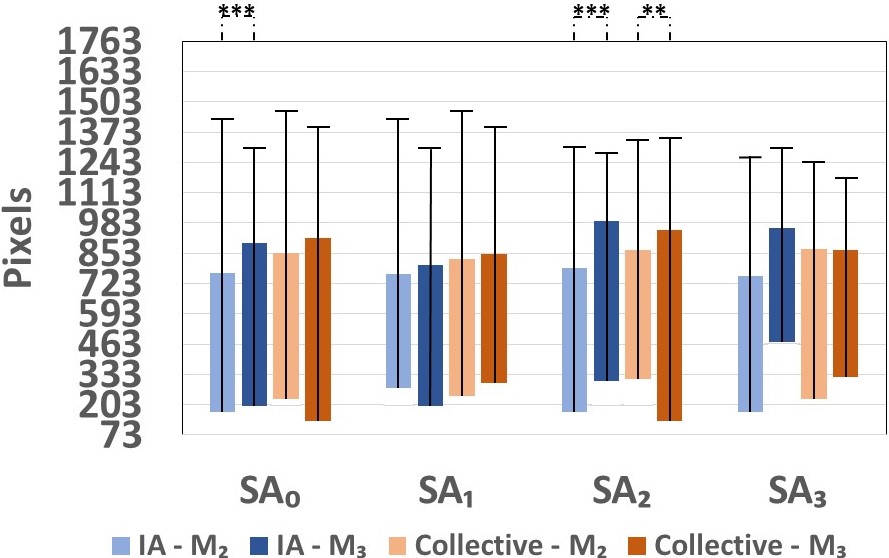}
		\captionsetup{width=\linewidth}
		\caption{15 seconds before asking a SA probe question.}
	\end{subfigure}
\begin{subfigure}{\figwidth\linewidth}
	\centering
	\includegraphics[keepaspectratio,height=1.71in]{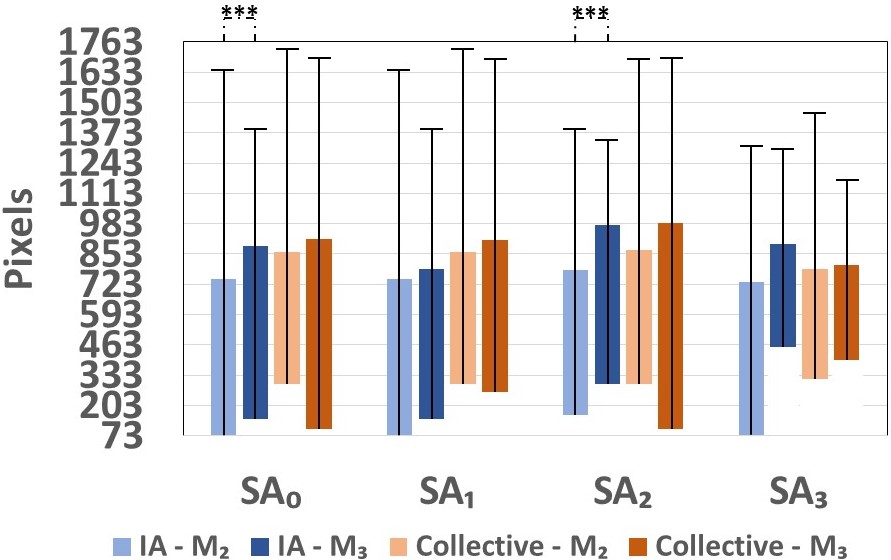}
	\captionsetup{width=\linewidth}
	\caption{While being asked a SA probe question.}
\end{subfigure}
\begin{subfigure}{\figwidth\linewidth}
	\centering
	\includegraphics[keepaspectratio,height=1.705in]{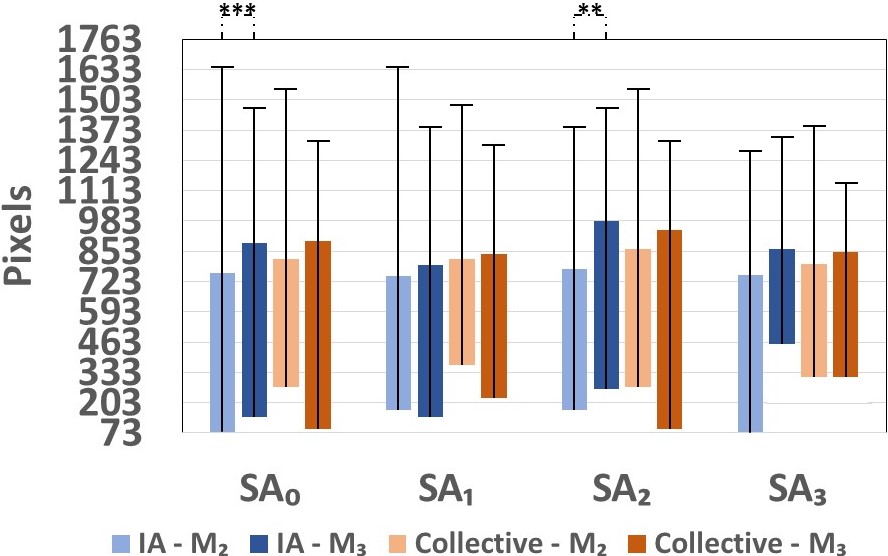}
	\captionsetup{width=\linewidth}
	\caption{During response to a SA probe question.}
\end{subfigure}
	\caption{Euclidean distance between SA probe interest and clicks median (min/max) and Mann-Whitney-Wilcoxin test by SA level between models a) 15 seconds before asking, b) while being asked, and c) during response to a SA probe question.}
	\label{fig: Dist Subj Click}
\end{figure*}

The Euclidean distance between SA probe interest and operator clicks median, min, max, and the Mann-Whitney-Wilcoxon significant effects between models are presented in Figure \ref{fig: Dist Subj Click}. IA operators had significantly different Euclidean distances between the SA probe interest and their current interaction between models 15 seconds before asking, while being asked, and during response to a SA probe question for $SA_{O}$ and $SA_{2}$. A significant difference for this metric between models occurred for Collective operators 15 seconds before asking a $SA_{2}$ probe question. Additional between visualizations Mann-Whitney-Wilcoxon tests identified significant effects when using the $M_{2}$ model 15 seconds before asking a SA probe question for $SA_{O}$ (n = 557, U = 43303, $\rho$ = 0.02) and $SA_{1}$ (n = 273, U = 10577, $\rho$ = 0.05). A moderate significant effect between visualizations when using the $M_{2}$ model while being asked a SA probe question was found for $SA_{O}$ (n = 464, U = 31052, $\rho$ $<$ 0.01) and a significant effect for $SA_{1}$ (n = 229, U = 7645, $\rho$ = 0.01). A significant effect between visualizations using the $M_{2}$ model during response to a SA probe question was also found for $SA_{O}$ (n = 499, U = 35029, $\rho$ = 0.02). Shorter Euclidean distances occurred when IA operators used the $M_{2}$ model compared to the Collective visualization, while Collective operators had shorter Euclidean distances when using the $M_{3}$ model. The Spearman correlation analysis revealed a weak correlation between the Euclidean distance of the SA probe's interest and the operators' current click and SA probe accuracy for the IA visualization when using the $M_{2}$ model 15 seconds before asking a SA probe question for $SA_{1}$ (r = -0.18, $\rho$ = 0.04). Weak correlations were revealed for the Collective visualization when using the $M_{3}$ model for $SA_{O}$ while being asked (r = 0.14, $\rho$ = 0.04) and during response to a SA probe question (r = 0.16, $\rho$ = 0.01).

The percentage of times an operator was in the \textit{middle of an action during a SA probe} question identified how often operators were interrupted by the secondary task. Distracted operators may have needed more time to focus their attention on the SA probe question, or may have prioritized their current interaction over answering the SA probe question immediately, or at all. Understanding how distractions may have negatively influenced operator behavior is needed to design the system to promote effective human-collective interactions. The percentage of times an operator was in the middle of an action during a SA probe question mean (SD) are presented in Table \ref{table:Usability,Middle of Action}. Operators using the $M_{2}$ model were interrupted less frequently by the SA probe question compared to those using the $M_{3}$ model irrespective of the visualization. The percentage of times operators using either visualization were in the middle of an action during a SA probe question median, min, max, and the Mann-Whitney-Wilcoxon significant effects between models are presented in Figure \ref{fig: Middle Action}. The percentage of times operators from both evaluations were in the middle of an action during a SA probe question was significantly different between models for $SA_{O}$, $SA_{1}$, and $SA_{2}$. Additional between visualizations Mann-Whitney-Wilcoxon tests identified highly significant effects when using the $M_{2}$ model for $SA_{O}$ (n = 670, U = 74938, $\rho$ $<$ 0.001), $SA_{1}$ (n = 294, U = 14595, $\rho$ $<$ 0.001), $SA_{2}$ (n = 224, U = 8344, $\rho$ $<$ 0.001), and $SA_{3}$ (n = 152, U = 3780, $\rho$ $<$ 0.001). Highly significant effects between visualizations using the $M_{3}$ model were found for $SA_{O}$ (n = 672, U = 78456, $\rho$ $<$ 0.001), $SA_{1}$ (n = 253, U = 10944, $\rho$ $<$ 0.001), $SA_{2}$ (n = 252, U = 11172, $\rho$ $<$ 0.001), and $SA_{3}$ (n = 167, U = 4678, $\rho$ $<$ 0.001). IA operators were interrupted less frequently by SA probe questions. The Spearman correlation analysis revealed weak correlations between the middle of an action during a SA probe and SA probe accuracy for the IA visualization using the $M_{2}$ model for $SA_{1}$ (r = -0.22, $\rho$ $<$ 0.01) as well as the $M_{3}$ model for $SA_{2}$ (r = 0.19, $\rho$ = 0.05) and $SA_{3}$ (r = -0.33, $\rho$ $<$ 0.01). A weak correlation was revealed for the Collective visualization using the $M_{2}$ model for $SA_{3}$ (r = 0.24, $\rho$ = 0.05). 

\begin{table}[h]
\begin{minipage}{0.5\linewidth}
\centering
\caption{Middle of an action during SA probe (\%) mean (SD) by SA level.}
\label{table:Usability,Middle of Action}
\begin{tabular}{c|c|c|c|}
\cline{2-4}
& \textbf{SA Level} & \textbf{IA} & \textbf{Collective} \\ \hline
\multicolumn{1}{|c|}{\multirow{4}{*}{$M_{2}$}} & $SA_{O}$ & 13.47 (34.19) & 47.02 (49.99) \\ \cline{2-4} 
\multicolumn{1}{|c|}{} & \cellcolor{gray1}$SA_{1}$ & \cellcolor{gray1}10.71 (31.04) & \cellcolor{gray1}46.1 (50.01) \\ \cline{2-4} 
\multicolumn{1}{|c|}{} & \cellcolor{gray2}$SA_{2}$ & \cellcolor{gray2}13.39 (34.21) & \cellcolor{gray2}46.43 (50.1) \\ \cline{2-4} 
\multicolumn{1}{|c|}{} & \cellcolor{gray3}$SA_{3}$ & \cellcolor{gray3}18.29 (38.9) & \cellcolor{gray3}50 (50.36) \\ \hline
\multicolumn{1}{|c|}{\multirow{4}{*}{$M_{3}$}} & $SA_{O}$ & 27.68 (44.81) & 66.67 (47.21) \\ \cline{2-4} 
\multicolumn{1}{|c|}{} & \cellcolor{gray1}$SA_{1}$ & \cellcolor{gray1}28.37 (45.24) & \cellcolor{gray1}66.96 (47.25) \\ \cline{2-4} 
\multicolumn{1}{|c|}{} & \cellcolor{gray2}$SA_{2}$ & \cellcolor{gray2}26.79 (44.48) & \cellcolor{gray2}69.29 (46.3) \\ \cline{2-4} 
\multicolumn{1}{|c|}{} & \cellcolor{gray3}$SA_{3}$ & \cellcolor{gray3}27.71 (45.03) & \cellcolor{gray3}61.9 (48.85) \\ \hline
\end{tabular}
\end{minipage} 
\hfill
\begin{minipage}{0.45\linewidth}
\centering
	\includegraphics[width=62mm]{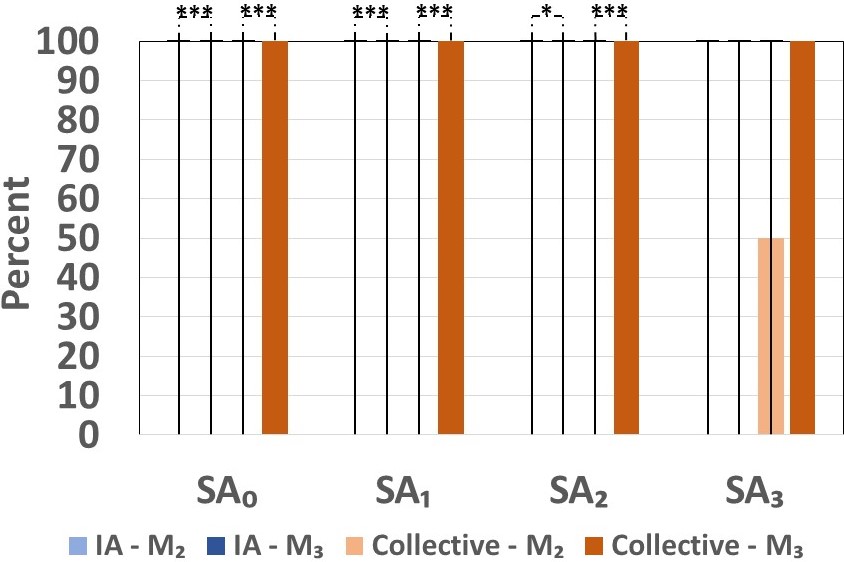}
	\captionof{figure}{The percentage of times a participant was in the middle of an action during a SA probe question median (min/max) and Mann-Whitney-Wilcoxin test by SA level between models.}
    \label{fig: Middle Action}
\end{minipage}
\end{table}

The percentage of times a participant \textit{completed an interrupted SA probe action} identified how often operators were able to return back to their previous task. A system designed to bring an operator back into-the-loop via engaging prompts, such as the dynamic individual entity behaviors or opacity of support for targets, can mitigate poor human-collective interactions and performance. A system that is easy to remember is desirable in order to attain optimal operator behavior \citep{Nielsen1993}. The percentage of completed interrupted SA probe actions mean (SD) are presented in Table \ref{table:Usability,Completed Interrupted Action}. IA operators using the $M_{3}$ model were able to complete 100\% of their interrupted actions compared to those using the $M_{2}$ model, while Collective operators using the $M_{2}$ model completed approximately 99\% of their interrupted actions. The percentage of completed interrupted SA probe actions median, min, max, and the Mann-Whitney-Wilcoxon significant effects between models are presented in Figure \ref{fig: Complete Action}. Significant differences existed between models for the IA operators for $SA_{O}$, while no differences existed for the Collective operators. Additional between visualizations Mann-Whitney-Wilcoxon tests identified a significant effect when using the $M_{3}$ model for $SA_{1}$ (n = 253, U = 55608, $\rho$ = 0.03). Operators using the IA visualization completed more interrupted actions compared those using the Collective visualization. No correlations were found between the completed interrupted SA probe actions and SA probe accuracy.

\begin{table}[h]
\begin{minipage}{0.5\linewidth}
\centering
\caption{Completed interrupted SA probe action (\%) mean (SD) by SA level.}
\label{table:Usability,Completed Interrupted Action}
\begin{tabular}{c|c|c|c|}
\cline{2-4}
& \textbf{SA Level} & \textbf{IA} & \textbf{Collective} \\ \hline
\multicolumn{1}{|c|}{\multirow{4}{*}{$M_{2}$}} & $SA_{O}$ & 98.8 (10.89) & 98.81 (10.86) \\ \cline{2-4} 
\multicolumn{1}{|c|}{} & \cellcolor{gray1}$SA_{1}$ & \cellcolor{gray1}100 (0) & \cellcolor{gray1}98.7 (11.36) \\ \cline{2-4} 
\multicolumn{1}{|c|}{} & \cellcolor{gray2}$SA_{2}$ & \cellcolor{gray2}99.11 (9.45) & \cellcolor{gray2}98.21 (13.3) \\ \cline{2-4} 
\multicolumn{1}{|c|}{} & \cellcolor{gray3}$SA_{3}$ & \cellcolor{gray3}96.34 (18.89) & \cellcolor{gray3}100 (0) \\ \hline
\multicolumn{1}{|c|}{\multirow{4}{*}{$M_{3}$}} & $SA_{O}$ & 100 (0) & 98.51 (12.13) \\ \cline{2-4} 
\multicolumn{1}{|c|}{} & \cellcolor{gray1}$SA_{1}$ & \cellcolor{gray1}100 (0) & \cellcolor{gray1}98.21 (13.3) \\ \cline{2-4} 
\multicolumn{1}{|c|}{} & \cellcolor{gray2}$SA_{2}$ & \cellcolor{gray2}100 (0) & \cellcolor{gray2}98.57 (11.91) \\ \cline{2-4} 
\multicolumn{1}{|c|}{} & \cellcolor{gray3}$SA_{3}$ & \cellcolor{gray3}100 (0) & \cellcolor{gray3}98.81 (10.91) \\ \hline
\end{tabular}
\end{minipage} 
\hfill
\begin{minipage}{0.45\linewidth}
\centering
	\includegraphics[width=62mm]{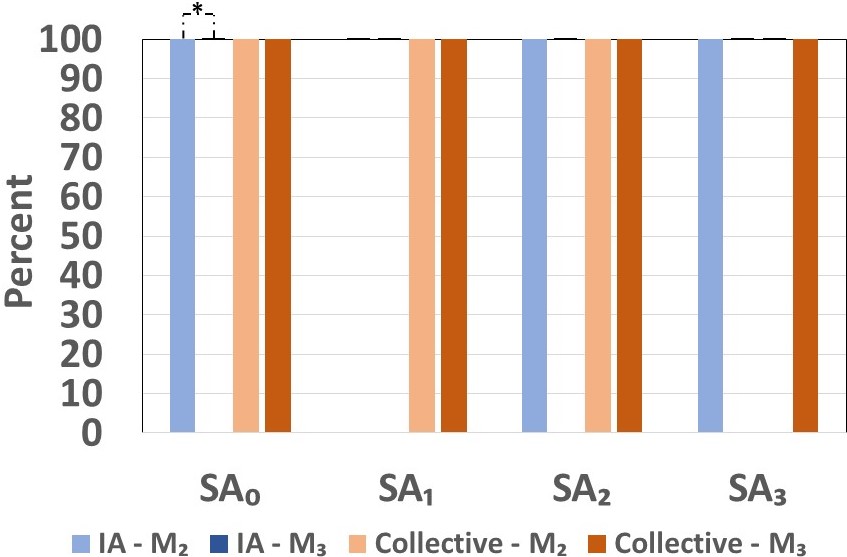}
	\captionof{figure}{Completed interrupted SA probe action median (min/max) and Mann-Whitney-Wilcoxin test by SA level between models.}
    \label{fig: Complete Action}
\end{minipage}
\end{table}

\begin{table}[!b]
\begin{minipage}{0.5\linewidth}
\centering
\caption{Investigate commands per decision mean (SD) by decision difficulty (Dec Diff).}
\label{table:Usability,Investigate}
\begin{tabular}{c|c|c|c|}
\cline{2-4}
& \textbf{Dec Diff} & \textbf{IA} & \textbf{Collective} \\ \hline
\multicolumn{1}{|c|}{\multirow{3}{*}{$M_{2}$}} & Overall & 2.1 (3.23) & 1.78 (1.62) \\ \cline{2-4} 
\multicolumn{1}{|c|}{} & \cellcolor{gray1}Easy & \cellcolor{gray1}2.06 (2.75) & \cellcolor{gray1}1.53 (1.49) \\ \cline{2-4}
\multicolumn{1}{|c|}{} & \cellcolor{gray2}Hard & \cellcolor{gray2}2.15 (3.79) & \cellcolor{gray2}2.06 (1.72) \\ \hline
\multicolumn{1}{|c|}{\multirow{3}{*}{$M_{3}$}} & Overall & 8.72 (3.82) & 4.74 (2.2) \\ \cline{2-4} 
\multicolumn{1}{|c|}{} & \cellcolor{gray1}Easy & \cellcolor{gray1}8.09 (3.95) & \cellcolor{gray1}4.23 (2.11) \\ \cline{2-4}
\multicolumn{1}{|c|}{} & \cellcolor{gray2}Hard & \cellcolor{gray2}9.64 (3.44) & \cellcolor{gray2}5.47 (2.12) \\ \hline
\end{tabular}
\end{minipage} 
\hfill
\begin{minipage}{0.45\linewidth}
\centering
	\includegraphics[width=62mm]{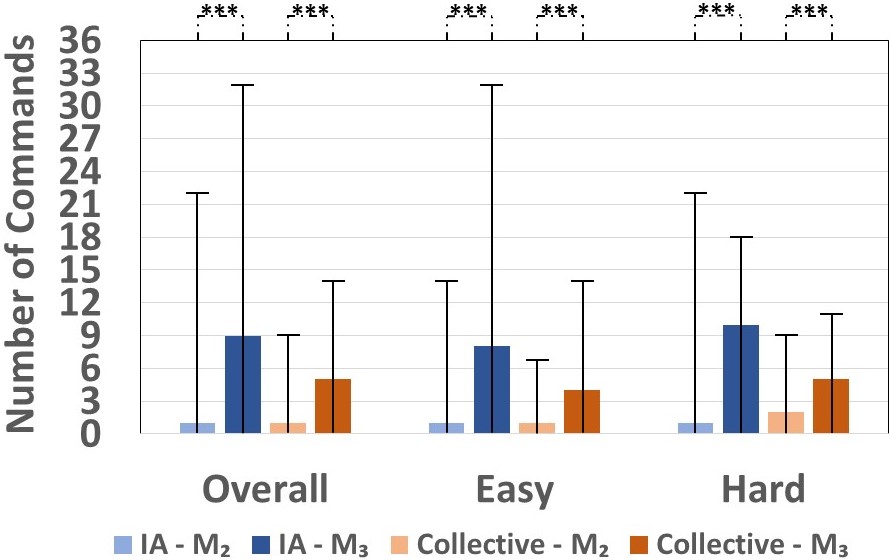}
	\captionof{figure}{The number of investigate commands issued per decision median (min/max) and Mann-Whitney-Wilcoxin test by decision difficulty between models.}
    \label{fig: Investigate}
\end{minipage}
\end{table}

The \textit{investigate} command permitted increasing a collective's support for an operator specified target. Additional support for the same target was achieved by reissuing the investigate command repeatedly. The number of investigate commands issued per decision mean (SD) are presented in Table \ref{table:Usability,Investigate} \citep{Roundtree20191}. Generally, operators using the $M_{2}$ model and Collective visualization issued fewer investigate commands. The number of investigate commands issued per decision median, min, max, and the Mann-Whitney-Wilcoxon significant effects between models are presented in Figure \ref{fig: Investigate}. Significant differences were found between models for the number of investigate commands issued per decision for both visualizations at all decision difficulties. Additional between visualizations Mann-Whitney-Wilcoxon tests identified a moderate significant effect when using the $M_{2}$ model for overall decisions (n = 672, U = 63866, $\rho$ $<$ 0.01) and a highly significant effect for hard decisions (n = 298, U = 14066, $\rho$ $<$ 0.001). Highly significant effects between visualizations when using the $M_{3}$ model were also found for overall (U = 17990, $\rho$ $<$ 0.001), easy (n = 396, U = 6279.5, $\rho$ $<$ 0.001), and hard decisions (n = 276, U = 2331.5, $\rho$ $<$ 0.001).

The \textit{abandon} command permitted decreasing a collective's support for a target. The abandon command only needed to be issued once in order for the collective to ignore a specified target for the duration of a decision. The number of abandon commands issued per decision mean (SD) are presented in Table \ref{table:Usability,Abandon} \citep{Roundtree20191}. Operators using the $M_{2}$ model in general issued fewer abandon commands compared to the $M_{3}$ model; however, IA operators using the $M_{3}$ model issued fewer abandon commands for hard decisions. The number of abandon commands issued per decision median, min, max, and the Mann-Whitney-Wilcoxon significant effects between models are presented in Figure \ref{fig: Abandon}. Significant differences were found between models for the number of abandon commands issued per decision with both visualizations for overall and easy decisions. No significant effects between visualizations were found. IA operators issued fewer abandon commands in general compared to those using the Collective visualization. Collective operators using the $M_{2}$ model issued fewer abandon commands for overall and hard decisions only. 

\begin{table}[h]
\begin{minipage}{0.5\linewidth}
\centering
\caption{Abandon commands per decision mean (SD) by decision difficulty (Dec Diff).}
\label{table:Usability,Abandon}
\begin{tabular}{c|c|c|c|}
\cline{2-4}
& \textbf{Dec Diff} & \textbf{IA} & \textbf{Collective} \\ \hline
\multicolumn{1}{|c|}{\multirow{3}{*}{$M_{2}$}} & Overall & 0.1 (0.54) & 0.09 (0.29) \\ \cline{2-4} 
\multicolumn{1}{|c|}{} & \cellcolor{gray1}Easy & \cellcolor{gray1}0.05 (0.22) & \cellcolor{gray1}0.06 (0.24) \\ \cline{2-4}
\multicolumn{1}{|c|}{} & \cellcolor{gray2}Hard & \cellcolor{gray2}0.16 (0.79) & \cellcolor{gray2}0.12 (0.34) \\ \hline 
\multicolumn{1}{|c|}{\multirow{3}{*}{$M_{3}$}} & Overall & 0.15 (0.43) & 0.17 (0.42) \\ \cline{2-4} 
\multicolumn{1}{|c|}{} & \cellcolor{gray1}Easy & \cellcolor{gray1}0.15 (0.45) & \cellcolor{gray1}0.16 (0.4) \\ \cline{2-4}
\multicolumn{1}{|c|}{} & \cellcolor{gray2}Hard & \cellcolor{gray2}0.15 (0.4) & \cellcolor{gray2}0.19 (0.45) \\ \hline
\end{tabular}
\end{minipage} 
\hfill
\begin{minipage}{0.45\linewidth}
\centering
	\includegraphics[width=62mm]{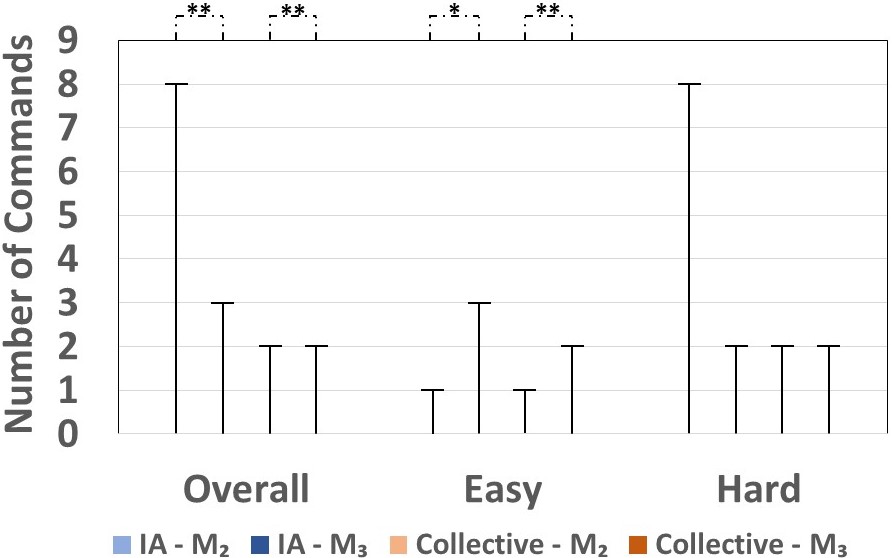}
	\captionof{figure}{The number of abandon commands issued per decision median (min/max) and Mann-Whitney-Wilcoxin test by decision difficulty between models.}
    \label{fig: Abandon}
\end{minipage}
\end{table}

A collective's entities stopped exploring alternative targets and moved to the operator selected target when the \textit{decide} command was issued. A decide request required at least 30\% of the collective support for the operator specified target. Collectives that reached 50\% support for a target transitioned into the executing state and the operator was no longer able to influence the collective behavior. The number of decide commands issued per decision mean (SD) are presented in Table \ref{table:Usability,Decide} \citep{Roundtree20191}. Operators using the $M_{2}$ model and the IA visualization issued fewer decide commands compared to those using the $M_{3}$ model or the Collective visualization. The number of decide commands issued per decision median, min, max, and the Mann-Whitney-Wilcoxon significant effects between models are presented in Figure \ref{fig: Decide}. Significant differences were found between models for the number of decide commands issued per decision for both visualizations at all decision difficulties. Additional between visualizations Mann-Whitney-Wilcoxon tests identified highly significant effects using the $M_{2}$ model for overall (n = 672, U = 63968, $\rho$ $<$ 0.01) and easy decisions (n = 374, U = 21014, $\rho$ $<$ 0.001). A moderately significant effect between visualizations when using the $M_{3}$ model was found for overall decisions (U = 57952, $\rho$ $<$ 0.01) and a significant effect existed for easy decisions (n = 377, U = 19997, $\rho$ = 0.05).

\begin{table}[h]
\begin{minipage}{0.5\linewidth}
\centering
\caption{Decide commands per decision mean (SD) by decision difficulty (Dec Diff).}
\label{table:Usability,Decide}
\begin{tabular}{c|c|c|c|}
\cline{2-4}
& \textbf{Dec Diff} & \textbf{IA} & \textbf{Collective} \\ \hline
\multicolumn{1}{|c|}{\multirow{3}{*}{$M_{2}$}} & Overall & 0.38 (0.49) & 0.52 (0.51) \\ \cline{2-4}
\multicolumn{1}{|c|}{} & \cellcolor{gray1}Easy & \cellcolor{gray1}0.38 (0.49) & \cellcolor{gray1}0.58 (0.51) \\ \cline{2-4}
\multicolumn{1}{|c|}{} & \cellcolor{gray2}Hard & \cellcolor{gray2}0.39 (0.49) & \cellcolor{gray2}0.44 (0.51) \\ \hline 
\multicolumn{1}{|c|}{\multirow{3}{*}{$M_{3}$}} & Overall & 0.99 (0.08) & 1.03 (0.26) \\ \cline{2-4}
\multicolumn{1}{|c|}{} & \cellcolor{gray1}Easy & \cellcolor{gray1}1 (0.07) & \cellcolor{gray1}1.03 (0.24) \\ \cline{2-4}
\multicolumn{1}{|c|}{} & \cellcolor{gray2}Hard & \cellcolor{gray2}0.99 (0.09) & \cellcolor{gray2}1.04 (0.29) \\ \hline
\end{tabular}
\end{minipage} 
\hfill
\begin{minipage}{0.45\linewidth}
\centering
	\includegraphics[width=62mm]{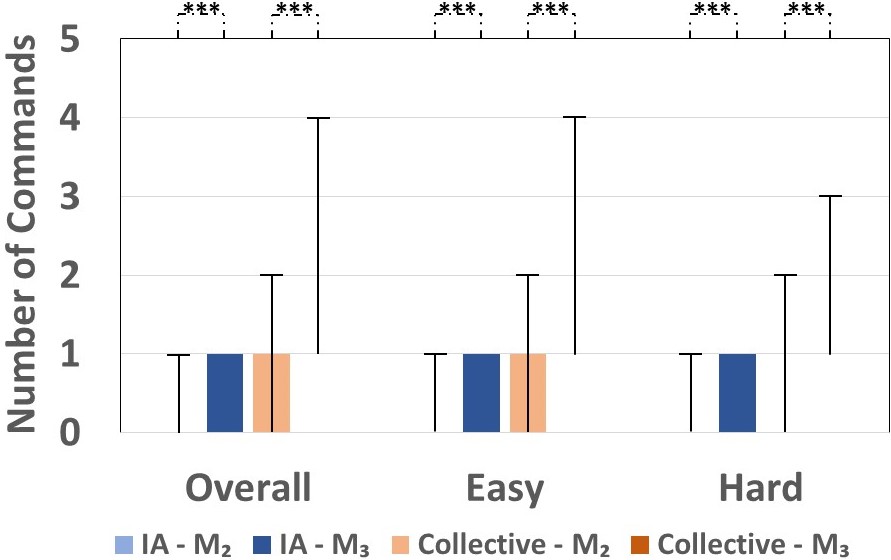}
	\captionof{figure}{The number of decide commands issued per decision median (min/max) and Mann-Whitney-Wilcoxin test by decision difficulty between models.}
    \label{fig: Decide}
\end{minipage}
\end{table}

\textit{Collective right-clicks} and \textit{target right-clicks} allowed the operator to access the respective information pop-up windows, which provided the number of individual collective entities in each particular state and the percentage of support each collective had for a respective target. The $M_{3}$ model in general had fewer collective and target right-clicks compared to $M_{2}$ model, while the Collective visualization had fewer target right-clicks compared to the IA visualization. The statistical analyses of both metrics were provided in Section \ref{sec: R2 metrics}.

\begin{table}[h]
\begin{minipage}{0.5\linewidth}
\centering
\caption{The percentage of times abandon commands exceeded abandoned targets per participant mean (SD) by decision difficulty (Dec Diff).}
\label{table:Usability,Abandon Exceeded}
\begin{tabular}{c|c|c|c|}
\cline{2-4}
& \textbf{Dec Diff} & \textbf{IA} & \textbf{Collective} \\ \hline
\multicolumn{1}{|c|}{\multirow{3}{*}{$M_{2}$}} & Overall & 1.18 (3.02) & 2.68 (6.27) \\ \cline{2-4}
\multicolumn{1}{|c|}{} & \cellcolor{gray1}Easy & \cellcolor{gray1}0.4 (1.55) & \cellcolor{gray1}2.05 (5.06) \\ \cline{2-4} 
\multicolumn{1}{|c|}{} & \cellcolor{gray2}Hard & \cellcolor{gray2}1.35 (4) & \cellcolor{gray2}3.08 (7.74) \\ \hline
\multicolumn{1}{|c|}{\multirow{3}{*}{$M_{3}$}} & Overall & 6.88 (6.62) & 6.54 (6.32) \\ \cline{2-4}
\multicolumn{1}{|c|}{} & \cellcolor{gray1}Easy & \cellcolor{gray1}1.48 (4.39) & \cellcolor{gray1}2.82 (5.73) \\ \cline{2-4} 
\multicolumn{1}{|c|}{} & \cellcolor{gray2}Hard & \cellcolor{gray2}13.26 (9.85) & \cellcolor{gray2}10.91 (9.38) \\ \hline
\end{tabular}
\end{minipage} 
\hfill
\begin{minipage}{0.45\linewidth}
\centering
	\includegraphics[width=62mm]{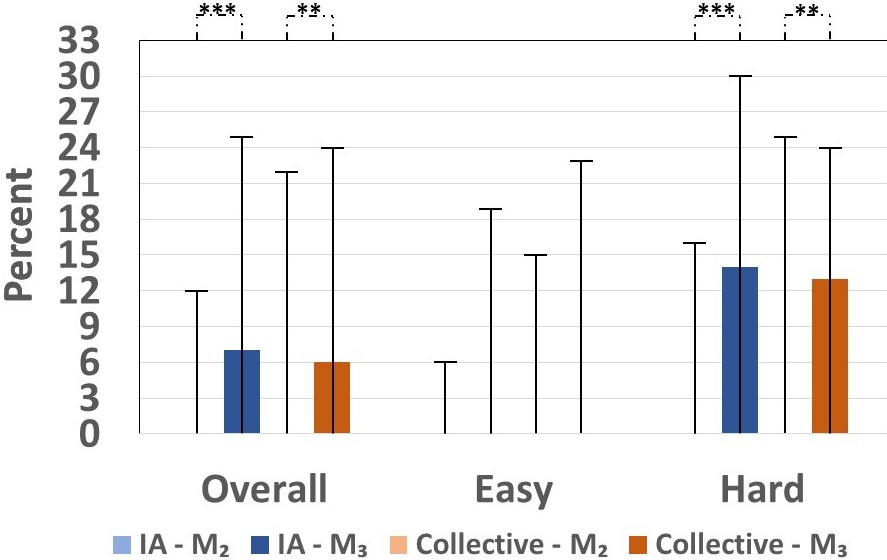}
	\captionof{figure}{The percent of times abandon commands exceeded abandoned targets median (min/max) and Mann-Whitney-Wilcoxin test by decision difficulty between models.}
    \label{fig: Exceed Abandon}
\end{minipage}
\end{table}

Metrics showing how operators used the abandon command were assessed. Operators using the $M_{3}$ model and IA visualization abandoned the \textit{highest value target} less frequently and had fewer \textit{abandoned target information pop-up windows open}. The statistical analyses of both metrics were provided in Section \ref{sec: R2 metrics}. Instances may have occurred when the operator accidentally issued an undesired abandon command or repeatedly issued the abandon command, although targets were abandoned after a single command was issued. The percent of times \textit{abandon commands exceeded abandoned targets} was examined and the mean (SD) are presented in Table \ref{table:Usability,Abandon Exceeded} \citep{Roundtree2020visual}. Operators using the $M_{2}$ model issued fewer repeated abandon commands compared to the $M_{3}$ model. The percent of times abandon commands exceeded abandoned targets median, min, max, and the Mann-Whitney-Wilcoxon significant effects between models are presented in Figure \ref{fig: Exceed Abandon}. Significant differences were found between models for the percent of times abandon commands exceeded abandoned targets with both visualizations for overall and hard decisions. No significant effects between visualizations were found. IA operators had fewer repeated abandon commands in general compared to those using the Collective visualization. Collective operators using the $M_{3}$ model had fewer repeated abandon commands for overall and hard decisions. 

The \textit{time difference (minutes) between the commit state and issued decide command} assessed the operator's ability to predict the collective's future state transition from the committed state (30\% support for a target) to executing (50\% support for a target). The time difference mean (SD) are shown in Table \ref{table:Usability,Timing of Commit} \citep{Roundtree2020visual}. Operators using the $M_{3}$ model issued decide commands faster than the $M_{2}$ model. The time difference between commit state and issued decide command median, min, max, and the Mann-Whitney-Wilcoxon significant effects between models are presented in Figure \ref{fig: Time Diff}. Significant differences existed between models for the time difference between the commit state and issued decide command for both visualizations at all decision difficulties. Collective operators in general had smaller time differences between the committed state and issued decide commands compared to those using the IA visualization; however, no significant effects between visualizations were found. IA operators using the $M_{2}$ model had smaller time differences between the commit state and decide command for hard decisions.

\begin{table}[h]
\begin{minipage}{0.5\linewidth}
\centering
\caption{The time difference (minutes) between commit state and issued decide command per participant mean (SD) by decision difficulty (Dec Diff).}
\label{table:Usability,Timing of Commit}
\begin{tabular}{c|c|c|c|}
\cline{2-4}
& \textbf{Dec Diff} & \textbf{IA} & \textbf{Collective} \\ \hline
\multicolumn{1}{|c|}{\multirow{3}{*}{$M_{2}$}} & Overall & 0.68 (0.27) & 0.65 (0.15) \\ \cline{2-4}
\multicolumn{1}{|c|}{} & \cellcolor{gray1}Easy & \cellcolor{gray1}0.7 (0.47) & \cellcolor{gray1}0.56 (0.14) \\ \cline{2-4} 
\multicolumn{1}{|c|}{} & \cellcolor{gray2}Hard & \cellcolor{gray2}0.72 (0.21) & \cellcolor{gray2}0.78 (0.3) \\ \hline
\multicolumn{1}{|c|}{\multirow{3}{*}{$M_{3}$}} & Overall & 0.6 (0.3) & 0.57 (0.2) \\ \cline{2-4} 
\multicolumn{1}{|c|}{} & \cellcolor{gray1}Easy & \cellcolor{gray1}0.57 (0.53) & \cellcolor{gray1}0.52 (0.32) \\ \cline{2-4} 
\multicolumn{1}{|c|}{} & \cellcolor{gray2}Hard & \cellcolor{gray2}0.62 (0.22) & \cellcolor{gray2}0.62 (0.18) \\ \hline
\end{tabular}
\end{minipage} 
\hfill
\begin{minipage}{0.45\linewidth}
\centering
	\includegraphics[width=62mm]{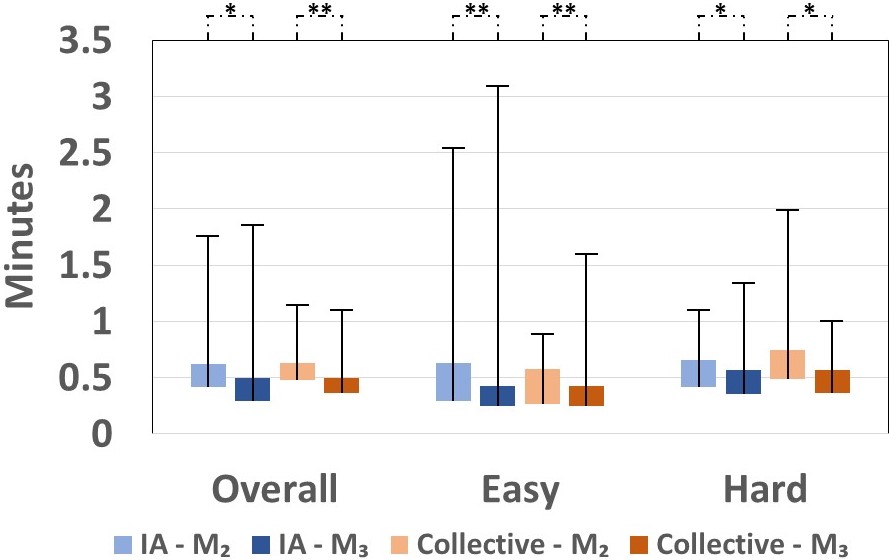}
	\captionof{figure}{The time difference between commit state and issued decide command median (min/max) and Mann-Whitney-Wilcoxin test by decision difficulty between models.}
    \label{fig: Time Diff}
\end{minipage}
\end{table}

Further analysis of how operators used the collective and target information pop-up windows was conducted. The average number of times target information pop-up windows were opened per target per decision identified the \textit{average frequency at which the information pop-up windows were accessed}. The average frequency of an accessed target information pop-up window per target per decision mean (SD) are shown in Table \ref{table:Usability,Number of Times Target Windows Opened}. Operators using the $M_{3}$ model in general accessed target information pop-up windows less frequently compared to the $M_{2}$ model. Target information pop-up windows were accessed less frequently for operators from both evaluations using the $M_{2}$ model for easy decisions. The average frequency of an accessed target information pop-up window median, min, max, and the Mann-Whitney-Wilcoxon significant effects between models are presented in Figure \ref{fig: Freq Target}. IA operators had significantly different average frequencies of accessed target information pop-up windows between models for hard decisions, while the Collective operators had no significant differences between models. Additional between visualizations Mann-Whitney-Wilcoxon tests identified a significant effect when using the $M_{2}$ model for overall decisions (n = 619, U = 42857, $\rho$ = 0.02) and a moderate significant effect for hard decisions (n = 282, U = 7908.5, $\rho$ $<$ 0.01). Operators using the Collective visualization accessed target information pop-up windows less frequently compared to the IA visualization. 

\begin{table}[h]
\begin{minipage}{0.5\linewidth}
\centering
\caption{Average frequency of accessed target information pop-up window per target per decision mean (SD) by decision difficulty (Dec Diff).}
\label{table:Usability,Number of Times Target Windows Opened}
\begin{tabular}{c|c|c|c|}
\cline{2-4}
& \textbf{Dec Diff} & \textbf{IA} & \textbf{Collective} \\ \hline
\multicolumn{1}{|c|}{\multirow{3}{*}{$M_{2}$}} & Overall & 1.93 (1.17) & 1.67 (0.94) \\ \cline{2-4}
\multicolumn{1}{|c|}{} & \cellcolor{gray1}Easy & \cellcolor{gray1}1.7 (0.98) & \cellcolor{gray1}1.57 (0.82) \\ \cline{2-4} 
\multicolumn{1}{|c|}{} & \cellcolor{gray2}Hard & \cellcolor{gray2}2.23 (1.33) & \cellcolor{gray2}1.79 (1.05) \\ \hline
\multicolumn{1}{|c|}{\multirow{3}{*}{$M_{3}$}} & Overall & 1.8 (1.33) & 1.67 (0.91) \\ \cline{2-4} 
\multicolumn{1}{|c|}{} & \cellcolor{gray1}Easy & \cellcolor{gray1}1.83 (1.48) & \cellcolor{gray1}1.62 (0.88) \\ \cline{2-4} 
\multicolumn{1}{|c|}{} & \cellcolor{gray2}Hard & \cellcolor{gray2}1.77 (1.1) & \cellcolor{gray2}1.73 (0.95) \\ \hline
\end{tabular}
\end{minipage} 
\hfill
\begin{minipage}{0.45\linewidth}
\centering
	\includegraphics[width=62mm]{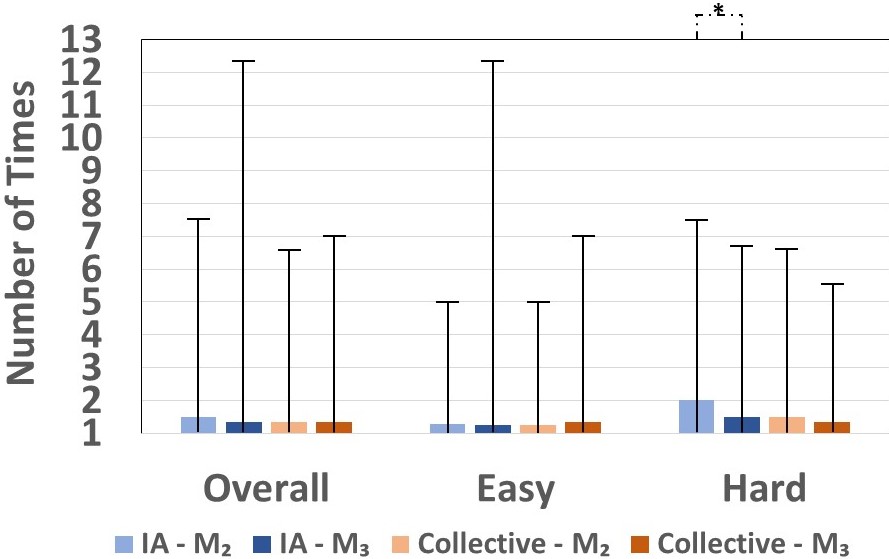}
	\captionof{figure}{Average frequency of accessed target information pop-up window per target median (min/max) and Mann-Whitney-Wilcoxin test by decision difficulty between models.}
    \label{fig: Freq Target}
\end{minipage}
\end{table}

Operators using the target information pop-up windows may have accessed them frequently for short time periods, or left them open for longer time periods. The average percentage of \textit{time a target information pop-up window was open per target} relative to the decision time mean (SD) are presented in Table \ref{table:Usability,Percentage of Time Target Windows Opened}. IA operators using the $M_{2}$ model left target information pop-up windows open for shorter time periods. The average time target information windows were opened median, min, max, and the Mann-Whitney-Wilcoxon significant effects between models are presented in Figure \ref{fig: Ave Time Target}. Significant differences were found between models for the average time target information pop-up windows were open for both visualizations at all decision difficulties; however, no significant effects between visualizations were found.

\begin{table}[h]
\begin{minipage}{0.5\linewidth}
\centering
\caption{Average time target information windows opened per target per decision (\%) mean (SD) by decision difficulty (Dec Diff).}
\label{table:Usability,Percentage of Time Target Windows Opened}
\begin{tabular}{c|c|c|c|}
\cline{2-4}
& \textbf{Dec Diff} & \textbf{IA} & \textbf{Collective} \\ \hline
\multicolumn{1}{|c|}{\multirow{3}{*}{$M_{2}$}} & Overall & 24.18 (26.65) & 28.38 (28.61) \\ \cline{2-4} 
\multicolumn{1}{|c|}{} & \cellcolor{gray1}Easy & \cellcolor{gray1}27.53 (28.76) & \cellcolor{gray1}30.48 (29.12) \\ \cline{2-4} 
\multicolumn{1}{|c|}{} & \cellcolor{gray2}Hard & \cellcolor{gray2}19.87 (23.05) & \cellcolor{gray2}26.05 (27.95) \\ \hline
\multicolumn{1}{|c|}{\multirow{3}{*}{$M_{3}$}} & Overall & 34.93 (25.29) & 36.58 (29.41) \\ \cline{2-4} 
\multicolumn{1}{|c|}{} & \cellcolor{gray1}Easy & \cellcolor{gray1}37.56 (27.01) & \cellcolor{gray1}37.63 (30.98) \\ \cline{2-4} 
\multicolumn{1}{|c|}{} & \cellcolor{gray2}Hard & \cellcolor{gray2}31.12 (22.12) & \cellcolor{gray2}35.09 (27.07) \\ \hline
\end{tabular}
\end{minipage} 
\hfill
\begin{minipage}{0.45\linewidth}
\centering
	\includegraphics[width=62mm]{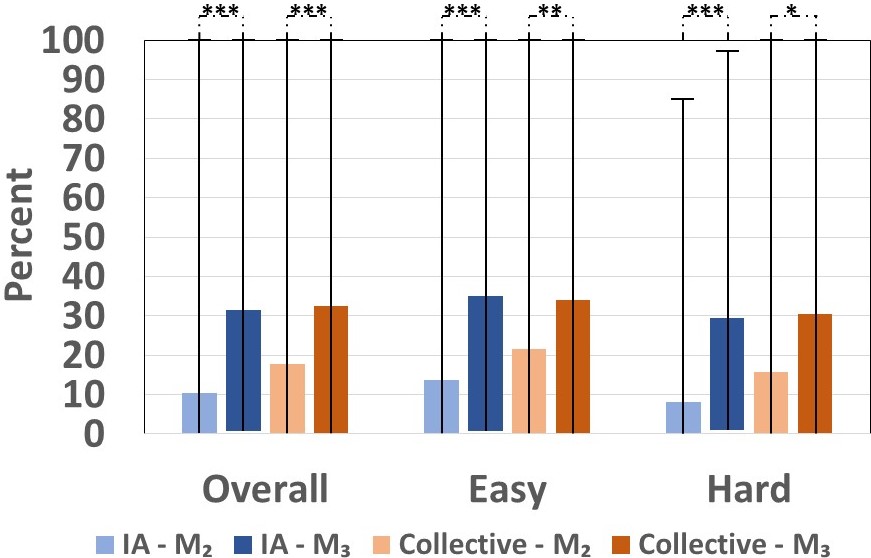}
	\captionof{figure}{Average time target information windows opened per target median (min/max) and Mann-Whitney-Wilcoxin test by decision difficulty between models.}
    \label{fig: Ave Time Target}
\end{minipage}
\end{table}

Operators may have accessed particular target information pop-up windows, such as the decision target, more frequently for longer time periods. The average percentage of \textit{time the decision target information pop-up window was open} relative to the decision time mean (SD) are shown in Table \ref{table:Usability,Percentage of Time Decision Target Window Open}. Operators using the $M_{2}$ model left the decision target information pop-up window open for shorter periods of time compared to the $M_{3}$ model. The time the decision target information window is open median, min, max, and the Mann-Whitney-Wilcoxon significant effects between models are shown in Figure \ref{fig: Time Dec Target}. Significant differences were found between models for the time the decision target information window was open for both visualizations at all decision difficulties. Additional between visualizations Mann-Whitney-Wilcoxon tests identified a highly significant effect using the $M_{2}$ model for overall decisions (n = 672, U = 65102, $\rho$ $<$ 0.001), as well as significant effects for easy (n = 374, U = 20114, $\rho$ = 0.01), and hard decisions (n = 298, U = 12832, $\rho$ = 0.02). A moderately significant effect between visualizations using the $M_{3}$ model was found for overall decisions (U = 48749, $\rho$ $<$ 0.01), with significant effects for easy (n = 396, U = 17095, $\rho$ = 0.03), and hard decisions (n = 276, U = 8157, $\rho$ = 0.04). IA operators using the $M_{2}$ model left the decision target information pop-up window open for shorter periods of time compared to the $M_{3}$ model, while the Collective operators had shorter time periods when using the $M_{3}$ model.  

\begin{table}[h]
\begin{minipage}{0.5\linewidth}
\centering
\caption{The time decision target information window open per decision (\%) mean (SD) by decision difficulty (Dec Diff).}
\label{table:Usability,Percentage of Time Decision Target Window Open}
\begin{tabular}{c|c|c|c|}
\cline{2-4}
& \textbf{Dec Diff} & \textbf{IA} & \textbf{Collective} \\ \hline
\multicolumn{1}{|c|}{\multirow{3}{*}{$M_{2}$}} & Overall & 21.64 (28.25) & 30.55 (32.6) \\ \cline{2-4} 
\multicolumn{1}{|c|}{} & \cellcolor{gray1}Easy & \cellcolor{gray1}23.69 (30.7) & \cellcolor{gray1}32.51 (34.43) \\ \cline{2-4} 
\multicolumn{1}{|c|}{} & \cellcolor{gray2}Hard & \cellcolor{gray2}18.84 (24.33) & \cellcolor{gray2}28.27 (30.31) \\ \hline
\multicolumn{1}{|c|}{\multirow{3}{*}{$M_{3}$}} & Overall & 50.56 (29.1) & 43.94 (31.69) \\ \cline{2-4} 
\multicolumn{1}{|c|}{} & \cellcolor{gray1}Easy & \cellcolor{gray1}50.71 (29.31) & \cellcolor{gray1}44.12 (33.33)  \\ \cline{2-4} 
\multicolumn{1}{|c|}{} & \cellcolor{gray2}Hard & \cellcolor{gray2}50.34 (28.91) & \cellcolor{gray2}43.67 (29.31) \\ \hline
\end{tabular}
\end{minipage} 
\hfill
\begin{minipage}{0.45\linewidth}
\centering
	\includegraphics[width=62mm]{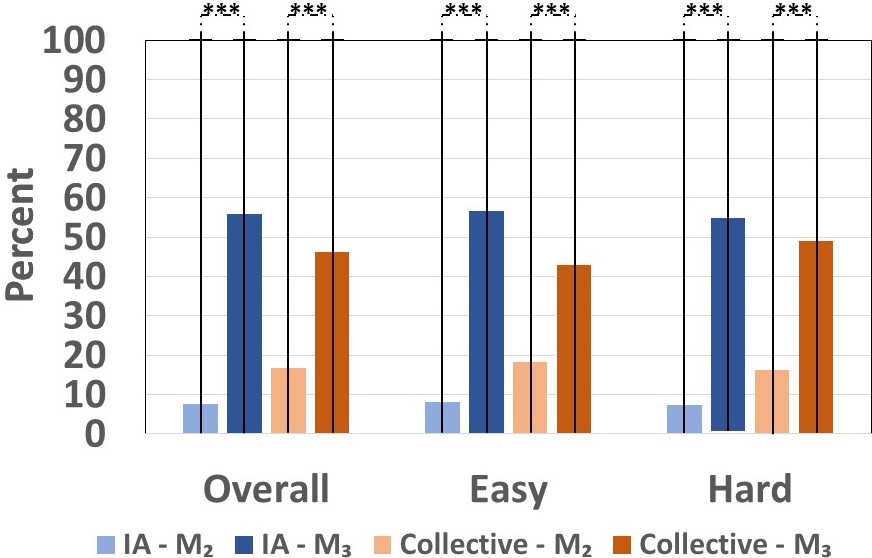}
	\captionof{figure}{The time decision target information window open median (min/max) and Mann-Whitney-Wilcoxin test by decision difficulty between models.}
    \label{fig: Time Dec Target}
\end{minipage}
\end{table}

\begin{table}[!b]
\begin{minipage}{0.5\linewidth}
\centering
\caption{The time decision collective information window open per decision (\%) mean (SD) by decision difficulty (Dec Diff).}
\label{table:Usability,Percentage of Time Decision Collective Window Open}
\begin{tabular}{c|c|c|}
\cline{2-3}
& \textbf{Dec Diff} & \textbf{IA} \\ \hline
\multicolumn{1}{|c|}{\multirow{3}{*}{$M_{2}$}} & Overall & 21.37 (35.24) \\ \cline{2-3} 
\multicolumn{1}{|c|}{} & \cellcolor{gray1}Easy & \cellcolor{gray1}20.16 (34.79) \\ \cline{2-3} 
\multicolumn{1}{|c|}{} & \cellcolor{gray2}Hard & \cellcolor{gray2}23.03 (35.91)) \\ \hline
\multicolumn{1}{|c|}{\multirow{3}{*}{$M_{3}$}} & Overall & 19.84 (35.28) \\ \cline{2-3} 
\multicolumn{1}{|c|}{} & \cellcolor{gray1}Easy & \cellcolor{gray1}19.74 (34.79) \\ \cline{2-3} 
\multicolumn{1}{|c|}{} & \cellcolor{gray2}Hard & \cellcolor{gray2}20 (36.12) \\ \hline
\end{tabular}
\end{minipage} 
\hfill
\begin{minipage}{0.45\linewidth}
\centering
	\includegraphics[width=62mm]{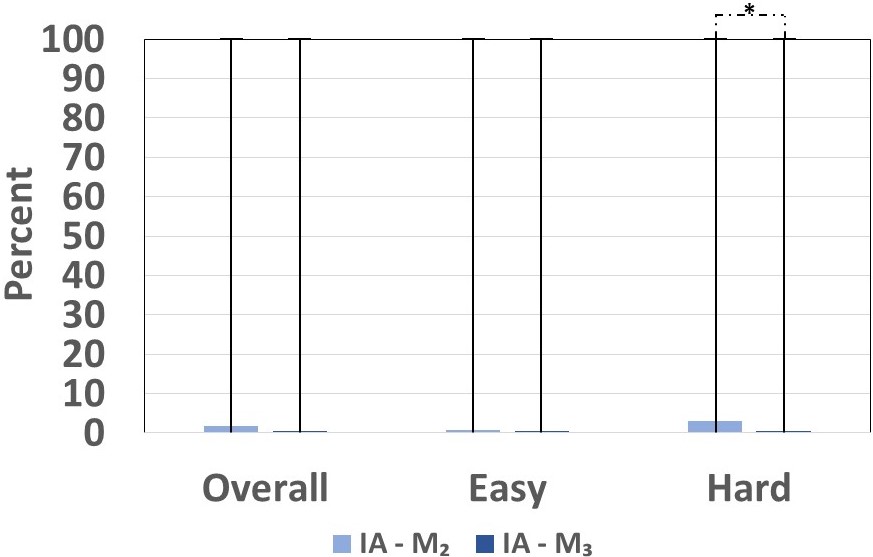}
	\captionof{figure}{The time decision collective information window open median (min/max) and Mann-Whitney-Wilcoxin test by decision difficulty between models.}
    \label{fig: Time Dec Collective}
\end{minipage}
\end{table}

The average percentage of \textit{time the decision collective information pop-up window was open} relative to the decision time mean (SD) are presented in Table \ref{table:Usability,Percentage of Time Decision Collective Window Open}. The time the decision collective information pop-up window was open was only assessed for the IA evaluation, because the Collective evaluation did not record which particular collective pop-up window was opened or closed. IA operators using the $M_{3}$ model left the decision collective information pop-up window open for shorter periods of time compared to the $M_{2}$ model. The time the decision collective information window is open median, min, max, and the Mann-Whitney-Wilcoxon significant effects between models are shown in Figure \ref{fig: Time Dec Collective}. IA operators had significantly different times for hard decisions.

The post-trial questionnaire assessed the \textit{perceived effectiveness of each request type} (investigate, abandon, and decide), not effective (1) to very effective (7). The post-trial effectiveness subjective ranking mean (SD) are presented in Table \ref{table:Usability,PT Effectiveness} \cite{Cody2018}. The investigate, abandon, and decide rankings were generally ranked higher for operators using the $M_{3}$ model when compared to those using the $M_{2}$ model. Collective operators using the $M_{2}$ model ranked abandon effectiveness higher. The post-trial effectiveness median, min, max, and the Mann-Whitney-Wilcoxon significant effects between models are shown in Figure \ref{fig: Post Trial Effective}. Significant differences between models were found in IA operator rankings for the decide command and for Collective operator rankings for both the abandon and decide commands. Additional between visualizations Mann-Whitney-Wilcoxon tests identified a moderate significant effect for the abandon effectiveness when using the $M_{2}$ model (n = 56, U = 554.5, $\rho$ $<$ 0.01). IA operators using the $M_{3}$ model ranked investigate, abandon, and decide effectiveness higher compared to those using the Collective visualization, while Collective operators ranked abandon effectiveness higher when using the $M_{2}$ model.

\begin{table}[h]
\begin{minipage}{0.5\linewidth}
\centering
\caption{Post-trial command effectiveness ranking mean (SD) (1-low, 7-high).}
\label{table:Usability,PT Effectiveness}
\begin{tabular}{c|c|c|c|}
\cline{2-4}
& \textbf{Metric} & \textbf{IA} & \textbf{Collective} \\ \hline
\multicolumn{1}{|c|}{\multirow{3}{*}{$M_{2}$}} & Investigate & 4.68 (1.56) & 4.75 (1.53) \\ \cline{2-4}
\multicolumn{1}{|c|}{} & \cellcolor{gray1}Abandon & \cellcolor{gray1}4.82 (1.96) & \cellcolor{gray1}6.18 (1.42) \\ \cline{2-4}
\multicolumn{1}{|c|}{} & \cellcolor{gray2}Decide & \cellcolor{gray2}5.29 (1.7) & \cellcolor{gray2}5.57 (1.99) \\ \hline
\multicolumn{1}{|c|}{\multirow{3}{*}{$M_{3}$}} & Investigate & 5.46 (1.4) & 5.18 (1.68) \\ \cline{2-4} 
\multicolumn{1}{|c|}{} & \cellcolor{gray1}Abandon & \cellcolor{gray1}5.29 (1.84) & \cellcolor{gray1}5.29 (1.76) \\ \cline{2-4}
\multicolumn{1}{|c|}{} & \cellcolor{gray2}Decide & \cellcolor{gray2}6.79 (0.5) & \cellcolor{gray2}6.54 (0.92) \\ \hline
\end{tabular}
\end{minipage} 
\hfill
\begin{minipage}{0.45\linewidth}
\centering
	\includegraphics[width=62mm]{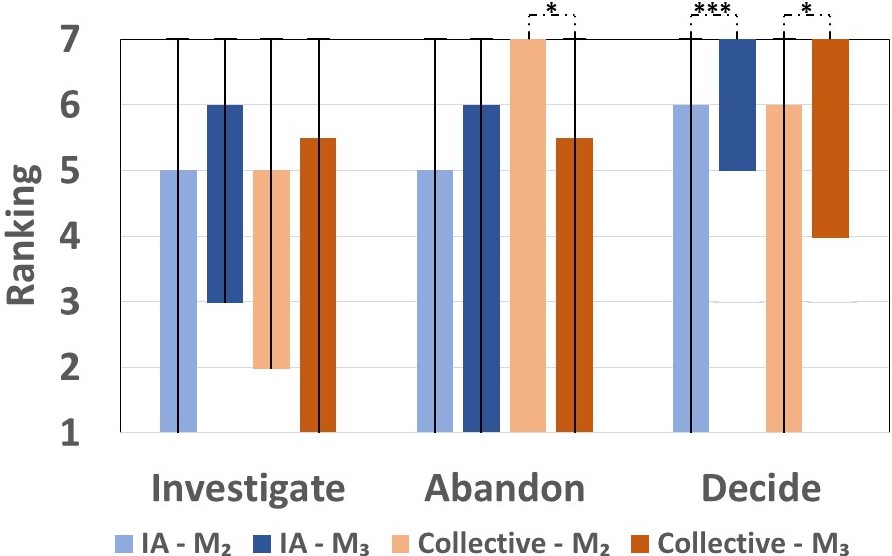}
	\captionof{figure}{Post-trial command effectiveness ranking median (min/max) and Mann-Whitney-Wilcoxin test between models.}
    \label{fig: Post Trial Effective}
\end{minipage}
\end{table}

The post-experiment questionnaire assessed the collective's \textit{responsiveness} to requests, the participants' \textit{ability} to choose the highest valued target, and their \textit{understanding} of the collective behavior. IA operators who used the $M_{2}$ model had the best collective responsiveness, operator ability, and understanding versus the $M_{3}$ model. Collective operators ranked the collective's responsiveness highest using the $M_{3}$ model, while operator ability and understanding were highest using the $M_{2}$ model. Details regarding the statistical tests were provided in the Metrics and Results Section \ref{section:R1 metrics}.

A summary of $R_{3}$'s results by the hypotheses, with significant results identified, is provided in Table \ref{table:Usability,Combined}. This summary table is intended to facilitate the discussion.

\begin{table}[!t]
\centering
\caption{A synopsis of $R_{3}$'s hypotheses associated with significant results. The SA probe timings are all timings (All), 15 seconds Before asking (B), While being asked (W), and During response (D) to a SA probe question.}
\label{table:Usability,Combined}
\begin{tabular}{?l|cc|cc|cc|c|c|c|c|c|c?}
\Cline{1pt}{1-10}
\multicolumn{1}{?c|}{\multirow{4}{*}{\textbf{Variable}}} & & \multicolumn{2}{?c?}{\textbf{Within}} & \multicolumn{2}{c?}{\textbf{Between}} & \multicolumn{4}{c?}{\multirow{2}{*}{\textbf{Correlation}}} \\
& \textbf{Sub-} & \multicolumn{2}{?c?}{\textbf{Model}} & \multicolumn{2}{c?}{\textbf{Visualization}} & \multicolumn{4}{c?}{} \\ \Cline{1pt}{3-10}
& \textbf{Variable} & \multicolumn{1}{?c|}{\multirow{2}{*}{IA}} & \multicolumn{1}{c?}{\multirow{2}{*}{Coll.}} & \multirow{2}{*}{$M_{2}$} & \multicolumn{1}{c?}{\multirow{2}{*}{$M_{3}$}} & \multicolumn{2}{c|}{IA} & \multicolumn{2}{c?}{Coll.} \\ \Cline{1pt}{7-10}
& & \multicolumn{1}{?c|}{} & \multicolumn{1}{c?}{} & & \multicolumn{1}{c?}{} & $M_{2}$ & $M_{3}$ & $M_{2}$ & \multicolumn{1}{c?}{$M_{3}$} \\ \Cline{1pt}{1-10}
\multirow{8}{*}{Global Clutter} & \multirow{2}{*}{$SA_{O}$} & \multicolumn{1}{?c|}{$H_{6}$} & \multicolumn{1}{c?}{} & $H_{6}$ & \multicolumn{1}{c?}{} & & & & \multicolumn{1}{c?}{$H_{6}$} \\
& & \multicolumn{1}{?c|}{-All} & \multicolumn{1}{c?}{} & -All & \multicolumn{1}{c?}{} & & & & \multicolumn{1}{c?}{-B} \\ \cline{2-10}
& \multirow{2}{*}{$SA_{1}$} & \multicolumn{1}{?c|}{$H_{6}$} & \multicolumn{1}{c?}{} & $H_{6}$ & \multicolumn{1}{c?}{} & & & $H_{6}$ & \multicolumn{1}{c?}{} \\
& & \multicolumn{1}{?c|}{-All} & \multicolumn{1}{c?}{} & -All & \multicolumn{1}{c?}{} & & & -B & \multicolumn{1}{c?}{} \\ \cline{2-10}
& \multirow{2}{*}{$SA_{2}$} & \multicolumn{1}{?c|}{$H_{6}$} & \multicolumn{1}{c?}{\color{gray7}\bm{$H_{6}$}} & $H_{6}$ & \multicolumn{1}{c?}{} & & & & \multicolumn{1}{c?}{} \\ 
& & \multicolumn{1}{?c|}{-All} & \multicolumn{1}{c?}{\color{gray7}\bm{$-D$}} & -B,W & \multicolumn{1}{c?}{} & & & & \multicolumn{1}{c?}{} \\ \cline{2-10}
& \multirow{2}{*}{$SA_{3}$} & \multicolumn{1}{?c|}{} & \multicolumn{1}{c?}{} & & \multicolumn{1}{c?}{} & & & & \multicolumn{1}{c?}{$H_{6}$} \\ 
& & \multicolumn{1}{?c|}{} & \multicolumn{1}{c?}{} & & \multicolumn{1}{c?}{} & & & & \multicolumn{1}{c?}{-B} \\ \Cline{1pt}{1-10}
\multirow{4}{*}{Euclidean Distance} & \multirow{2}{*}{$SA_{O}$} & \multicolumn{1}{?c|}{$H_{7}$} & \multicolumn{1}{c?}{} & \color{gray7}\bm{$H_{7}$} & \multicolumn{1}{c?}{} & & & & \multicolumn{1}{c?}{$H_{7}$} \\ 
\multirow{4}{*}{Between SA Probe} & & \multicolumn{1}{?c|}{-All} & \multicolumn{1}{c?}{} & \color{gray7}\bm{$-All$} & \multicolumn{1}{c?}{} & & & & \multicolumn{1}{c?}{-W,D} \\ \cline{2-10}
\multirow{4}{*}{Interests and Clicks} & \multirow{2}{*}{$SA_{1}$} & \multicolumn{1}{?c|}{} & \multicolumn{1}{c?}{} & {$H_{7}$} & \multicolumn{1}{c?}{} & \multicolumn{1}{c|}{\color{gray7}\bm{$H_{7}$}} & & & \multicolumn{1}{c?}{} \\ 
& & \multicolumn{1}{?c|}{} & \multicolumn{1}{c?}{} & -B,W & \multicolumn{1}{c?}{} & \multicolumn{1}{c|}{\color{gray7}\bm{$-B$}} & & & \multicolumn{1}{c?}{} \\ \cline{2-10}
& \multirow{2}{*}{$SA_{2}$} & \multicolumn{1}{?c|}{{$H_{7}$}} & \multicolumn{1}{c?}{{$H_{7}$}} & & \multicolumn{1}{c?}{} & & & & \multicolumn{1}{c?}{} \\
& & \multicolumn{1}{?c|}{-All} & \multicolumn{1}{c?}{-B} & & \multicolumn{1}{c?}{} & & & & \multicolumn{1}{c?}{} \\ \Cline{1pt}{1-10}
\multirow{3}{*}{Middle of Action During} & $SA_{O}$ & \multicolumn{1}{?c|}{$H_{7}$} & \multicolumn{1}{c?}{$H_{7}$} & $H_{7}$ & \multicolumn{1}{c?}{\color{gray7}\bm{$H_{7}$}} & & & & \multicolumn{1}{c?}{} \\ \cline{2-10}
\multirow{3}{*}{SA Probe} & $SA_{1}$ & \multicolumn{1}{?c|}{$H_{7}$} & \multicolumn{1}{c?}{$H_{7}$} & $H_{7}$ & \multicolumn{1}{c?}{\color{gray7}\bm{$H_{7}$}} & {$H_{7}$} & & & \multicolumn{1}{c?}{} \\ \cline{2-10}
& $SA_{2}$ & \multicolumn{1}{?c|}{$H_{7}$} & \multicolumn{1}{c?}{$H_{7}$} & $H_{7}$ & \multicolumn{1}{c?}{\color{gray7}\bm{$H_{7}$}} & & \color{gray7}\bm{$H_{7}$} & & \multicolumn{1}{c?}{} \\ \cline{2-10}
& $SA_{3}$ & \multicolumn{1}{?c|}{} & \multicolumn{1}{c?}{} & $H_{7}$ & \multicolumn{1}{c?}{\color{gray7}\bm{$H_{7}$}} & & $H_{7}$ & \color{gray7}\bm{$H_{7}$} & \multicolumn{1}{c?}{} \\ \Cline{1pt}{1-10}
\multirow{3}{*}{Completed Interrupted SA} & \multirow{2}{*}{{$SA_{O}$}} & \multicolumn{1}{?c|}{\color{gray7}\bm{$H_{6},$}} & \multicolumn{1}{c?}{} & & \multicolumn{1}{c?}{} & & & & \multicolumn{1}{c?}{} \\
\multirow{3}{*}{Probe Action} & & \multicolumn{1}{?c|}{\color{gray7}\bm{$H_{7}$}} & \multicolumn{1}{c?}{} & & \multicolumn{1}{c?}{} & & & & \multicolumn{1}{c?}{} \\ \cline{2-10}
& \multirow{2}{*}{$SA_{1}$} & \multicolumn{1}{?c|}{} & \multicolumn{1}{c?}{} & & \multicolumn{1}{c?}{\color{gray7}\bm{$H_{6},$}} & & & & \multicolumn{1}{c?}{} \\ 
& & \multicolumn{1}{?c|}{} & \multicolumn{1}{c?}{} & & \multicolumn{1}{c?}{\color{gray7}\bm{$H_{7}$}} & & & & \multicolumn{1}{c?}{} \\ \Cline{1pt}{1-10}
\multirow{3}{*}{Investigate Commands} & Overall & \multicolumn{1}{?c|}{$H_{7}$} & \multicolumn{1}{c?}{\bm{$H_{7}$}} & \bm{$H_{7}$} & \multicolumn{1}{c?}{$H_{7}$} & \multicolumn{4}{c?}{\multirow{15}{*}{-----------------}} \\ \cline{2-6}
& Easy & \multicolumn{1}{?c|}{$H_{7}$} & \multicolumn{1}{c?}{\bm{$H_{7}$}} & & \multicolumn{1}{c?}{$H_{7}$} & \multicolumn{1}{c}{} & \multicolumn{1}{c}{} & \multicolumn{1}{c}{} & \multicolumn{1}{c?}{} \\ \cline{2-6}
& Hard & \multicolumn{1}{?c|}{$H_{7}$} & \multicolumn{1}{c?}{\bm{$H_{7}$}} & \bm{$H_{7}$} & \multicolumn{1}{c?}{$H_{7}$} & \multicolumn{1}{c}{} & \multicolumn{1}{c}{} & \multicolumn{1}{c}{} & \multicolumn{1}{c?}{} \\  \Cline{1pt}{1-6}
\multirow{2}{*}{Abandon Commands} & Overall & \multicolumn{1}{?c|}{$H_{7}$} & \multicolumn{1}{c?}{$H_{7}$} & & \multicolumn{1}{c?}{} & \multicolumn{1}{c}{} & \multicolumn{1}{c}{} & \multicolumn{1}{c}{} & \multicolumn{1}{c?}{} \\ \cline{2-6}
& Easy & \multicolumn{1}{?c|}{$H_{7}$} & \multicolumn{1}{c?}{$H_{7}$} & & \multicolumn{1}{c?}{} & \multicolumn{1}{c}{} & \multicolumn{1}{c}{} & \multicolumn{1}{c}{} & \multicolumn{1}{c?}{} \\ \Cline{1pt}{1-6}
\multirow{6}{*}{Decide Commands} & \multirow{2}{*}{Overall} & \multicolumn{1}{?c|}{$H_{6},$} & \multicolumn{1}{c?}{$H_{6},$} & $H_{6}$, & \multicolumn{1}{c?}{\color{gray7}\bm{$H_{6},$}} & \multicolumn{1}{c}{} & \multicolumn{1}{c}{} & \multicolumn{1}{c}{} & \multicolumn{1}{c?}{} \\
& & \multicolumn{1}{?c|}{$H_{7}$} & \multicolumn{1}{c?}{$H_{7}$} & $H_{7}$ & \multicolumn{1}{c?}{\color{gray7}\bm{$H_{7}$}} & \multicolumn{1}{c}{} & \multicolumn{1}{c}{} & \multicolumn{1}{c}{} & \multicolumn{1}{c?}{} \\ \cline{2-6}
& \multirow{2}{*}{Easy} & \multicolumn{1}{?c|}{$H_{6},$} & \multicolumn{1}{c?}{$H_{6}$,} & $H_{6}$, & \multicolumn{1}{c?}{\color{gray7}\bm{$H_{6},$}} & \multicolumn{1}{c}{} & \multicolumn{1}{c}{} & \multicolumn{1}{c}{} & \multicolumn{1}{c?}{} \\ 
& & \multicolumn{1}{?c|}{$H_{7}$} & \multicolumn{1}{c?}{$H_{7}$} & $H_{7}$ & \multicolumn{1}{c?}{\color{gray7}\bm{$H_{7}$}} & \multicolumn{1}{c}{} & \multicolumn{1}{c}{} & \multicolumn{1}{c}{} & \multicolumn{1}{c?}{} \\ \cline{2-6}
& \multirow{2}{*}{Hard} & \multicolumn{1}{?c|}{$H_{6}$,} & \multicolumn{1}{c?}{$H_{6}$,} & & \multicolumn{1}{c?}{} & \multicolumn{1}{c}{} & \multicolumn{1}{c}{} & \multicolumn{1}{c}{} & \multicolumn{1}{c?}{} \\
& & \multicolumn{1}{?c|}{$H_{7}$} & \multicolumn{1}{c?}{$H_{7}$} & & \multicolumn{1}{c?}{} & \multicolumn{1}{c}{} & \multicolumn{1}{c}{} & \multicolumn{1}{c}{} & \multicolumn{1}{c?}{} \\ \Cline{1pt}{1-6}
\multirow{2}{*}{Collective Right-Clicks} & Overall & \multicolumn{1}{?c|}{\color{gray7}\bm{$H_{7}$}} & \multicolumn{1}{c}{} & \multicolumn{1}{c}{} & \multicolumn{1}{c}{} & \multicolumn{1}{c}{} & \multicolumn{1}{c}{} & \multicolumn{1}{c}{} & \multicolumn{1}{c?}{} \\ \cline{2-3}
& Hard & \multicolumn{1}{?c|}{\color{gray7}\bm{$H_{7}$}} & \multicolumn{1}{c}{} & \multicolumn{1}{c}{} & \multicolumn{1}{c}{} & \multicolumn{1}{c}{} & \multicolumn{1}{c}{} & \multicolumn{1}{c}{} & \multicolumn{1}{c?}{} \\ \Cline{1pt}{1-6}
{Target Right-Clicks per} & \multirow{2}{*}{Easy} & \multicolumn{1}{?c|}{\multirow{2}{*}{$H_{7}$}} & \multicolumn{1}{c?}{} & & \multicolumn{1}{c?}{} & \multicolumn{1}{c}{} & \multicolumn{1}{c}{} & \multicolumn{1}{c}{} & \multicolumn{1}{c?}{} \\
{Decision} & & \multicolumn{1}{?c|}{} & \multicolumn{1}{c?}{} & & \multicolumn{1}{c?}{} & \multicolumn{1}{c}{} & \multicolumn{1}{c}{} & \multicolumn{1}{c}{} & \multicolumn{1}{c?}{} \\ \Cline{1pt}{1-10}
\multicolumn{10}{c}{} \\
\end{tabular}
\end{table}

\begin{table}[!t]
\centering
\begin{tabular}{?l|cc|cc|cc|c|c|c|c|c|c?}
\Cline{1pt}{1-10}
\multicolumn{1}{?c|}{\multirow{4}{*}{\textbf{Variable}}} & \multicolumn{1}{c}{\multirow{3}{*}{\textbf{Sub-}}} & \multicolumn{2}{?c?}{\textbf{Within}} & \multicolumn{2}{c?}{\textbf{Between}} & \multicolumn{4}{c?}{\multirow{2}{*}{\textbf{Correlation}}} \\
& \multicolumn{1}{c}{\multirow{3}{*}{\textbf{variable}}} & \multicolumn{2}{?c?}{\textbf{Model}} & \multicolumn{2}{c?}{\textbf{Visualization}} & \multicolumn{4}{c?}{} \\ \Cline{1pt}{3-10}
& & \multicolumn{1}{?c|}{\multirow{2}{*}{IA}} & \multicolumn{1}{c?}{\multirow{2}{*}{Coll.}} & \multirow{2}{*}{$M_{2}$} & \multicolumn{1}{c?}{\multirow{2}{*}{$M_{3}$}} & \multicolumn{2}{c|}{IA} & \multicolumn{2}{c?}{Coll.} \\ \Cline{1pt}{7-10}
& & \multicolumn{1}{?c|}{} & \multicolumn{1}{c?}{} & & \multicolumn{1}{c?}{} & $M_{2}$ & $M_{3}$ & $M_{2}$ & \multicolumn{1}{c?}{$M_{3}$} \\ \Cline{1pt}{1-10}
{Highest Value Target} & Overall & \multicolumn{1}{?c|}{} & \multicolumn{1}{c?}{\color{gray7}\bm{$H_{6}$}} & & \multicolumn{1}{c?}{} & \multicolumn{4}{c?}{\multirow{24}{*}{-----------------}} \\ \cline{2-6}
{Abandoned} & {Easy} & \multicolumn{1}{?c|}{{\color{gray7}\bm{$H_{6}$}}} & \multicolumn{1}{c?}{} & & \multicolumn{1}{c?}{{}} & \multicolumn{1}{c}{} & \multicolumn{1}{c}{} & \multicolumn{1}{c}{} & \multicolumn{1}{c?}{} \\ \Cline{1pt}{1-6}
Abandoned Target & Overall & \multicolumn{1}{?c|}{} & \multicolumn{1}{c?}{} & & \multicolumn{1}{c?}{\color{gray7}\bm{$H_{6}$}} & \multicolumn{1}{c}{} & \multicolumn{1}{c}{} & \multicolumn{1}{c}{} & \multicolumn{1}{c?}{} \\ \cline{2-6}
Information Window Open & {Easy} & \multicolumn{1}{?c|}{} & \multicolumn{1}{c?}{} & & \multicolumn{1}{c?}{{\color{gray7}\bm{$H_{6}$}}} & \multicolumn{1}{c}{} & \multicolumn{1}{c}{} & \multicolumn{1}{c}{} & \multicolumn{1}{c?}{} \\ \Cline{1pt}{1-6}
Abandon Requests & {Overall} & \multicolumn{1}{?c|}{{$H_{6}$}} & \multicolumn{1}{c?}{{$H_{6}$}} & & \multicolumn{1}{c?}{} & \multicolumn{1}{c}{} & \multicolumn{1}{c}{} & \multicolumn{1}{c}{} & \multicolumn{1}{c?}{} \\ \cline{2-6}
Exceeded Abandon Targets & {Hard} & \multicolumn{1}{?c|}{{$H_{6}$}} & \multicolumn{1}{c?}{{$H_{6}$}} & & \multicolumn{1}{c?}{} & \multicolumn{1}{c}{} & \multicolumn{1}{c}{} & \multicolumn{1}{c}{} & \multicolumn{1}{c?}{} \\ \Cline{1pt}{1-6}
Time Between Commit & Overall & \multicolumn{1}{?c|}{\color{gray7}\bm{$H_{6}$}} & \multicolumn{1}{c?}{$H_{6}$} & & \multicolumn{1}{c?}{} & \multicolumn{1}{c}{} & \multicolumn{1}{c}{} & \multicolumn{1}{c}{} & \multicolumn{1}{c?}{} \\ \cline{2-6}
State Issued Decide & Easy & \multicolumn{1}{?c|}{\color{gray7}\bm{$H_{6}$}} & \multicolumn{1}{c?}{$H_{6}$} & & \multicolumn{1}{c?}{} & \multicolumn{1}{c}{} & \multicolumn{1}{c}{} & \multicolumn{1}{c}{} & \multicolumn{1}{c?}{} \\ \cline{2-6}
Command & {Hard} & \multicolumn{1}{?c|}{{\color{gray7}\bm{$H_{6}$}}} & \multicolumn{1}{c?}{{$H_{6}$}} &  & \multicolumn{1}{c?}{} & \multicolumn{1}{c}{} & \multicolumn{1}{c}{} & \multicolumn{1}{c}{} & \multicolumn{1}{c?}{} \\ \Cline{1pt}{1-6}
\multirow{2}{*}{Frequency of Accessed} & \multirow{2}{*}{Overall} & \multicolumn{1}{?c|}{} & \multicolumn{1}{c?}{} & $H_{6}$, & \multicolumn{1}{c?}{} & \multicolumn{1}{c}{} & \multicolumn{1}{c}{} & \multicolumn{1}{c}{} & \multicolumn{1}{c?}{} \\ 
\multirow{2}{*}{Target Information} & & \multicolumn{1}{?c|}{} & \multicolumn{1}{c?}{} & $H_{7}$ & \multicolumn{1}{c?}{} & \multicolumn{1}{c}{} & \multicolumn{1}{c}{} & \multicolumn{1}{c}{} & \multicolumn{1}{c?}{} \\ \cline{2-6}
\multirow{2}{*}{Windows} & \multirow{2}{*}{Hard} & \multicolumn{1}{?c|}{\color{gray7}\bm{$H_{6},$}} & \multicolumn{1}{c?}{} & {$H_{6}$,} & \multicolumn{1}{c?}{} & \multicolumn{1}{c}{} & \multicolumn{1}{c}{} & \multicolumn{1}{c}{} & \multicolumn{1}{c?}{} \\ 
& & \multicolumn{1}{?c|}{\color{gray7}\bm{$H_{7}$}} & \multicolumn{1}{c?}{} & {$H_{7}$} & \multicolumn{1}{c?}{} & \multicolumn{1}{c}{} & \multicolumn{1}{c}{} & \multicolumn{1}{c}{} & \multicolumn{1}{c?}{} \\ \Cline{1pt}{1-6}
\multirow{2}{*}{Time Target Information} & Overall & \multicolumn{1}{?c|}{$H_{6}$} & \multicolumn{1}{c?}{$H_{6}$} & & \multicolumn{1}{c?}{} & \multicolumn{1}{c}{} & \multicolumn{1}{c}{} & \multicolumn{1}{c}{} & \multicolumn{1}{c?}{} \\ \cline{2-6}
\multirow{2}{*}{Windows Open} & Easy & \multicolumn{1}{?c|}{$H_{6}$} & \multicolumn{1}{c?}{$H_{6}$} & & \multicolumn{1}{c?}{} & \multicolumn{1}{c}{} & \multicolumn{1}{c}{} & \multicolumn{1}{c}{} & \multicolumn{1}{c?}{} \\ \cline{2-6}
& Hard & \multicolumn{1}{?c|}{$H_{6}$} & \multicolumn{1}{c?}{$H_{6}$} & & \multicolumn{1}{c?}{} & \multicolumn{1}{c}{} & \multicolumn{1}{c}{} & \multicolumn{1}{c}{} & \multicolumn{1}{c?}{} \\ \Cline{1pt}{1-6}
Time Decision Collective & \multirow{2}{*}{Hard} & \multicolumn{1}{?c|}{\multirow{2}{*}{\color{gray7}\bm{$H_{6}$}}} & \multicolumn{1}{c}{} & \multicolumn{1}{c}{} & \multicolumn{1}{c}{} & \multicolumn{1}{c}{} & \multicolumn{1}{c}{} & \multicolumn{1}{c}{} & \multicolumn{1}{c?}{} \\ 
Information Window Open & & \multicolumn{1}{?c|}{} & \multicolumn{1}{c}{} & \multicolumn{1}{c}{} & \multicolumn{1}{c}{} & \multicolumn{1}{c}{} & \multicolumn{1}{c}{} & \multicolumn{1}{c}{} & \multicolumn{1}{c?}{} \\ \Cline{1pt}{1-6}
{Time Decision Target} & Overall & \multicolumn{1}{?c|}{{$H_{6}$}} & \multicolumn{1}{c?}{$H_{6}$} & $H_{6}$ & \multicolumn{1}{c?}{$H_{6}$} & \multicolumn{1}{c}{} & \multicolumn{1}{c}{} & \multicolumn{1}{c}{} & \multicolumn{1}{c?}{} \\ \cline{2-6}
{Information Window} & Easy & \multicolumn{1}{?c|}{$H_{6}$} & \multicolumn{1}{c?}{$H_{6}$} & $H_{6}$ & \multicolumn{1}{c?}{$H_{6}$} & \multicolumn{1}{c}{} & \multicolumn{1}{c}{} & \multicolumn{1}{c}{} & \multicolumn{1}{c?}{} \\ \cline{2-6}
Open & Hard & \multicolumn{1}{?c|}{$H_{6}$} & \multicolumn{1}{c?}{$H_{6}$} & $H_{6}$ & \multicolumn{1}{c?}{$H_{6}$} & \multicolumn{1}{c}{} & \multicolumn{1}{c}{} & \multicolumn{1}{c}{} & \multicolumn{1}{c?}{} \\ \Cline{1pt}{1-6}
\multirow{2}{*}{Post-Trial} & Abandon & \multicolumn{1}{?c|}{} & \multicolumn{1}{c?}{$H_{6}$} & $H_{6}$ & \multicolumn{1}{c?}{} & \multicolumn{1}{c}{} & \multicolumn{1}{c}{} & \multicolumn{1}{c}{} & \multicolumn{1}{c?}{} \\ \cline{2-6}
& Decide & \multicolumn{1}{?c|}{\color{gray7}\bm{$H_{6}$}} & \multicolumn{1}{c?}{$H_{6}$} & & \multicolumn{1}{c?}{} & \multicolumn{1}{c}{} & \multicolumn{1}{c}{} & \multicolumn{1}{c}{} & \multicolumn{1}{c?}{} \\ \Cline{1pt}{1-6}
Post-Experiment & Responsive & \multicolumn{1}{?c|}{$H_{6}$} & \multicolumn{1}{c?}{$H_{6}$} & \multicolumn{1}{c}{} & \multicolumn{1}{c}{} & \multicolumn{1}{c}{} & \multicolumn{1}{c}{} & \multicolumn{1}{c}{} & \multicolumn{1}{c?}{} \\ \Cline{1pt}{1-10}
\end{tabular}
\end{table}

\subsection{Discussion}

The analysis of which model and visualization promoted better \textit{usability} suggests that the IA visualization promoted transparency more \textit{effectively} than the Collective visualization, while both models had their respective advantages and disadvantages. Operators using the $M_{2}$ model had less global clutter, due to target \textit{information} pop-up windows being open for less time, smaller Euclidean distances between the interest of a \textit{SA} probe question and their current interaction, were able to complete interrupted actions after answering a \textit{SA} probe question, and issued fewer abandon and decide commands. $H_{6}$, which hypothesized that the $M_{2}$ model and Collective visualization will promote better \textit{usability} by being more \textit{predictable} and \textit{explainable}, was not supported by the $M_{2}$ model results. Operators from both evaluations using the $M_{2}$ model abandoned the highest value target more frequently, which may have occurred due to mis\textit{understanding} or poor \textit{SA}. IA operators using the $M_{2}$ model were not as \textit{timely} (i.e., faster) at \textit{predicting} when a collective was committed to a target and had the decision collective \textit{information} pop-up window open for a longer duration of time (i.e., lower \textit{explainability}) compared to when using the $M_{3}$ model. The Collective evaluation did not record which collectives were right-clicked on, which impeded the ability to associate right-clicks to a collective per decision; however, a similar reliance of having the decision collective \textit{information} pop-up window visible, like the IA operators, may have occurred considering how the Collective operators used the target \textit{information} pop-up windows. Further evaluations are needed to validate Collective operator \textit{usability} behavior.

The Collective visualization enabled operators to complete actions prior to a SA probe question and aided operators to issue decide commands shortly after a collective was committed to a target. $H_{6}$ was not supported by the Collective visualization findings, since more of the highest value targets were abandoned. The continuous display of collective and target \textit{information} pop-up windows promoted higher \textit{SA performance} for the Collective operators when using both models. The reliance of the supplementary \textit{information} provided in the pop-up windows suggests that the \textit{information} was more \textit{explainable} and \textit{reliable} than the information provided on the collective icons. Incorporating the numerical percentage of support from the respective Collectives on a target icon or identifying the most favored target on a collective hub may help reduce the reliance of the \textit{information} pop-up windows and simultaneously improve \textit{SA} by mitigating potential \textit{observability} issues if the operator must interact with more collectives. 

IA operators using the $M_{3}$ model and Collective operators using the $M_{2}$ model were able to complete actions that were interrupted by a SA probe question 99\% of the time. The \textit{memorability} of both models and visualizations enabled operators to return to a previous task after answering the \textit{SA} probe question, because of the required operator engagement ($M_{3}$ model) and established expectations of collective behaviors ($M_{2}$ model). The \textit{predictability} of the $M_{3}$ model and Collective visualization \textit{justified} issuing decide commands shortly after collectives were in a committed state; however, this finding may be biased for the $M_{3}$ model, because of the required operator influence to achieve the decision-making task. The same bias can attribute to the command \textit{effectiveness} rankings, which were higher for the $M_{3}$ model. The IA operators' ability to identify objects on the visualization may have been impeded by displaying all of the individual collective entities, collective and target icons, and collective and target \textit{information} pop-up windows when the \textit{SA} probe question inquired about an object further away from the center of the operator's current attentional focus. Asking SA probe questions about objects at various distances from the operator's current focal point is necessary in order to \textit{understand} how clutter, or moving individual collective entities, may affect the operator's ability to identify the object of interest and impact \textit{SA performance}. 

$H_{7}$, which hypothesized that operators using the $M_{2}$ model and Collective visualization will require fewer interactions, was not supported. The $M_{2}$ model enabled fewer commands compared to the $M_{3}$ model. The IA visualization enabled fewer abandon and decide commands. Collective operators using the $M_{2}$ model had better decision-making \textit{performance} when more investigate commands were issued. Issuing more investigate commands for high-value targets located further away from the collective hub may suggest that the interaction delay embedded in the $M_{2}$ model, which was designed to reduce the impacts of environmental bias and improve the success of choosing the ground truth best targets, may have not accommodated operators' expectations if lower valued targets were being favored solely because they were closer to the hub. Collective operators who issued more commands may have wanted \textit{control} and \textit{directability} over the decision-making, which may have occurred due to lower trust, or mis\textit{understanding} collective behavior. Investigations are needed to determine if and how trust may influence operators. Operators implemented different strategies to fulfill the decision-making task; however, the most successful strategy promoted more consensus decision-making (i.e., investigate commands), as opposed to prohibiting exploration of targets (i.e., abandon commands). \textit{Understanding} how operators used commands is necessary to promote \textit{effective} interactions and produce desired human-collective \textit{performance}. 

The transparency embedded in the $M_{2}$ model and Collective visualization combination did not support the best overall system \textit{usability}. The IA visualization promoted less clutter, by alleviating the dependence of the collective and target \textit{information} pop-up windows, and promoted fewer interactions. Modifications to both the $M_{2}$ model and Collective visualization must be made in order to mitigate the highest value target being abandoned more frequently, as well as reduce the reliance on the \textit{information} windows. The assumption that fewer interactions are optimal may not be accurate for all decision difficulties, such as hard decisions. \textit{Understanding} strategies and \textit{justifications} for more interactions is necessary in order to promote transparency that aids operators during particular situations and results in higher human-collective \textit{performance}. 

\section{$R_{4}$: System Design Element Influence on Team Performance}

\begin{figure}[h]
\begin{center}
	\includegraphics[width=\textwidth]{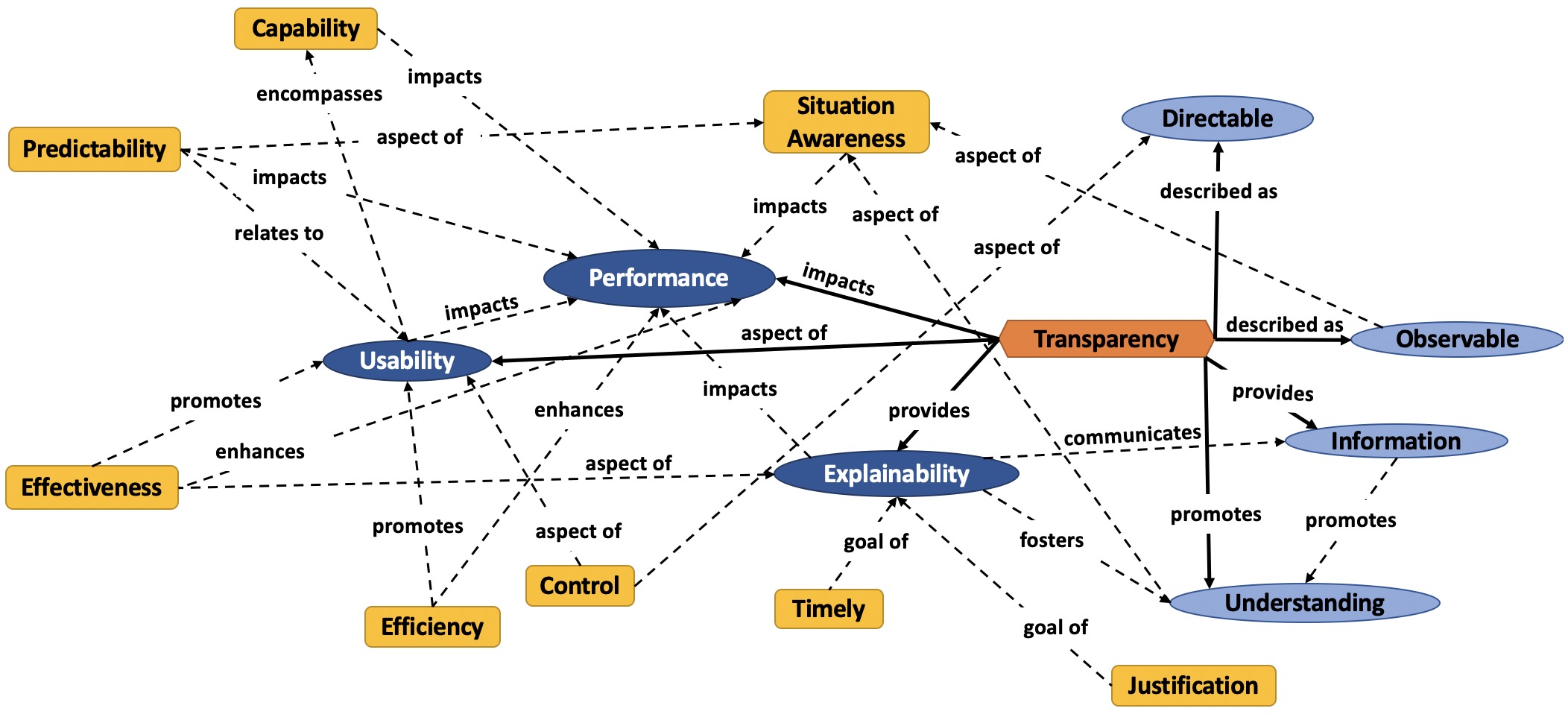}
	\caption{$R_{4}$ concept map of the assessed direct and indirect transparency factors.}
	\label{fig: Model Vis Concept Map R8}
	\end{center}
\end{figure}

Assessing \textit{which model and visualization promoted better human-collective performance}, $R_{4}$, is necessary to determine whether the human-collective system transparency aided task completion. An ideal system performs a task quickly, safely, and successfully. The associated objective dependent variables were (1) decision time, (2) selection success rate, and (3) SA probe accuracy. Additional objective metrics were included to support the correlation analyses. The specific direct and indirect transparency factors related to $R_{4}$ are identified in Figure \ref{fig: Model Vis Concept Map R8}. The relationship between the variables and the corresponding hypotheses, as well as the direct and indirect transparency factors, are identified in Table \ref{table:Performance,Variables}. Additional relationships between the variable and the direct or indirect transparency factors, not identified in Figure \ref{fig: Concept Map}, are provided via correlation analyses.

Performance of the human-collective team can be used to assess the effects of the model and visualization transparency on the team's ability to fulfill tasks. An ideal system design desires high performance rates. It was hypothesized ($H_{8}$) that the human-collective performance, effectiveness, efficiency, and timing will be better using the $M_{2}$ model and Collective visualization.

\begin{table}[h]
\centering
\caption{Interaction of system design elements influence on human-collective performance objective (obj) and subjective (subj) variables (vars), relationship to hypothesis $H_{16}$, as well as the associated direct and indirect transparency factors, are presented in Figure \ref{fig: Concept Map}.}
\label{table:Performance,Variables}
\begin{tabular}{?l?c|c|c|c|c|c|c?c|c|c|c|c|c|c|c?}
\Cline{1pt}{2-16}
\multicolumn{1}{c?}{} & \multicolumn{15}{c?}{\textbf{Transparency Factors}} \\ \Cline{1pt}{2-16}
\multicolumn{1}{c?}{} & \multicolumn{7}{c?}{\textbf{Direct}} & \multicolumn{8}{c?}{\textbf{Indirect}} \\ \Cline{1pt}{2-16}
\multicolumn{1}{c?}{} &
 {\multirow[b]{6}{*}{\rotatebox{90}{\textbf{Directable}}}} & {\multirow[b]{6}{*}{\rotatebox{90}{\textbf{Explainability}}}} &
 {\multirow[b]{6}{*}{\rotatebox{90}{\textbf{Information}}}} &
 {\multirow[b]{6}{*}{\rotatebox{90}{\textbf{Observable}}}}&
 {\multirow[b]{6}{*}{\rotatebox{90}{\textbf{Performance}}}} &  {\multirow[b]{6}{*}{\rotatebox{90}{\textbf{Understanding}}}} &
 {\multirow[b]{6}{*}{\rotatebox{90}{\textbf{Usability}}}} &
 {\multirow[b]{6}{*}{\rotatebox{90}{\textbf{Capability}}}} &
 {\multirow[b]{6}{*}{\rotatebox{90}{\textbf{Control}}}} & {\multirow[b]{6}{*}{\rotatebox{90}{\textbf{Effectiveness}}}} & {\multirow[b]{6}{*}{\rotatebox{90}{\textbf{Efficiency}}}} & {\multirow[b]{6}{*}{\rotatebox{90}{\textbf{Justification}}}} & {\multirow[b]{6}{*}{\rotatebox{90}{\textbf{Predictability}}}} & {\multirow[b]{6}{*}{\rotatebox{90}{\textbf{SA}}}} &  {\multirow[b]{6}{*}{\rotatebox{90}{\textbf{Timing}}}} \\
\multicolumn{1}{c?}{} & & & & & & & & & & & & & & & \\ 
\multicolumn{1}{c?}{} & & & & & & & & & & & & & & & \\
\multicolumn{1}{c?}{} & & & & & & & & & & & & & & & \\
\multicolumn{1}{c?}{} & & & & & & & & & & & & & & & \\ \Cline{1pt}{1-1}
\multicolumn{1}{?c?}{\textbf{Obj Vars}} & & & & & & & & & & & & & & & \\ \Cline{1pt}{1-16}
{Decision Time} & & & & & {\checkmark} & & & & & {\checkmark} & {\checkmark} & & & & {\checkmark} \\ \hline
{Selection Success Rate} & & & & & {\checkmark} & & & & & {\checkmark} & & & & & \\ \hline
{SA Probe Accuracy} & & & & {\checkmark} & {\checkmark} & {\checkmark} & & & & {\checkmark} & & & {\checkmark} & {\checkmark} & \\ \hline
{Collective Observation} & & {\checkmark} & {\checkmark} & & & & {\checkmark} & & & {\checkmark} & & {\checkmark} & & & \\ \hline
{Target Observations} & & {\checkmark} & & & & {\checkmark} & {\checkmark} & & & & & & & & \\ \hline
{Collective Right-Clicks} & & {\checkmark} & {\checkmark} & & & & {\checkmark} & & & {\checkmark} & & {\checkmark} & & & \\ \hline
{Target Right-Clicks} & & {\checkmark} & {\checkmark} & & & & {\checkmark} & & & {\checkmark} & & {\checkmark} & & & \\ \hline
{Investigate Commands} & & & & & {\checkmark} & & {\checkmark} & & {\checkmark} & {\checkmark} & & & & & \\ \hline
{Abandon Commands} & & & & & {\checkmark} & & {\checkmark} & & {\checkmark} & {\checkmark} & & & & & \\ \hline
{Decide Commands} & & & & & {\checkmark} & & {\checkmark} & & {\checkmark} & {\checkmark} & & & {\checkmark} & & \\ \hline
{Time Decision Target} & & & & & & & \multirow{3}{*}{\checkmark} & & & \multirow{3}{*}{\checkmark} & & & & & \multirow{3}{*}{\checkmark} \\ 
{Information Window} & & & & & & & & & & & & & & & \\
{Open} & & & & & & & & & & & & & & & \\ \hline
{Time Decision} & & & & & & & \multirow{3}{*}{\checkmark} & & & \multirow{3}{*}{\checkmark} & & & & & \multirow{3}{*}{\checkmark} \\ 
{Collective Information} & & & & & & & & & & & & & & & \\
{Window Open} & & & & & & & & & & & & & & & \\ \hline
{Mental Rotation} & & & & & \multirow{2}{*}{\checkmark} & & & \multirow{2}{*}{\checkmark} & & & & & & & \\ 
{Assessment} & & & & & & & & & & & & & & & \\ \hline
{Working Memory} & & & & & \multirow{3}{*}{\checkmark} & & & \multirow{3}{*}{\checkmark} & & & & & & & \\ 
{Capacity} & & & & & & & & & & & & & & & \\ \Cline{1pt}{1-16}
\multicolumn{1}{?c}{\textbf{Subj Vars}} & \multicolumn{15}{c?}{} \\ \Cline{1pt}{1-16}
{Weekly Hours on a} & & & & & \multirow{2}{*}{\checkmark} & & & \multirow{2}{*}{\checkmark} & & & & & & & \\ 
{Desktop or Laptop} & & & & & & & & & & & & & & & \\ \hline
{Post-Trial Performance} & & \multirow{2}{*}{\checkmark} & & & \multirow{2}{*}{\checkmark} & \multirow{2}{*}{\checkmark} & & \multirow{2}{*}{\checkmark} & & & & & & & \\ 
{and Understanding} & & & & & & & & & & & & & & & \\ \Cline{1pt}{1-16}
\end{tabular}
\end{table}

\subsection{Metrics and Results}

The length of time it took the human-collective team to reach a decision, \textit{decision time} (minutes), was examined. The decision time mean (SD) per decision are presented in Table \ref{table:Performance,Decision Time} \cite{Roundtree20191, Roundtree2020visual, Cody2020}. Collective operators using the $M_{2}$ model had the fastest decision times. The decision time median, min, max, and the Mann-Whitney-Wilcoxon significant effects between models are shown in Figure \ref{fig: Decision Time}. Significant differences in decision time were found between models for both visualizations at all decision difficulties. Additional between visualizations Mann-Whitney-Wilcoxon tests identified significant effects when using the $M_{2}$ model for overall (n = 672, U = 50921, $\rho$ = 0.03), easy (n = 375, U = 15452, $\rho$ = 0.04), and hard decisions (n = 297, U = 9521, $\rho$ = 0.04). A significant effect between visualizations using the $M_{3}$ model was also found for easy decisions (n = 396, U = 17376, $\rho$ = 0.05). 

\begin{table}[h]
\begin{minipage}{0.5\linewidth}
\centering
\caption{Decision time (minutes) mean (SD) per decision difficulty (Dec Diff).}
\label{table:Performance,Decision Time}
\begin{tabular}{c|c|c|c|}
\cline{2-4}
& \textbf{Dec Diff} & \textbf{IA} & \textbf{Collective} \\ \hline
\multicolumn{1}{|c|}{\multirow{3}{*}{$M_{2}$}} & Overall & 4.32 (1.83) & 3.97 (1.37) \\ \cline{2-4} 
\multicolumn{1}{|c|}{} & \cellcolor{gray1}Easy & \cellcolor{gray1}3.77 (1.63) & \cellcolor{gray1}3.37 (1.23) \\ \cline{2-4} 
\multicolumn{1}{|c|}{} & \cellcolor{gray2}Hard & \cellcolor{gray2}5.09 (1.82) & \cellcolor{gray2}4.67 (1.2) \\ \hline 
\multicolumn{1}{|c|}{\multirow{3}{*}{$M_{3}$}} & Overall & 5.67 (2.86) & 5.32 (2.22) \\ \cline{2-4} 
\multicolumn{1}{|c|}{} & \cellcolor{gray1}Easy & \cellcolor{gray1}5.22 (3.06) & \cellcolor{gray1}4.67 (1.96) \\ \cline{2-4} 
\multicolumn{1}{|c|}{} & \cellcolor{gray2}Hard & \cellcolor{gray2}6.32 (2.42) & \cellcolor{gray2}6.24 (2.24) \\ \hline
\end{tabular}
\end{minipage} 
\hfill
\begin{minipage}{0.45\linewidth}
\centering
	\includegraphics[width=62mm]{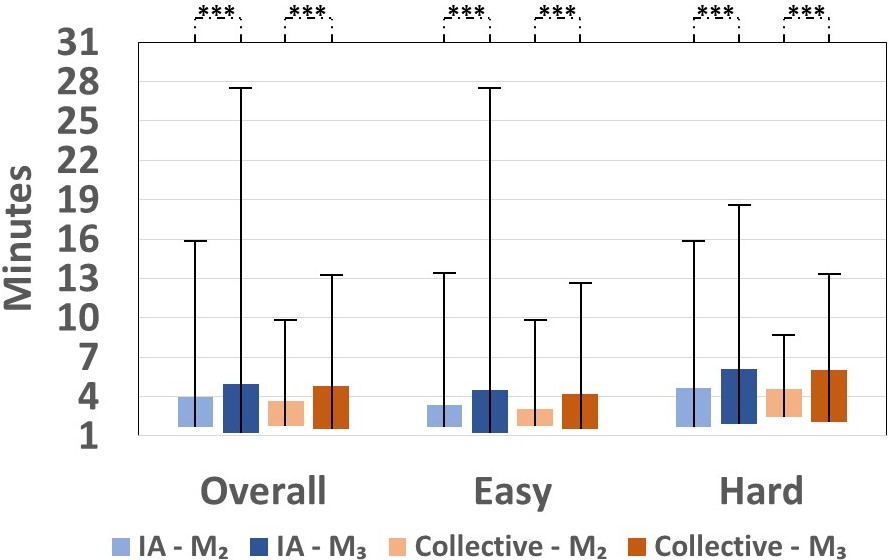}
	\captionof{figure}{Decision time median (min/max) and Mann-Whitney-Wilcoxin test by decision difficulty between models.}
    \label{fig: Decision Time}
\end{minipage}
\end{table}

The \textit{selection success rate} was the number of correct decisions (the collective moved to the highest valued target) relative to the total number of decisions. Selection success rate mean (SD) per decision are shown in Table \ref{table:Performance,Success Rate} \cite{Roundtree20191, Roundtree2020visual, Cody2020}. Operators using the $M_{3}$ model and Collective visualization in general had higher selection success rates, while IA operators using the $M_{2}$ model had higher selection success rates for hard decisions. The selection success rate median, min, max, and the Mann-Whitney-Wilcoxon significant effects between models are shown in Figure \ref{fig: Success Rate}. Collective operators had significant differences in selection success rate between models for overall decisions, while no significant differences between models were found for IA operators at any decision difficulty. Additional between visualizations Mann-Whitney-Wilcoxon tests identified highly significant effects when using the $M_{2}$ model for overall (n = 672, U = 64008, $\rho$ $<$ 0.001) and easy decisions (n = 375, U = 19845, $\rho$ $<$ 0.001), as well as a moderate significant effect for hard decisions (n = 297, U = 12761, $\rho$ $<$ 0.01). Highly significant effects between visualizations using the $M_{3}$ model for overall (U = 66360, $\rho$ $<$ 0.001, easy (n = 396, U = 21662, $\rho$ $<$ 0.001), and hard decisions (n = 276, U = 12178, $\rho$ $<$ 0.01). The Spearman correlation analysis revealed a moderate correlation between decision time and selection success rate using the IA visualization and $M_{2}$ model for easy decisions (r = -0.42, $\rho$ $<$ 0.001) and a weak correlation for overall decisions (r = -0.27, $\rho$ $<$ 0.001). Weak correlations existed when using the Collective visualization and $M_{2}$ model for overall (r = -0.11, $\rho$ = 0.05), easy (r = -0.18, $\rho$ = 0.02), and hard decisions (r = 0.18, $\rho$ = 0.03). A weak correlation was found for hard problems when using the $M_{3}$ model with the IA (r = 0.32, $\rho$ $<$ 0.001) and Collective visualizations (r = 0.25, $\rho$ $<$ 0.01).

\begin{table}[h]
\begin{minipage}{0.5\linewidth}
\centering
\caption{Selection success rate (\%) mean (SD) per decision difficulty (Dec Diff).}
\label{table:Performance,Success Rate}
\begin{tabular}{c|c|c|c|}
\cline{2-4}
& \textbf{Dec Diff} & \textbf{IA} & \textbf{Collective} \\ \hline
\multicolumn{1}{|c|}{\multirow{3}{*}{$M_{2}$}} & Overall & 75 (43.37) & 88.39 (32.08) \\ \cline{2-4} 
\multicolumn{1}{|c|}{} & \cellcolor{gray1}Easy & \cellcolor{gray1}81.44 (38.98) & \cellcolor{gray1}94.44 (22.97) \\ \cline{2-4} 
\multicolumn{1}{|c|}{} & \cellcolor{gray2}Hard & \cellcolor{gray2}66.2 (47.47) & \cellcolor{gray2}81.41 (39.03) \\ \hline
\multicolumn{1}{|c|}{\multirow{3}{*}{$M_{3}$}} & Overall & 75.3 (43.19) & 92.86 (25.79) \\ \cline{2-4} 
\multicolumn{1}{|c|}{} & \cellcolor{gray1}Easy & \cellcolor{gray1}85.43 (35.37) & \cellcolor{gray1}95.94 (19.79) \\ \cline{2-4} 
\multicolumn{1}{|c|}{} & \cellcolor{gray2}Hard & \cellcolor{gray2}60.58 (49.05) & \cellcolor{gray2}88.49 (32.03) \\ \hline
\end{tabular}
\end{minipage} 
\hfill
\begin{minipage}{0.45\linewidth}
\centering
	\includegraphics[width=62mm]{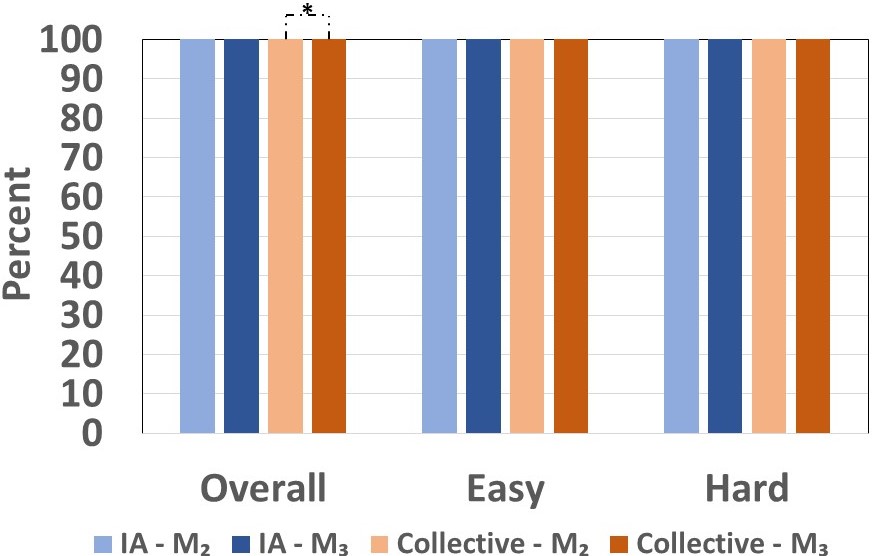}
	\captionof{figure}{Selection success rate median (min/max) and Mann-Whitney-Wilcoxin test by decision difficulty between models.}
    \label{fig: Success Rate}
\end{minipage}
\end{table}

The IA and Collective operators' \textit{SA probe accuracy} when using the $M_{2}$ model was higher for $SA_{3}$, while the IA operators had higher $SA_{2}$ and the Collective operators had higher $SA_{O}$. Collective operators had higher SA probe accuracy, compared to the IA operators. The detailed statistical analyses were provided in Section \ref{section:R1 metrics}.

Additional Spearman correlation analyses analyzed if any correlations existed between selection success rate and some objective metrics, including collective and target observations and right-clicks, investigate, abandon, and decide commands, as well as the time a decision collective and target information pop-up window was open. A weak correlation existed for collective observations using the Collective visualization with the $M_{2}$ model for overall decisions (r = -0.12, $\rho$ = 0.03). Weak correlations were found for target observations when using the Collective visualization with the $M_{3}$ model for overall (r = 0.14, $\rho$ = 0.01) and hard decisions (r = 0.16, $\rho$ = 0.05). Weak correlation were found for the number of target right-clicks using the IA visualization with the $M_{2}$ model for overall decisions (r = -0.13, $\rho$ = 0.02), and with the $M_{3}$ model for overall (r = 0.1, $\rho$ = 0.05) and hard decisions (r = 0.18, $\rho$ = 0.03), as well as when using the Collective visualization with the $M_{2}$ model for hard decisions (r = 0.17, $\rho$ = 0.04). Weak correlations were found for the number of investigate commands when using the Collective visualization with the $M_{2}$ model for hard decisions (r = 0.2, $\rho$ = 0.01), as well as when using the IA visualization with $M_{3}$ model for easy (r = -0.16, $\rho$ = 0.02) and hard decisions (r = 0.24, $\rho$ $<$ 0.01). Weak correlations were found for the number of abandon commands when using the IA visualization with the $M_{2}$ model for easy decisions (r = -0.19, $\rho$ $<$ 0.01), and with the $M_{3}$ model for hard decisions (r = 0.2, $\rho$ = 0.02). A weak correlation existed for the number of decide commands using the Collective visualization with the $M_{3}$ model for overall decisions (r = 0.11, $\rho$ = 0.05). Weak correlations were found for the time a decision target information pop-up window was open when using the Collective visualization with the $M_{2}$ model for overall (r = 0.11, $\rho$ = 0.04) and hard decisions (r = 0.16, $\rho$ = 0.04). No significant effects were found for collective right-clicks and the time a decision collective information pop-up window was open.

Spearman correlation analyses were also conducted to identify correlations between selection success rate and some subjective metrics, including the weekly hours that participants' used a desktop or laptop, the mental rotations assessment, and working memory capacity. Weak correlations were found for the weekly hours participants' used a desktop or laptop when using the IA visualization with the $M_{2}$ model for easy decisions (r = 0.16, $\rho$ = 0.02), and with the $M_{3}$ model for easy decisions (r = -0.15, $\rho$ = 0.04), as well as when using the Collective visualization with the $M_{3}$ model for hard decisions (r = 0.17, $\rho$ = 0.05). A weak correlation was found for the mental rotations assessment using the IA visualization with the $M_{3}$ model for hard decisions (r = 0.18, $\rho$ = 0.04). Weak correlations were found for working memory capacity and easy decisions when using the IA visualization with the $M_{2}$ model (r = -0.17, $\rho$ = 0.02), and with the $M_{3}$ model (r = -0.15, $\rho$ = 0.04).

The \textit{post-trial performance and understanding} questionnaire results assessed the participants' understanding of the collectives' behavior and their ability to chose the best target for each decision. The Collective operators ranked performance and understanding higher when using the $M_{3}$ model. The statistical analysis details were provided in Section \ref{sec: R2 metrics}.

A summary of $R_{4}$'s results that show the hypotheses with associated significant results is provided in Table \ref{table:Performance,Combined}. This summary table is intended to facilitate the discussion.

\begin{table}[h]
\centering
\caption{A synopsis of $R_{4}$'s hypotheses associated with significant results. The SA probe timings are all timings (All), 15 seconds Before asking (B), While being asked (W), and During response (D) to a SA probe question.}
\label{table:Performance,Combined}
\begin{tabular}{?l|cc|cc|cc|c|c|c|c|c|c?}
\Cline{1pt}{1-10}
\multicolumn{1}{?c|}{\multirow{4}{*}{\textbf{Variable}}} & \multicolumn{1}{c}{\multirow{3}{*}{\textbf{Sub-}}} & \multicolumn{2}{?c?}{\textbf{Within}} & \multicolumn{2}{c?}{\textbf{Between}} & \multicolumn{4}{c?}{\multirow{2}{*}{\textbf{Correlation}}} \\
& \multicolumn{1}{c}{\multirow{3}{*}{\textbf{variable}}} & \multicolumn{2}{?c?}{\textbf{Model}} & \multicolumn{2}{c?}{\textbf{Visualization}} & \multicolumn{4}{c?}{} \\ \Cline{1pt}{3-10}
& & \multicolumn{1}{?c|}{\multirow{2}{*}{IA}} & \multicolumn{1}{c?}{\multirow{2}{*}{Coll.}} & \multirow{2}{*}{$M_{2}$} & \multicolumn{1}{c?}{\multirow{2}{*}{$M_{3}$}} & \multicolumn{2}{c|}{IA} & \multicolumn{2}{c?}{Coll.} \\ \Cline{1pt}{7-10}
& & \multicolumn{1}{?c|}{} & \multicolumn{1}{c?}{} & & \multicolumn{1}{c?}{} & $M_{2}$ & $M_{3}$ & $M_{2}$ & \multicolumn{1}{c?}{$M_{3}$} \\ \Cline{1pt}{1-10}
\multirow{3}{*}{Decision Time} & Overall & \multicolumn{1}{?c|}{$H_{8}$} & \multicolumn{1}{c?}{\bm{$H_{8}$}} & \bm{$H_{8}$} & \multicolumn{1}{c?}{} & $H_{8}$ & & \bm{$H_{8}$} & \multicolumn{1}{c?}{} \\ \cline{2-10}
& Easy & \multicolumn{1}{?c|}{$H_{8}$} & \multicolumn{1}{c?}{\bm{$H_{8}$}} & \bm{$H_{8}$} & \multicolumn{1}{c?}{$H_{8}$} & $H_{8}$ & & \bm{$H_{8}$} & \multicolumn{1}{c?}{} \\ \cline{2-10}
& Hard & \multicolumn{1}{?c|}{$H_{8}$} & \multicolumn{1}{c?}{\bm{$H_{8}$}} & \bm{$H_{8}$} & \multicolumn{1}{c?}{} & & \color{gray7}\bm{$H_{8}$} & \bm{$H_{8}$} & \multicolumn{1}{c?}{\color{gray7}\bm{$H_{8}$}} \\ \Cline{1pt}{1-10}
\multirow{3}{*}{Selection Success Rate} & Overall & \multicolumn{1}{?c|}{} & \multicolumn{1}{c?}{$H_{8}$} & $H_{8}$ & \multicolumn{1}{c?}{$H_{8}$} & \multicolumn{4}{c?}{\multirow{3}{*}{-----------------}} \\ \cline{2-6}
& Easy & \multicolumn{1}{?c|}{} & \multicolumn{1}{c?}{} & $H_{8}$ & \multicolumn{1}{c?}{$H_{8}$} & \multicolumn{1}{c}{} & \multicolumn{1}{c}{} & \multicolumn{1}{c}{} & \multicolumn{1}{c?}{} \\ \cline{2-6}
& Hard & \multicolumn{1}{?c|}{} & \multicolumn{1}{c?}{} & $H_{8}$ & \multicolumn{1}{c?}{$H_{8}$} & \multicolumn{1}{c}{} & \multicolumn{1}{c}{} & \multicolumn{1}{c}{} & \multicolumn{1}{c?}{} \\ \Cline{1pt}{1-10}
\end{tabular}
\end{table}

\begin{table}[!t]
\centering
\begin{tabular}{?l|cc|cc|cc|c|c|c|c|c|c?}
\Cline{1pt}{1-10}
\multicolumn{1}{?c|}{\multirow{4}{*}{\textbf{Variable}}} & \multicolumn{1}{c}{\multirow{3}{*}{\textbf{Sub-}}} & \multicolumn{2}{?c?}{\textbf{Within}} & \multicolumn{2}{c?}{\textbf{Between}} & \multicolumn{4}{c?}{\multirow{2}{*}{\textbf{Correlation}}} \\
& \multicolumn{1}{c}{\multirow{3}{*}{\textbf{variable}}} & \multicolumn{2}{?c?}{\textbf{Model}} & \multicolumn{2}{c?}{\textbf{Visualization}} & \multicolumn{4}{c?}{} \\ \Cline{1pt}{3-10}
& & \multicolumn{1}{?c|}{\multirow{2}{*}{IA}} & \multicolumn{1}{c?}{\multirow{2}{*}{Coll.}} & \multirow{2}{*}{$M_{2}$} & \multicolumn{1}{c?}{\multirow{2}{*}{$M_{3}$}} & \multicolumn{2}{c|}{IA} & \multicolumn{2}{c?}{Coll.} \\ \Cline{1pt}{7-10}
& & \multicolumn{1}{?c|}{} & \multicolumn{1}{c?}{} & & \multicolumn{1}{c?}{} & $M_{2}$ & $M_{3}$ & $M_{2}$ & \multicolumn{1}{c?}{$M_{3}$} \\ \Cline{1pt}{1-10}
\multirow{4}{*}{SA Probe Accuracy} & $SA_{O}$ & \multicolumn{1}{?c|}{} & \multicolumn{1}{c?}{} & \bm{$H_{8}$} & \multicolumn{1}{c?}{$H_{8}$} & \multicolumn{4}{c?}{\multirow{4}{*}{-----------------}} \\ \cline{2-6}
& $SA_{1}$ & \multicolumn{1}{?c|}{$H_{8}$} & \multicolumn{1}{c?}{} & \bm{$H_{8}$} & \multicolumn{1}{c?}{$H_{8}$} & \multicolumn{1}{c}{} & \multicolumn{1}{c}{} & \multicolumn{1}{c}{} & \multicolumn{1}{c?}{} \\ \cline{2-6}
& $SA_{2}$ & \multicolumn{1}{?c|}{} & \multicolumn{1}{c?}{} & \bm{$H_{8}$} & \multicolumn{1}{c?}{$H_{8}$} & \multicolumn{1}{c}{} & \multicolumn{1}{c}{} & \multicolumn{1}{c}{} & \multicolumn{1}{c?}{} \\ \cline{2-6}
& $SA_{3}$ & \multicolumn{1}{?c|}{} & \multicolumn{1}{c?}{} & \bm{$H_{8}$} & \multicolumn{1}{c?}{$H_{8}$} & \multicolumn{1}{c}{} & \multicolumn{1}{c}{} & \multicolumn{1}{c}{} & \multicolumn{1}{c?}{} \\ \Cline{1pt}{1-10}
\multirow{3}{*}{Collective Observations} & Overall & \multicolumn{1}{?c|}{\color{gray7}\bm{$H_{8}$}} & \multicolumn{1}{c?}{} & \color{gray7}\bm{$H_{8}$} & \multicolumn{1}{c?}{\color{gray7}\bm{$H_{8}$}} & & & \bm{$H_{8}$} &  \multicolumn{1}{c?}{} \\ \cline{2-10}
& Easy & \multicolumn{1}{?c|}{\color{gray7}\bm{$H_{8}$}} & \multicolumn{1}{c?}{$H_{8}$} & \color{gray7}\bm{$H_{8}$} & \multicolumn{1}{c?}{\color{gray7}\bm{$H_{8}$}} & & & &  \multicolumn{1}{c?}{} \\ \cline{2-10}
& Hard & \multicolumn{1}{?c|}{\color{gray7}\bm{$H_{8}$}} & \multicolumn{1}{c?}{} & & \multicolumn{1}{c?}{\color{gray7}\bm{$H_{8}$}} & & & &  \multicolumn{1}{c?}{} \\ \Cline{1pt}{1-10}
\multirow{3}{*}{Target Observations} & Overall & \multicolumn{1}{?c|}{$H_{8}$} & \multicolumn{1}{c?}{\bm{$H_{8}$}} & \bm{$H_{8}$} & \multicolumn{1}{c?}{$H_{8}$} & & & &  \multicolumn{1}{c?}{$H_{8}$} \\ \cline{2-10}
& Easy & \multicolumn{1}{?c|}{} & \multicolumn{1}{c?}{\bm{$H_{8}$}} & \bm{$H_{8}$} & \multicolumn{1}{c?}{$H_{8}$} & & & & \multicolumn{1}{c?}{} \\ \cline{2-10}
& Hard & \multicolumn{1}{?c|}{} & \multicolumn{1}{c?}{\bm{$H_{8}$}} & \bm{$H_{8}$} & \multicolumn{1}{c?}{$H_{8}$} & & & & \multicolumn{1}{c?}{$H_{8}$} \\ \Cline{1pt}{1-10}
\multirow{2}{*}{Collective Right-Clicks} & Overall & \multicolumn{1}{?c|}{\color{gray7}\bm{$H_{8}$}} & \multicolumn{3}{c?}{\multirow{2}{*}{-----------------}} & & & \multicolumn{2}{c?}{\multirow{2}{*}{--------}} \\ \cline{2-3} \cline{7-8}
& Hard & \multicolumn{1}{?c|}{\color{gray7}\bm{$H_{8}$}} & \multicolumn{1}{c}{} & \multicolumn{1}{c}{} & \multicolumn{1}{c?}{} & & & \multicolumn{1}{c}{} & \multicolumn{1}{c?}{} \\ \Cline{1pt}{1-10}
\multirow{2}{*}{Target Right-Clicks per} & Overall & \multicolumn{1}{?c|}{{}} & \multicolumn{1}{c?}{} & & \multicolumn{1}{c?}{} & $H_{8}$ & \color{gray7}\bm{$H_{8}$} & & \multicolumn{1}{c?}{} \\ \cline{2-10}
\multirow{2}{*}{Decision} & Easy & \multicolumn{1}{?c|}{$H_{8}$} & \multicolumn{1}{c?}{} & & \multicolumn{1}{c?}{} & & & & \multicolumn{1}{c?}{} \\ \cline{2-10}
& Hard & \multicolumn{1}{?c|}{} & \multicolumn{1}{c?}{} & & \multicolumn{1}{c?}{} & & \color{gray7}\bm{$H_{8}$} & \bm{$H_{8}$} & \multicolumn{1}{c?}{} \\ \Cline{1pt}{1-10}
\multirow{3}{*}{Investigate Commands} & Overall & \multicolumn{1}{?c|}{$H_{8}$} & \multicolumn{1}{c?}{\bm{$H_{8}$}} & \bm{$H_{8}$} & \multicolumn{1}{c?}{$H_{8}$} & & & & \multicolumn{1}{c?}{} \\ \cline{2-10}
& Easy & \multicolumn{1}{?c|}{$H_{8}$} & \multicolumn{1}{c?}{\bm{$H_{8}$}} & & \multicolumn{1}{c?}{$H_{8}$} & & \color{gray7}\bm{$H_{8}$} & & \multicolumn{1}{c?}{} \\ \cline{2-10}
& Hard & \multicolumn{1}{?c|}{$H_{8}$} & \multicolumn{1}{c?}{\bm{$H_{8}$}} & \bm{$H_{8}$} & \multicolumn{1}{c?}{$H_{8}$} & & \color{gray7}\bm{$H_{8}$} & \bm{$H_{8}$} & \multicolumn{1}{c?}{} \\ \Cline{1pt}{1-10}
\multirow{3}{*}{Abandon Commands} & Overall & \multicolumn{1}{?c|}{$H_{8}$} & \multicolumn{1}{c?}{$H_{8}$} & & \multicolumn{1}{c?}{} & & & & \multicolumn{1}{c?}{} \\ \cline{2-10}
& Easy & \multicolumn{1}{?c|}{$H_{8}$} & \multicolumn{1}{c?}{$H_{8}$} & & \multicolumn{1}{c?}{} & $H_{8}$ & & & \multicolumn{1}{c?}{} \\ \cline{2-10}
& Hard & \multicolumn{1}{?c|}{} & \multicolumn{1}{c?}{} & & \multicolumn{1}{c?}{} & & \color{gray7}\bm{$H_{8}$} & & \multicolumn{1}{c?}{} \\ \Cline{1pt}{1-10}
\multirow{3}{*}{Decide Commands} & Overall & \multicolumn{1}{?c|}{$H_{8}$} & \multicolumn{1}{c?}{$H_{8}$} & $H_{8}$ & \multicolumn{1}{c?}{\color{gray7}\bm{$H_{8}$}} & & & & \multicolumn{1}{c?}{$H_{8}$} \\ \cline{2-10}
& Easy & \multicolumn{1}{?c|}{$H_{8}$} & \multicolumn{1}{c?}{$H_{8}$} & $H_{8}$ & \multicolumn{1}{c?}{\color{gray7}\bm{$H_{8}$}} & & & & \multicolumn{1}{c?}{} \\ \cline{2-10}
& Hard & \multicolumn{1}{?c|}{$H_{8}$} & \multicolumn{1}{c?}{$H_{8}$} & & \multicolumn{1}{c?}{} & & & & \multicolumn{1}{c?}{} \\ \Cline{1pt}{1-10}
Time Decision Collective & \multirow{2}{*}{Hard} & \multicolumn{1}{?c|}{\multirow{2}{*}{\color{gray7}\bm{$H_{8}$}}} & \multicolumn{3}{c?}{\multirow{2}{*}{-----------------}} & & & \multicolumn{2}{c?}{\multirow{2}{*}{--------}} \\ 
Information Window Open & & \multicolumn{1}{?c|}{} & \multicolumn{1}{c}{} & \multicolumn{1}{c}{} & \multicolumn{1}{c?}{} & & & \multicolumn{1}{c}{} & \multicolumn{1}{c?}{} \\ \Cline{1pt}{1-10}
\multirow{2}{*}{Time Decision Target} & Overall & \multicolumn{1}{?c|}{{$H_{8}$}} & \multicolumn{1}{c?}{$H_{8}$} & $H_{8}$ & \multicolumn{1}{c?}{$H_{8}$} & & & \bm{$H_{8}$} & \multicolumn{1}{c?}{} \\ \cline{2-10}
\multirow{2}{*}{Information Window Open} & Easy & \multicolumn{1}{?c|}{$H_{8}$} & \multicolumn{1}{c?}{$H_{8}$} & $H_{8}$ & \multicolumn{1}{c?}{$H_{8}$} & & & & \multicolumn{1}{c?}{} \\ \cline{2-10}
& Hard & \multicolumn{1}{?c|}{$H_{8}$} & \multicolumn{1}{c?}{$H_{8}$} & $H_{8}$ & \multicolumn{1}{c?}{$H_{8}$} & & & \bm{$H_{8}$} & \multicolumn{1}{c?}{} \\ \Cline{1pt}{1-10}
\multirow{4}{*}{Mental Rotations Assessment} & $SA_{O}$ & \multicolumn{4}{?c?}{\multirow{13}{*}{-----------------}} & \color{gray7}\bm{$H_{8}$} & \color{gray7}\bm{$H_{8}$} & & \multicolumn{1}{c?}{} \\ \cline{2-2} \cline{7-10}
& $SA_{1}$ & \multicolumn{1}{?c}{} & \multicolumn{1}{c}{} & \multicolumn{1}{c}{} & \multicolumn{1}{c?}{} & \color{gray7}\bm{$H_{8}$} & \color{gray7}\bm{$H_{8}$} & & \multicolumn{1}{c?}{} \\ \cline{2-2} \cline{7-10}
& $SA_{2}$ & \multicolumn{1}{?c}{} & \multicolumn{1}{c}{} & \multicolumn{1}{c}{} & \multicolumn{1}{c?}{} & & \color{gray7}\bm{$H_{8}$} & & \multicolumn{1}{c?}{} \\ \cline{2-2} \cline{7-10}
& Hard & \multicolumn{1}{?c}{} & \multicolumn{1}{c}{} & \multicolumn{1}{c}{} & \multicolumn{1}{c?}{} & & \color{gray7}\bm{$H_{8}$} & & \multicolumn{1}{c?}{} \\ \Cline{1pt}{1-2} \Cline{1pt}{7-10}
\multirow{4}{*}{Working Memory Capacity} & $SA_{O}$ & \multicolumn{1}{?c}{} & \multicolumn{1}{c}{} & \multicolumn{1}{c}{} & \multicolumn{1}{c?}{} & \color{gray7}\bm{$H_{8}$} & \color{gray7}\bm{$H_{8}$} & & \multicolumn{1}{c?}{} \\ \cline{2-2} \cline{7-10}
& $SA_{1}$ & \multicolumn{1}{?c}{} & \multicolumn{1}{c}{} & \multicolumn{1}{c}{} & \multicolumn{1}{c?}{} & \color{gray7}\bm{$H_{8}$} & & & \multicolumn{1}{c?}{} \\ \cline{2-2} \cline{7-10}
& $SA_{3}$ & \multicolumn{1}{?c}{} & \multicolumn{1}{c}{} & \multicolumn{1}{c}{} & \multicolumn{1}{c?}{} & \color{gray7}\bm{$H_{8}$} & & & \multicolumn{1}{c?}{} \\ \cline{2-2} \cline{7-10}
& Easy & \multicolumn{1}{?c}{} & \multicolumn{1}{c}{} & \multicolumn{1}{c}{} & \multicolumn{1}{c?}{} & \color{gray7}\bm{$H_{8}$} & \color{gray7}\bm{$H_{8}$} & & \multicolumn{1}{c?}{} \\ \Cline{1pt}{1-2} \Cline{1pt}{7-10}
& $SA_{O}$ & \multicolumn{1}{?c}{} & \multicolumn{1}{c}{} & \multicolumn{1}{c}{} & \multicolumn{1}{c?}{} & \color{gray7}\bm{$H_{8}$} & & & \multicolumn{1}{c?}{} \\ \cline{2-2} \cline{7-10}
\multirow{2}{*}{Weekly Hours on a Desktop} & $SA_{1}$ & \multicolumn{1}{?c}{} & \multicolumn{1}{c}{} & \multicolumn{1}{c}{} & \multicolumn{1}{c?}{} & \color{gray7}\bm{$H_{8}$} & & & \multicolumn{1}{c?}{} \\ \cline{2-2} \cline{7-10}
\multirow{2}{*}{or Laptop} & $SA_{2}$ & \multicolumn{1}{?c}{} & \multicolumn{1}{c}{} & \multicolumn{1}{c}{} & \multicolumn{1}{c?}{} & & & \color{gray7}\bm{$H_{8}$} & \multicolumn{1}{c?}{} \\ \cline{2-2} \cline{7-10}
& Easy & \multicolumn{1}{?c}{} & \multicolumn{1}{c}{} & \multicolumn{1}{c}{} & \multicolumn{1}{c?}{} & \color{gray7}\bm{$H_{8}$} & \color{gray7}\bm{$H_{8}$} & & \multicolumn{1}{c?}{} \\ \cline{2-2} \cline{7-10}
& Hard & \multicolumn{1}{?c}{} & \multicolumn{1}{c}{} & \multicolumn{1}{c}{} & \multicolumn{1}{c?}{} & & & & \multicolumn{1}{c?}{\color{gray7}\bm{$H_{8}$}} \\ \Cline{1pt}{1-10}
\end{tabular}
\end{table}

\subsection{Discussion}

The analysis suggests that the Collective visualization promoted better human-collective \textit{performance}; however, the models had their respective advantages and disadvantages. The $M_{2}$ model promoted faster decision \textit{times}, while the Collective visualization promoted faster decision \textit{times}, higher selection success rates, and higher subjective \textit{performance}. \textit{SA performance} varied across the models and visualizations. $H_{8}$, which hypothesized that the human-collective \textit{performance}, \textit{effectiveness}, \textit{efficiency}, and \textit{timing} will be better using the $M_{2}$ model with the Collective visualization, was partially supported. Collective operators using the $M_{2}$ model had faster decision \textit{times}; however, the $M_{3}$ model enabled higher selection success rates. Embedding transparency into the $M_{2}$ model requires (1) balancing \textit{control} between the operator and the collectives so that the operators can positively contribute and \textit{direct} decision-making, (2) promoting positive human-collective interactions so that the operator's and the collective's strengths are maximized, and (3) alleviating the operator's workload. 

Understanding \textit{usability} and what interactions were used by operators to \textit{justify} actions that contributed to performance are necessary in order to identify the most \textit{effective} and \textit{efficient} strategies. Operators using the $M_{2}$ model issued fewer commands to the collectives, which was desired in order to maximize the usage of the collectives' consensus decision-making process; however, more particular interactions, such as investigate commands, resulted in higher selection success rate performance. Requiring operators to influence the task ensured better \textit{performance}, because the operators were in-the-loop, versus operators who were supervising the collective behaviors and potentially correcting actions towards task success. Further analysis is required to determine how to improve target selection when using the $M_{2}$ model. Improvements during training may help emphasize the necessity of selecting the highest-value targets. 

Realistic human-collective scenarios will require high \textit{performance} with short decision \textit{times}, especially in uncertain and dynamic environments. The design of an \textit{effective} human-collective system must enable the human-collective team to fulfill primary objectives, without hindering other metrics, such as decision \textit{time} and accuracy. Devoting more \textit{time} to ensure high task \textit{performance} is a common trade-off \cite{Golman2015}. Expedited decisions may have occurred if higher valued targets were more \textit{observable} further away from other objects (less clutter), making them more salient, or if impatient operators \textit{predicted} future collective behaviors and influenced collectives more to make faster decisions. Using target outlines, collective and target \textit{information} pop-up windows, and issuing investigate commands were necessary to fulfill the primary task and can be used to ensure an \textit{explainable} and \textit{usable} system. The $M_{2}$ model with the Collective visualization enabled operators with different spatial \textit{capabilities} to perform relatively the same, unlike IA operators, specifically those with lower Working Memory Capacity and more weekly desktop or laptop exposure, who had higher selection success rates.

The transparency embedded in the Collective visualization with the $M_{2}$ model promoted the fastest decision times; however, modifications are needed in order to improve the other human-collective \textit{performance} metrics. \textit{Understanding} what interactions contributed to higher \textit{performance} is necessary to determine what operator strategies are most \textit{effective} and \textit{efficient}. The $M_{2}$ model subjective \textit{performance} rankings may have had a consistent negative bias due to learning effects, since this model was always presented before using the $M_{3}$ model. Improving the transparency embedded in the Collective visualization to promote better \textit{SA performance} must be considered. \textit{Understanding} what IA visualization aspects, such as streamlines between collectives and targets, promoted better \textit{SA performance} can be emulated in the more abstract Collective visualization.

\section{Discussion}

The first research objective was to expand on the existing transparency literature by assessing how different models and visualizations influenced human-collective behavior. The analysis assessed \textit{understanding} how the transparency embedded in the models and visualizations influenced operators with individual differences (i.e., \textit{capabilities}), operator comprehension (i.e, \textit{capability} of \textit{understanding}), system design element \textit{usability} (i.e.,  model and visualization usability), and human-collective \textit{performance}. The second research objective was to determine whether using the best model and visualization, derived from two previous results, provided the best transparency. Previous results indicated that the $M_{2}$ model enabled faster decisions and relied less on operator influence \citep{Cody2020}, while the Collective visualization provided better transparency \citep{Roundtree2020visual}, because operators with different individual capabilities \textit{performed} similarly for both tasks, and the human-collective team \textit{performed} better. The $M_{2}$ model, independently, did not enable operators with individual differences to perform similarly; however, it did promote fewer interactions and less clutter, which enabled operators to complete interrupted actions, promoted faster decision times, and higher \textit{SA performance}. The Collective visualization independently enabled operators with different individual differences to perform similarly, promoted higher \textit{understanding} and \textit{SA}, enabled operators to complete interrupted actions and issue decide commands shortly after a collective was committed to a target, promoted faster decision times, higher selection success rates, and higher subjective \textit{performance}. Together the $M_{2}$ model and Collective visualization promoted lower overall \textit{workload}, required less physical demand, had fewer investigate commands and target observations (i.e., extra clicks), and enabled the fastest decision \textit{time}. The different outcomes between the findings in this evaluation versus the findings from Cody \textit{et al.} \cite{Cody2020} and Roundtree \textit{et al.} \cite{Roundtree2020visual} suggest that transparency cannot be quantified by using the best system design elements, but instead must be quantified by considering how the transparency of the different system design elements interact with one another along with the implications of how that system transparency influences human-collective interactions and \textit{performance}.

Fewer operator interactions was a desired behavior in order to minimize negative influence on collective behaviors and reduce the reliance on supplementary \textit{information}; however, operator influence was anticipated to aid the decision-making process and \textit{time} to complete decisions. This analysis identified positive and negative interactions associated with both models and visualizations. Collective operators relied on visible target \textit{information} pop-up windows more than 25\% of the decision time, resulting in more global clutter. Clutter, from a system design perspective, can hinder \textit{effective} task performance. Collective operators with more clutter were able to answer more \textit{SA} probe questions correctly and had higher selection success rates. The dependence on visible collective and target \textit{information} pop-up windows may have been influenced by the type of \textit{SA} probe questions asked and the visualization not being observable without the supplementary \textit{information}. Sixteen of twenty-four \textit{SA} probe questions depended on numerical values of collective state and target support \textit{information} provided in the collective and target \textit{information} pop-up windows. Collective state \textit{information} was provided via the different color individual collective entities on the IA visualization and the opacity of the Collective visualization's hub quadrants, while color and opacity were used to indicate the highest supporting collective on the target icon. The use of opacity may have been in\textit{effective} and less salient; however, using different colors to indicate state \textit{information} may be a possible design modification to the Collective visualization. Experimental design modifications can also be implemented in order to ensure a more even distribution of \textit{SA} probe questions that rely on other \textit{information}, such as the icons, system messages, or collective assignments versus \textit{information} pop-up windows. 

The use of target \textit{information} pop-up windows aided Collective operators to abandon targets more than 25\% of the \textit{time}. Operators who used the target \textit{information} pop-up windows to \textit{justify} that a target was abandoned by a collective, may have been confused if the reported target support was not equal to zero. Additional abandon commands may have been issued in an attempt to reduce the collective support to zero. IA operators may have experienced a similar confusion if they observed individual collective entities still travelling to an abandoned target. Implementing design changes, such as showing zero support when an abandon request has been committed, or not displaying lost entities after a specific period of \textit{time}, once a collective hub has moved to a new location, may reduce the number of reissued abandon commands. Collective operators using the $M_{2}$ model abandoned the highest value target more frequently than IA operators. Further analysis is required to determine if the entire target icon must represent the target value, as was the case with the IA visualization, to be more salient. Opacity levels must also be validated to ensure an unique distinction between low-, medium-, and high-valued targets. Reiterating the task objective, to choose and move each collective to the highest value target for each decision, numerous times during training may also help mitigate operator mis\textit{understanding}.

Target observations, which were additional target left-clicks that did not influence collective behavior or aid in accessing supplemental \textit{information}, and interventions were additional undesired interactions. IA operators may have confused the target integer identifiers with the collective roman numeral identifiers causing additional target observations. Using distinct identifiers, such as integers versus letters, can potentially reduce the number of observations. IA operators' \textit{capability} to identify objects far from their current attentional focal point may have been impeded by displaying all of the individual collective entities, collective and target icons, as well as the collective and target \textit{information} pop-up windows. Asking \textit{SA} probe questions about objects at various distances from the operator's current focal point is necessary to \textit{understand} how clutter, or moving individual collective entities, may affect the operator's ability to identify the \textit{SA} probe object of interest and answer the question correctly. The use of eye-tracking technology can provide improved insight regarding operator \textit{understanding} and \textit{usability} by recording where the operator was looking. \textit{Understanding} what types of \textit{information} the operator was potentially perceiving and comprehending, the difficulty of identifying the desired \textit{information} due to clutter, and the duration of \textit{time} looking for \textit{information} will illuminate why operators interacted with the system in a particular way.

The $M_{2}$ model enabled fewer commands, which was expected. Requiring operators to influence the decision-making process ensured better \textit{performance}, because the operator was in-the-loop, versus operators who were supervising the collective behaviors and potentially correcting actions towards task success. Different strategies were used to fulfill the decision-making task; however, the most successful promoted more consensus decision-making (i.e., investigate commands) versus prohibiting exploration of particular targets (i.e., abandon commands). The \textit{memorability} of both the models and visualizations enabled operators to come back into-the-loop after answering the SA probe question, because of the required involvement of the operator ($M_{3}$ model) and established expectations of collective behaviors ($M_{2}$ model). The \textit{predictability} of the $M_{3}$ model and Collective visualization enabled operators to issue decide commands shortly after collectives were in a committed state. Collective operators using the $M_{3}$ model reported the best \textit{control} mechanism responsiveness, which was anticipated due to the amount of operator influence and gained experience using the \textit{control} mechanisms in the prior trial that used the $M_{2}$ model.  

Additional design guidance recommendations, provided in Table \ref{table:Discussion,Design}, were created to expand on those from Roundtree \textit{et al.} \cite{Roundtree2020visual}. These new recommendations are applicable irrespective of a specific model or visualization type with a focus on \textit{control} mechanism and model features. Providing \textit{control} mechanisms that can influence the collective decision-making process positively are ideal for ensuring task completion. Further investigations are required to determine how to improve the \textit{efficacy} of \textit{control} mechanisms, such as abandon, that can negatively influence task completion. Providing \textit{control} mechanisms to undo undesired abandon commands is recommended. Additional analyses and investigations are needed to verify the \textit{effectiveness} of the proposed design strategies for real-world scenarios where bandwidth limitations occur. \textit{Understanding} how \textit{information} latency and inaccurate collective state \textit{information} influence human-collective behavior negatively is essential to designing a resilient transparent collective system.

\begin{table}[h]
\centering
\caption{Additional human-collective system design guidance.}
\label{table:Discussion,Design}
\begin{tabular}{|l|}
\hline
\multicolumn{1}{|c|}{\textbf{Design Guidance}} \\ \hline
1. Provide indicators that identify which particular objects are currently selected, such as \\
the Collective and Target fields in the Collective Request area. \\ \hline
2. Provide control mechanisms that influence the collective consensus decision-making \\
process positively, such as the investigate command. \\ \hline
3. Provide control mechanisms that can undo negative influence, such as cancel assignment. \\ \hline
4. Limit the use of decision making control mechanism only after a particular certainty \\
 value, such as 30\% support for a specific target. \\ \hline
5. Limit the amount of times operators can issue particular commands, such as one time for \\
the abandon or decide command. \\ \hline
6. Use underlying intelligent algorithms (e.g., sequential best-of-n decision making) capable \\
of fulfilling the task without operator influence. \\ \hline 
7. Ensure that the underlying intelligent algorithms compensate for environmental biases and \\
other influential factors on the collective processes. \\ \hline
\end{tabular}
\end{table}

Transparency for human-collective systems can be achieved via different design strategies for specific system design elements and must be assessed holistically by \textit{understanding} how the different factors impact transparency and are influenced by transparency. The four research questions assessed four categories of transparency factors that contribute to an \textit{effective} system: (1) different operator individual \textit{capabilities}, (2) operator comprehension, (3) system \textit{usability}, and (4) human-collective team \textit{performance}. Ideal collective systems will enable operators with different individual \textit{capabilities} to perform relatively the same, promote operator comprehension, be \textit{usable}, and promote high human-collective \textit{performance}. As collective systems grow in complexity (e.g., size, heterogeneity), visualizations that show the individual collective entities will cause perceptual and comprehension challenges, as well as influence operator actions negatively. The same advantageous observation (i.e., dynamically seeing collective behaviors and support) from this analysis may not occur with large collectives ($>$ 10000). A collective system designed using the provided guidelines can help promote better transparency and enable effective human-collective teams. 

\section{Conclusion}

Designers of human-collective systems continue to debate what models, control mechanisms, and visualizations are needed to provide transparency of collective behaviors to operators. This manuscript evaluates two models, one consensus decision-making model and another that required operator influence in order to achieve the task, and two visualizations, a traditional and abstract collective representation, for a sequential best-of-\textit{n} decision-making task with four collectives, each consisting of 200 individual collective entities. The model and visualization transparency were evaluated with respect to how the system design elements impacted the human operators, operator comprehension, usability, and human-collective performance. Both models and visualizations provided transparency differently. The $M_{2}$ model and Collective visualization combination did not support any of the research questions collectively, but did partially support specific research questions independently. Quantifying system transparency requires evaluating the transparency embedded in the various system design elements, which has not been done in previous analyses, in order to determine how they interact with one another and influence human-collective interactions and performance. Designers must build collective systems that are effective regardless of how how heterogeneous or large the collective size may become, how simple or complex the collective behaviors are, and how real-world use scenarios, such as bandwidth limitations. Models (e.g., intelligent algorithms) that can aid operators to fulfill the sequential decision-making task that require operator influence and collective visualizations that are observable may be more resilient to real-world scenarios, and provide transparency to enable effective human-collective teams. 

\begin{acks}
The US Office of Naval Research Awards N000141613025, N000141210987, and N00014161302, supported this effort. The work of Jason R. Cody was fully supported by the United States Military Academy and the United States Army Advanced Civil Schooling program. 
\end{acks}



\appendix

\end{document}